\documentclass[12pt]{article}

\usepackage{xr}    

\usepackage{natbib}
\usepackage{url} 
\usepackage{amsmath,amssymb,amstext} 
\usepackage[pdftex]{graphicx} 
\usepackage{adjustbox} 
\usepackage{subcaption}

\usepackage{amsmath,amssymb,amstext,amsthm,amsfonts}
\usepackage{dsfont}
\usepackage[pdftex]{graphicx}
\usepackage{caption}
\usepackage[x11names]{xcolor}
\usepackage{dcolumn}
\usepackage{bm}
\usepackage{float}
\usepackage{multirow}
\usepackage{listings} 
\usepackage{xtab}
\usepackage{authblk}

\usepackage{algorithm} 
\usepackage{algorithmicx, algpseudocode}

\usepackage[pdftex,pagebackref=false]{hyperref} 

\newcommand{\code}[1]{\texttt{#1}}

\newcommand*{\R}{\textsf{R}$~$}

\newcommand{\pkg}[1]{\textsf{#1}}

\algblock{Input}{EndInput}
\algnotext{EndInput}
\algblock{Structures}{EndStructures}
\algnotext{EndStructures}

\newtheorem{proposition}{Proposition}

\newcommand{\by}{\times} 
\newcommand{\ve}[1]{\mathbf{#1}}           
\newcommand{\m}[1]{\mathbf{#1}}               
\newcommand{\sm}[1]{\boldsymbol{#1}}   
\newcommand{\tr}[1]{{#1}^{\mkern-1.5mu\mathsf{T}}}              
\newcommand{\iden}[1]{\m{I}_{#1}}

\newcommand*{\diag}{\operatorname{diag}}

\newcommand{\degree}{$^{\circ}$}

\newcommand{\widebar}[1]{\overline{#1}}

\newcommand{\follows}{\sim}  

\newcommand{\given}{~\vert~}
\newcommand{\biggiven}{~\vline~}
\newcommand{\indep}{\bot\hspace{-.6em}\bot}
\newcommand{\notindep}{\bot\hspace{-.6em}\bot\hspace{-0.75em}/\hspace{.4em}}

\newcommand{\bigChi}{\texttt{\Large X}^2}

\newcommand{\bigGstat}{\texttt{\Large G}^2}

\newcommand{\mutInf}{\texttt{\Large I}}
\newcommand{\Hyp}{\mathcal{H}} 
\newcommand{\samp}[1]{\mathcal{#1}}

\DeclareMathOperator*{\argmax}{arg\,max}

\newcommand*{\intersect}{\cap}
\newcommand*{\union}{\cup}

\newcommand{\set}[1]{\mathcal{#1}}
\newcommand{\Nset}[1]{[#1]}
\newcommand{\cardinality}[1]{|{#1}|}
\newcommand{\field}[1]{\mathbb{#1}}
\newcommand{\Reals}{\field{R}}

\graphicspath{{./img/}}

\definecolor{background-color}{gray}{0.95}
\definecolor{steelblue}{rgb}{0.27, 0.51, 0.71}
\definecolor{brickred}{rgb}{0.8, 0.25, 0.33}
\definecolor{bluegray}{rgb}{0.4, 0.6, 0.8}
\definecolor{amethyst}{rgb}{0.6, 0.4, 0.8}

\hypersetup{
    plainpages=false,       
    unicode=false,          
    pdftoolbar=true,        
    pdfmenubar=true,        
    pdffitwindow=false,     
    pdfstartview={FitH},    
    pdftitle={Dependence\ detection,\ display},    
    pdfauthor={Chris Salahub},    
    pdfsubject={Statistics},  
    pdfnewwindow=true,      
    colorlinks=true,        
    linkcolor=steelblue,         
    citecolor=brickred,        
    filecolor=magenta,      
    urlcolor=cyan           
}

\def\spacingset#1{\renewcommand{\baselinestretch}%
{#1}\small\normalsize} \spacingset{1}

\begin{document}


\newcommand{\blind}{0}

\if0\blind
{
  \title{\bf Recursive random binning to detect and display pairwise dependence}
  \author{Chris Salahub  and  R.Wayne Oldford \\
    Department of Statistics and Actuarial Science\\ University of Waterloo}
  \maketitle
} \fi

\if1\blind
{
  \bigskip
  \bigskip
  \bigskip
  \begin{center}
    {\LARGE\bf  Recursive random binning to detect and display pairwise dependence} 
\end{center}
  \medskip
} \fi

\bigskip
\begin{abstract}
Random binnings generated via recursive binary splits are introduced as a way
to detect, measure the strength of, and to display the pattern of association between any two variates, whether one or both are continuous or categorical.
This provides a single approach to ordering large numbers of variate pairs by their measure of dependence and then to examine any pattern of dependence via a common display, the departure display (colouring bins by a standardized Pearson residual).
Continuous variates are first ranked and their rank pairs binned.  The Pearson's goodness of fit statistic is applicable but the classic $\chi^2$ approximation to its null distribution is not.  Theoretical and empirical investigations motivate several approximations, including a simple $\chi^2$ approximation with real-valued, yet intuitive, degrees of freedom.  Alternatively, applying an inverse probability transform from the ranks before binning returns a simple Pearson statistic with the classic degrees of freedom.  Recursive random binning with different approximations is compared to recent grid-based methods on a variety of non-null dependence patterns; the method with any of these approximations is found to be well-calibrated and relatively powerful against common test alternatives.
Method and displays are illustrated by applying the screening methodology to a publicly available data set having several continuous and categorical measurements of each of 6,497 Portuguese wines.  The software is publicly available as the \R package \pkg{AssocBin}.
\end{abstract}

\noindent%
{\it Keywords:}  association, visualization, exploratory data analysis, scagnostics, test of independence, goodness of fit, rank based statistics, copulas
\vfill

\newpage

\spacingset{1.75} 

\section{Introduction}
\label{sec:intro}

Modern data present enormous logistical and comprehension challenges to data exploration.  
There can be tens, hundreds,  even thousands, of variates of all types -- continuous real-valued, nominal, binary, or ordered categorical.  Almost any pattern observed could be of scientific interest, potentially involving any number of variates (marginally, jointly, or conditionally).  
Consequently, practical data exploration is initially
constrained to search for patterns in only one or two variates at a time. For a small number, $d$, of variates  all $d(d-1)/2$ scatterplots might be examined individually; for a moderate number, $d \le 30$, all scatterplots can be examined at once in a scatterplot matrix; for a moderately large number, $d \le  60$, the ``zenplots'' of \cite{hofertoldford2020zigzag} can lay out all $1770$ (when $d = 60$) scatterplots for examination in a single display.
For larger numbers, it may still be possible to present all pairs at once in some form, but the analysis becomes little more than a visual search for the most unusual patterns  (e.g.,
\citealp{hofert2018visualizing} describe a case with $d=465$ where zenplots arrange all 107,880 scatterplots in a 164 page document that can be usefully scanned in about thirty minutes).

A  time honoured solution (since \citealp{TukeyTukeyPaper2}) is to compute some measure of interest for every pair and to display only those pairs having values that are most, or least, ``interesting''. 
These include classical statistics such as Pearson's correlation as a measure of the nearness of the data to a line or  Spearman's rho as a measure of nearness to an arbitrary monotonic relationship. More complex functional dependencies could be measured using nonparametric methods such as alternating conditional expectations (\citealp{ACE85}) or principal curves (\citealp{PrincipalCurves89}). 
Much research has focused on developing various association measures tuned to some dependency type of interest (e.g., see \citealp{liuetal2018kernel} for a review; or the methods compared in \citealp{reshefetal2018empirical}).
\cite{TukeyGraphicsEDA85}  proposed scatterplot diagnostic measures (scagnostics), designed to detect interesting visual features of a scatterplot such as point clusters, the presence of outliers, and whether the point cloud appears to be convex or not; these and other scagnostic measures were given graph-theoretic implementation by  \cite{wilkinson2005graph}.
Any, or all,  of the above measures could be computed over all pairs and used to select scatterplots worth further exploration.

In what follows, we propose the $p$-value from a chi-squared goodness of fit statistic based on novel recursive binnings of the ranked data as a measure of interest.  
The $p$-values, from small to large, order the pairs from most to least evidence against the hypothesis of independence.  
Moreover, and in contrast to many proposed grid-based methods (e.g., \citealp{reshef2011MIC, helleretal2016consistent, caoetal2021improved}), the binning is data-independent so that a $\chi^2$ approximation holds for the $p$-value under the null hypothesis.
The measure is entirely agnostic about the form of the dependence, if any, between the pair of variates and the $p$-value provides a common scale on which to make comparisons, whatever the combination of variate type.
Indeed, by choosing the bins randomly, recursive binning simultaneously achieves the power that maximizing methods strive for while keeping a null distribution amenable to a simple $\chi^2$ approximation.
A simple graphic, similar to the residual mosaic plots of \cite{FriendlyMosaic94} for categorical data, reveals the structure of any observed dependence.

The next section describes the goodness of fit test for a pair of variates of either type -- continuous or categorical.  The proposed recursive binning method is described in Section \ref{sec:recursiveRandomBinning} -- including a discussion of different data independent stop criteria in Section \ref{sec:StopCriteria} -- and demonstrated on several non-null distributions in Section \ref{sec:dataConfigs}. These non-null configurations also aid in a demonstration of the proposed display of departures from independence in Section \ref{sec:graphic}. The resulting statistic's null distribution for different binning parameters is investigated in Section \ref{sec:null_dist}, resulting in a simple approximation for the goodness of fit test of randomly binned ranks that has general application. 
Recursive random binning is compared to two other state-of-the-art binning methods in \ref{sec:power}.
Section \ref{sec:wineEx} illustrates the method on some real data using the methods in the \R{} package \pkg{AssocBin} available now on CRAN (\citealp{AssocBin, Rlang}).
The paper closes with a summary discussion of the method and display as well as possible extensions.

\section{Testing independence}
\label{sec:independence}

Suppose the independence between two random variates, $X$ and $Y$, is to be tested based on $N$ observed pairs $(x_1, y_1), \ldots, (x_N, y_N)$ independently sampled from the joint distribution $F_{X, Y}(x,y)$.   Either one, or both, of $X$ and $Y$ could be categorical or continuous random variates.  

To assess the goodness of fit of a statistical model to observed data,  a universally applicable method is Pearson's $\chi^2$-statistic 
\[ 
\bigChi = \sum_{k = 1}^K  \frac{(o_{k} - e_{k})^2}{e_{k}}
~~~~~~~~~~~~ 
\left( \text{or, equivalently, } ~~ \bigChi  = N  \sum_{k = 1}^K  \frac{(f_{k} - p_{k})^2}{p_{k}} \right)
\]
where $o_k$ and $e_k$ are, respectively, the frequencies observed and expected (under the assumed model) of  $N$ observations distributed over $K$ cells, or categories defined in advance.  Parenthetically, it is written in terms of the sample size $N$, the fractions $f_k$ observed, and proportions $p_k$ expected. If the model hypothesis holds and $N$ is large, $\bigChi$ will be approximately distributed as a
$\chi^2$ random variate with degrees of freedom equal to $K$ less the number of constraints on the expected values.

When $X$ and $Y$ are both categorical, having $I$ and $J$ categories respectively, their frequencies may be arranged as an $I \by J$  contingency table to give the classic 
\[ 
\bigChi = \sum_{i = 1}^I  \sum_{j = 1}^J  \frac{(o_{ij} - e_{ij})^2}{e_{ij}}
 ~~~~~~~~~~~~
 \left( \text{or, equivalently, } ~~ \bigChi  = N  \sum_{i = 1}^I  \sum_{j = 1}^J   \frac{(f_{ij} - p_{ij})^2}{p_{ij}} \right)
 \]
having a $\chi^2_{(I-1)(J-1)}$ distribution under the null hypothesis  $\Hyp_0: X \indep Y$.

When both $X$ and $Y$ are absolutely continuous,  probability integral transforms  (viz.,   $U = F_X(x)$ and $V = F_Y(y)$) will turn each into a Uniform$(0,1)$ random variate.  A scatterplot of observed pairs $(u_1, v_1), \ldots, (u_N, v_N)$ should appear as uniform scatter on the unit square when $\Hyp_0: U \indep V$ holds, or equivalently $\Hyp_0: X \indep Y$.
Unfortunately, both $F_X(x)$ and $F_Y(y)$ are unknown, rendering the transform unavailable.  
A practical alternative is to use the ranks instead.  That is, plot the pairs $(s_1, t_1), \ldots, (s_N, t_N)$ where $s_i$ is the rank of $x_i$ amongst the $x$s, and $t_i$ that of $y_i$ amongst the $y$s.

\begin{figure}[!h]
	\begin{center}
		\begin{tabular}{cccc}
			\includegraphics[scale = 0.5]{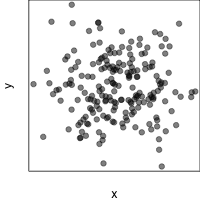} & 
			\includegraphics[scale = 0.5]{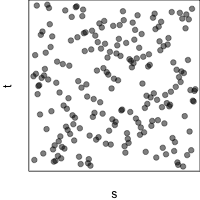} & 
			\includegraphics[scale = 0.5]{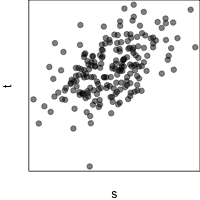} & 
			\includegraphics[scale = 0.5]{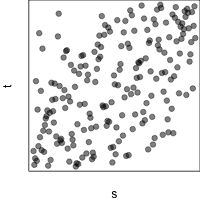} \\
			{\footnotesize (a) Independent $(x, y)$} & 
			{\footnotesize (b) Independent $(s, t)$} & 
			{\footnotesize (c) Dependent $(x, y)$} & 
			{\footnotesize (d) Dependent $(s, t)$}
		\end{tabular}
		\caption{Comparing scatterplots of data $(x,y)$ versus of ranks $(s,t)$. }
		\label{fig:plotRanksExample}
	\end{center}
\end{figure}

Figure \ref{fig:plotRanksExample}
provides examples of the scatterplots of the raw data and their ranks when  $X \indep Y$ and when $X \notindep Y$.
For independent data, both $X$ and $Y$ are standard normal; for the dependent data, they are jointly normal with $\rho = 0.5$.
As can be seen, a uniform scatter in the plot of the paired ranks $(s, t)$ is indicative of independence.

\begin{figure}[!h]
	\begin{center}
		\begin{tabular}{cccc}
			\includegraphics[scale = 0.5]{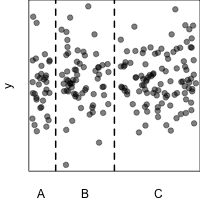} & 
			\includegraphics[scale = 0.5]{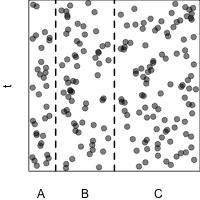} & 
			\includegraphics[scale = 0.5]{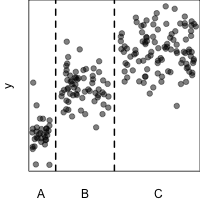} & 
			\includegraphics[scale = 0.5]{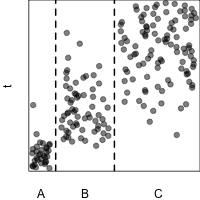} \\
			{\footnotesize (a) Independent $(x, y)$} & 
			{\footnotesize (b) Independent $(x, t)$} & 
			{\footnotesize (c) Dependent $(x, y)$} & 
			{\footnotesize (d) Dependent $(x, t)$}
		\end{tabular}
		\caption{Categorical $X \in \{A, B, C\}$ -- no ranks or $s = x$;  Continuous $Y$ with ranks $t$.}
		\label{fig:plotRanksMixExample}
	\end{center}
\end{figure}

Figure \ref{fig:plotRanksMixExample}
shows the case when $X$ is categorical, taking values `A', `B', or `C'  with probabilities $\frac{1}{6}$, $\frac{1}{3}$,  and $\frac{1}{2}$, respectively, and $Y$ is continuous.
The distances between dashed vertical lines indicate the proportion for each value of $X$ which are displayed with $X$ located uniformly within category.  Only the continuous $Y$ is ranked.  For the independent case, $Y$ values are standard normals; for the dependent case the $Y$ means depend on $X$.

In this case, independence is indicated when the ranks $t$ of $Y$ appear to be uniformly scattered for each and every value of $X \in \left\{A, B, C \right\}$.
When, as in these displays, the horizontal locations are deliberately uniformly distributed within category, independence appears as uniform scatter across the unit square.
Again, lack of uniformity over the whole vertical  range  is easily seen in the dependent ranks of Figure \ref{fig:plotRanksMixExample}(d) while the ranks of Figure \ref{fig:plotRanksMixExample}(b) appear to be uniformly distributed.

When at least one of $X$ or $Y$ is continuous, as in the last two cases, bins must be defined into which the ranks can be placed and counted to produce Pearson's $\chi^2$-statistic.
When both
$X$ and $Y$ 
are continuous, as in Figures \ref{fig:plotRanksExample}(b) and (d), essentially any tessellation of the plane will do.
When only one is continuous, as in Figures \ref{fig:plotRanksMixExample}(b) and (d),  the ranks need only be binned separately for each value of the categorical variate. In either case, $e_k = Np_k$ and $p_k$ is proportional to the area of bin $k$, but the distributions differ under independence as discussed in Appendix \ref{app:paired_ranks}. 

\subsection{Defining the bins}
\label{sec:binDefs}
Different problems suggest different binning strategies depending on what departure from independence is of interest. 

\begin{figure}[!h]
	\begin{center}
		\begin{tabular}{cccc}
			\includegraphics[scale = 0.5]{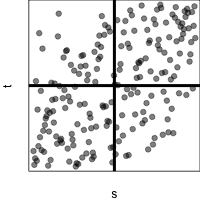} & 
			\includegraphics[scale = 0.5]{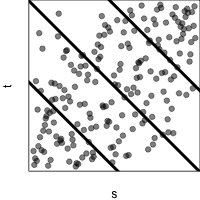} & 
			\includegraphics[scale = 0.5]{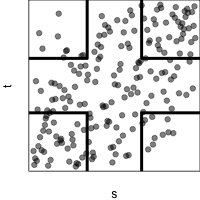} & 
			\includegraphics[scale = 0.5]{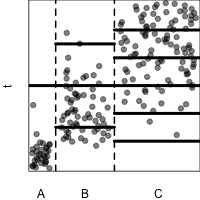}  \\
			{\footnotesize (a) Simple grid} & 
			{\footnotesize (b) Diagonal bins} & 
			{\footnotesize (c) Isolated corners} & 
			{\footnotesize (d) Bin within category}
		\end{tabular}
		\caption{Possible bins:  Continuous $Y$ and $X$ in (a), (b), (c); categorical $X$ in (d).}
		\label{fig:plotRanksBinExample}
	\end{center}
\end{figure}

One simple definition would be a grid of squares as in Figure \ref{fig:plotRanksBinExample}(a).
For each of the four bins, $p_k = \frac{1}{4}$, and $\bigChi$ would have $e_k \ge 5$ whenever $N \ge 20$.  For much larger $N$, the bins could be made smaller still, limited only by rules such as $e_k \ge 5$ for all $k$.
Algorithmically, one could halve the width of each square until $e_k$ is small enough -- giving a total of $k = 2^{2h}$ bins for $h$ halvings of each axis.
For example, to have $e_k$ as close to $c \le e_k$ for some $c \ge 5$, would require $p = \lfloor \frac{1}{2} \log_2 \left(N / c\right) \rfloor$ or, equivalently, $N \ge c 2^{2h}$.  While finer grids might be more sensitive to departures from independence, such sensitivity comes at increased computational cost.

Of course, neither equiprobable nor common shaped bins are required, as Figures \ref{fig:plotRanksBinExample}(b) and (c) show.
All three bin structures of Figures \ref{fig:plotRanksBinExample}(a-c) produce a large $\bigChi$  
and strong evidence against $\Hyp_0$ for the same data.  Figure \ref{fig:plotRanksBinExample}(d) shows a case when $X$ is categorical with bins selected within category to have equal probabilities across categories which also finds strong evidence against $\Hyp_0$.
Were different bins chosen, evidence against $\Hyp_0$ may or may not have been found (e.g., remove the horizontal line in Figure  \ref{fig:plotRanksBinExample}(a) or the two short diagonal lines in  Figure \ref{fig:plotRanksBinExample}(b)).

\begin{figure}[!h]
	\begin{center}
		\begin{tabular}{cccc}
			\includegraphics[scale = 0.5]{depNormsRanksTailDependencyBins.png} & 
			\includegraphics[scale = 0.5]{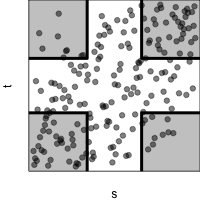} & 
			\includegraphics[scale = 0.5]{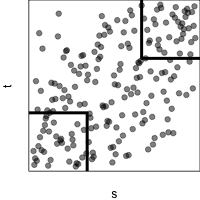} & 
			\includegraphics[scale = 0.5]{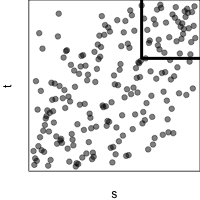}  \\
			{\footnotesize (a) Corners and middle} & 
			{\footnotesize (b) Corners versus middle} & 
			{\footnotesize (c) Tail dependence} & 
			{\footnotesize (d) Upper tail dependence}
\\
			{\footnotesize  $K= 5$;  $p = 1.5 \times 10^{-6}$} & 
			{\footnotesize  $K= 2$;  $p = 0.76$} & 
			{\footnotesize  $K = 3$;  $p = 1.1 \times 10^{-5}$} & 
			{\footnotesize  $K = 2$; $p = 1.5\times 10^{-4}$}
		\end{tabular}
		\caption{Bin choice depends on interest and $p$-value on bin choice:    (a) All five bins;  (b) Two bins (one grey, one not); (c) Tails versus the rest; (d) Upper tail versus rest.}
		\label{fig:plotTailDependenciesExample}
	\end{center}
\end{figure}

Bin design strategy can be tailored to the kind of departure from $\Hyp_0$ it is important to detect.
Isolating corners as in
Figure \ref{fig:plotRanksBinExample}(c), for example,  allows patterns in extreme values to be detected by $\bigChi$.
Figure \ref{fig:plotTailDependenciesExample} 
shows some variations on this binning strategy together with the number of bins, $K$, and a $p$-value produced using $\bigChi \follows \chi^2_{K-1}$.  The five bins of Figure  \ref{fig:plotTailDependenciesExample}(a) produce a very significant result whereas the two bins consisting of the four corners and the middle in Figure  \ref{fig:plotTailDependenciesExample}(b) do not.  The low concentration of points in the top left and bottom right corners is balanced against the high concentrations in the top right and bottom left corners -- the observed count is near the expected count in the four corners combined but not individually.
Figures \ref{fig:plotTailDependenciesExample}(c) and (d) show it is enough to isolate the high concentration corners, either both or even just one.  This might be done, for example, when detecting tail dependency as in copula models for financial series \citep[e.g., see][]{hofert2018visualizing}.
Note also that Figures \ref{fig:plotTailDependenciesExample}(d), (c), and (a) constitute a series of nested binnings whose $p$-values become smaller as the number of bins increase.  The evidence against uniformity (and hence independence) increases as more areas of high and low concentration are isolated.

In practice, one is rarely certain which departures from uniformity might arise, what size of area they might occupy, and where those areas might be on the unit square.
This suggests that bins not be so large as to have both high and low concentration areas within the same bin (e.g., as in Figure \ref{fig:plotTailDependenciesExample}b) so that smaller, and hence more, bins might be preferred.
It also suggests that the binning not be too predictable in that any tessellation can fail on non-uniform patterns that match it  \cite[e.g., the `X' patterns of Figure \ref{fig:rBEX_patterns} introduced by][can be difficult for regular grids to detect]{zhang2019BET}.
Chances of a non-uniform (possibly regular) pattern matching a bin tessellation is reduced if the tessellation is irregular and randomly determined, and if the tessellation is chosen anew for each pair of variates whose independence is being assessed.
These considerations suggest using a random construction of many bins (which tessellate the unit square) that is different for every pair of variates.

\section{Recursive random binning}
\label{sec:recursiveRandomBinning}
We propose a recursive random binning where at each step of the recursion every bin from the previous step is split into two, either horizontally or vertically, at a position randomly chosen along the horizontal or vertical edge of the bin.  
(Other bin shapes might be chosen, but rectangles aligned with the axes have the advantage of simplicity.)
Which side to split could be chosen with equal probability, or, the recommended default here, the longer side is always split resulting in ``squarified'' bins -- rectangular bins that are more nearly square.   

\begin{figure}[!h]
	\begin{center}
		\begin{tabular}{cccc}
			\includegraphics[scale = 0.5]{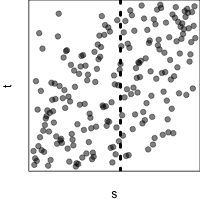} & 
			\includegraphics[scale = 0.5]{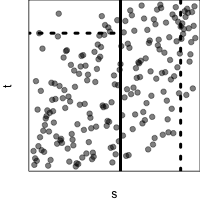} & 
			\includegraphics[scale = 0.5]{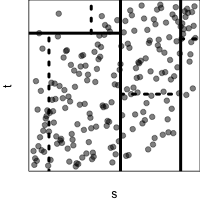}   & 
			\includegraphics[scale = 0.5]{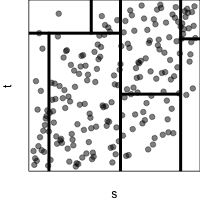}  \\
			{\footnotesize (a) Step 1 } & 
			{\footnotesize (b) Step 2} & 
			{\footnotesize (c) Step 3} & 
			{\footnotesize (d) Final bins}
		\end{tabular}
		\caption{Recursive random binning.  Splits at each step are shown as dashed lines, previous steps bin boundaries are shown as solid.  Stopping after three steps yields $K = 8$ final bins.}
		\label{fig:plotRecursiveRandomSplits}
	\end{center}
\end{figure}

Figure \ref{fig:plotRecursiveRandomSplits}
shows three recursive steps where splits occur with equiprobability on either side. 
The first step splits the starting square into two at a random location along one of the randomly selected $s$ or $t$ axes.   
In the next step, each of the two bins are split randomly in the same fashion.  
At every step,  every bin from the previous step has an edge randomly selected --  horizontal $s$ or vertical $t$ -- and then is split at a randomly chosen location along that edge. 
Figure  \ref{fig:plotRecursiveRandomSplits}(c) shows the results at step three.

The principal parameter in this algorithm is the \textit{depth} of the recursion,  or number of splitting steps taken, used to define the final bins. 
This could be set \textit{a priori} as with the regular grid halving algorithm  -- for depth $d$, there would be $2^d$ bins here though, many fewer than the number of the grid algorithm. 
Alternatively, it could be chosen based on the bin areas -- perhaps to ensure $e_k = N p_k > 5$ for all bins \citep[e.g., as traditionally recommended, see][]{cochran1952chi2}.
In the latter case, one might choose to reject a split (and randomly choose another if possible) whenever it would result in one of the two new bins having too small an $e_k$.
The final binning of \ref{fig:plotRecursiveRandomSplits}(d), for example, would then be rejected since the top right square of the final binning has $e_k < 5$; that split of step 3 would be repeated (or designed such as) to ensure an $e_k \ge 5$ for each new bin.
A consequence of such a limit is that only large enough bins will be split, resulting in a final number of bins that is not necessarily a power of two.
Whatever the stopping criterion, it must be made independent of the observed data to ensure the binning is not data-dependent.

As simple as this algorithm appears to be, there are still several choices to be made. 

\subsection{Data independent stop criteria}
\label{sec:StopCriteria}
Three different parameters describe the final binning: $K$, the number of bins, $d$, the maximum depth of any bin, and the smallest allowable expected count $e_k$.  Stopping criteria for the recursive splitting could be any one, or any combination, of these.

\begin{figure}[!h]
	\begin{center}
	\includegraphics[scale = 0.25]{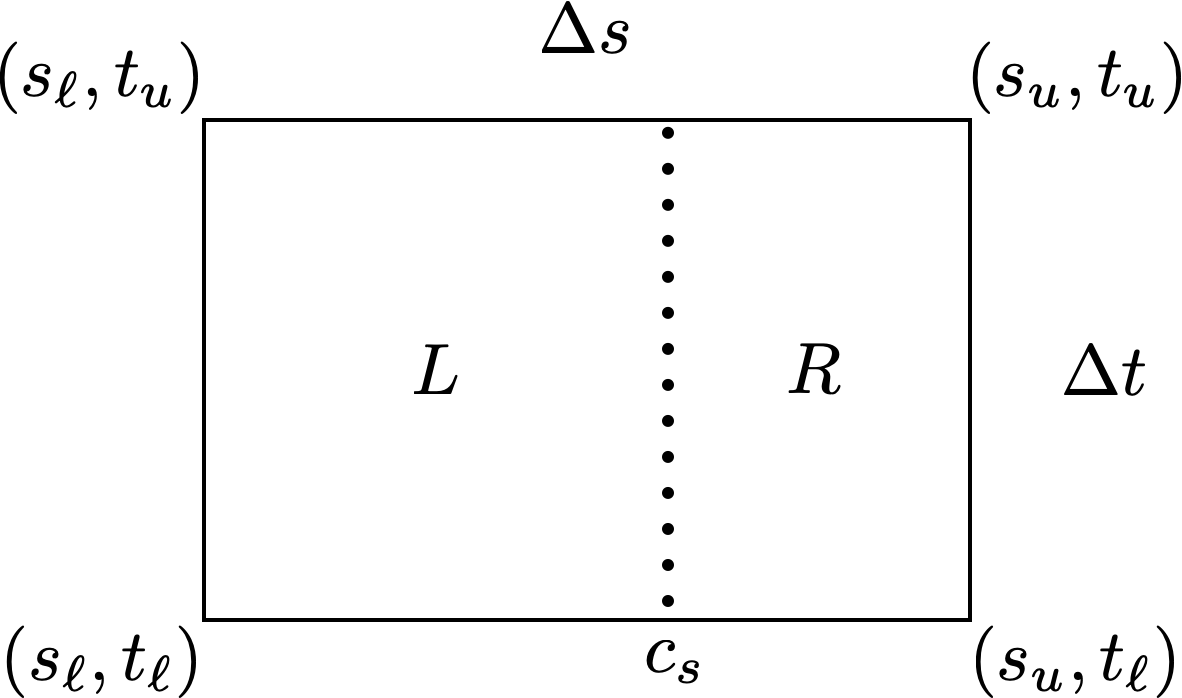}
        	\caption{{\footnotesize Splitting a bin horizontally at $c_s$.  
	Bin is of width  $\Delta s = s_u - s_\ell$ and height $\Delta t = t_u - t_\ell$.}}
\label{fig:boundingExpected}
	\end{center}
\end{figure}

For Pearson's $\chi^2$, it is generally recommended that $e_k \ge 5$ for all $k = 1, \ldots, K$.  To be more general, suppose the restriction is $e_k \ge z$ for some positive integer $z$.
This constrains when and where a bin may be split.
Suppose, for example, that the bin shown in Figure \ref{fig:boundingExpected}
is to be split horizontally at $c_s$ into two smaller bins $L$ and $R$.  Requiring $e_k \ge z$ for all $k$ constrains the area of both smaller bins.   To have $e_L \ge z$ and $e_R \ge z$  the cut location $c_s$ must satisfy
\[s_\ell + \frac{Nz}{\Delta t} \leq c_s ~~ \text{ and } ~~~ c_s \leq  s_u - \frac{Nz}{\Delta t}\]
with $\Delta t = t_u - t_\ell$.
This follows, for example, from the expected number of observations in region $R$ being $e_R = N \times \frac{Area(R)}{N^2} \geq z$ and similarly for $e_{L}$.  Note that, since $s_\ell$, $s_u$, and $c_s$ are all taken to be integer and all bin boundaries are open on the left, $c_s \in \{s_\ell + 1, s_\ell + 2, \ldots, s_u-1\}$.  Restricted to this set, the above restriction can be rewritten as
\[ \max\left(s_\ell + 1, s_\ell + \left\lceil \frac{Nz}{\Delta t} \right\rceil \right) \le c_s ~~ \text{ and } ~~~ c_s  \le \min\left(s_u - 1, s_u - \left\lceil \frac{Nz}{\Delta t} \right\rceil \right)\]
where $\lceil\cdot\rceil$ denotes the ceiling, or smallest integer greater than or equal to its argument. For it to be possible to split the bin (i.e. $s_u - Nz/\Delta t > s_\ell + Nz/\Delta t$) it must be that 
\[ \Delta s \Delta t \ge 2 N z.\]
If the above does not hold, the bin cannot be split (in either direction) as at least one of $R$ or $L$ would have area less than $z$. 
If it does hold, then it can be split horizontally provided $ \left\lfloor \frac{2Nz}{\Delta t} \right\rfloor \ge 1$; in this case, 
$c_s$ is selected randomly from a discrete uniform distribution over the integer set
\[c_s \in \left\{ s_\ell + \left\lceil \frac{Nz}{\Delta t} \right\rceil, \ldots, s_u - \left\lceil \frac{Nz}{\Delta t} \right\rceil \right\}.\]
The analogous result holds when splitting vertically.
The stopping criterion based on making all bins as near to $e_k = z$ would stop splitting a bin whenever $\Delta s \Delta t \leq 2 Nz$ or $\Delta t > 2N z$.

The depth is the number of recursive splits that have been applied to an individual bin.  If every bin has a depth $d$, then the total number of bins is $K = 2^d$. 
However this is not generally reached as a bound on $e_k$ stops splitting at a depth less than $d$ for some bins.
Moreover, to introduce a single data dependency, empty bins are not split any further as this causes no change in $\bigChi$. 
Consequently, $d$ constrains $K \le 2^d$. 

\subsection{Exemplar data configurations}
\label{sec:dataConfigs}
\begin{figure}[!ht]
  \begin{center}
    \begin{tabular}{ccccccc}
      \includegraphics[scale = 0.75]{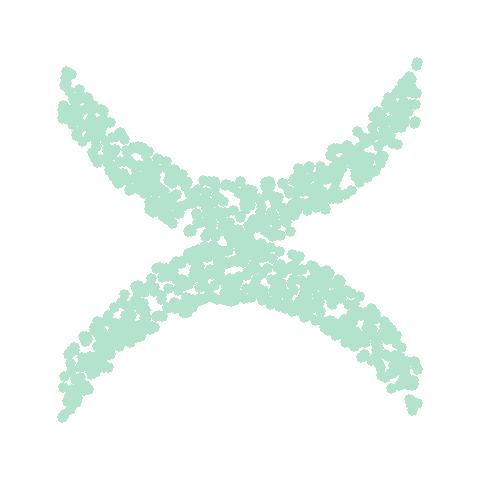} & \includegraphics[scale = 0.75]{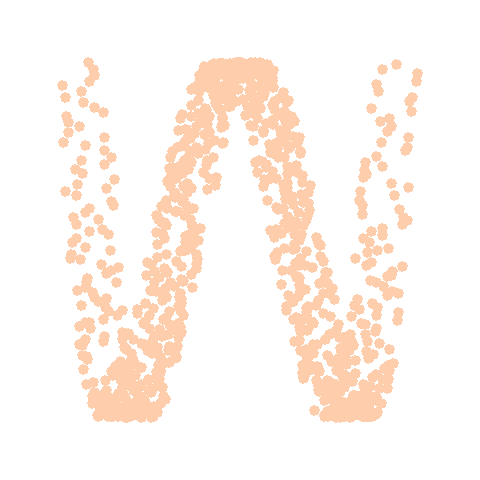} & \includegraphics[scale = 0.75]{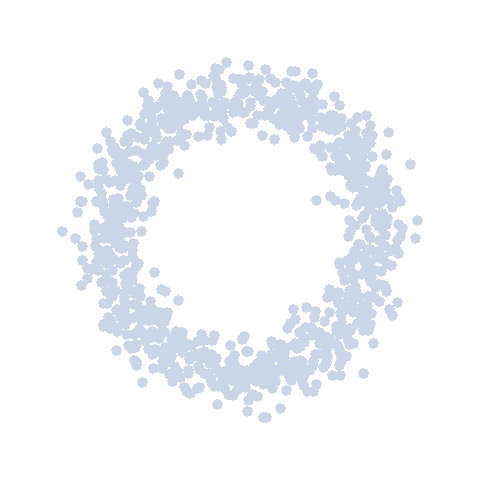} & \includegraphics[scale = 0.75]{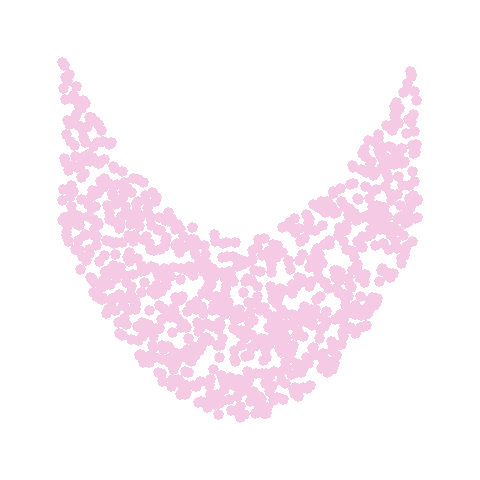} & \includegraphics[scale = 0.75]{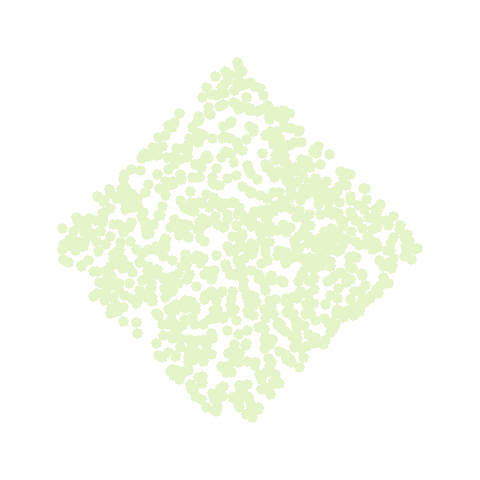} &
      \includegraphics[scale = 0.75]{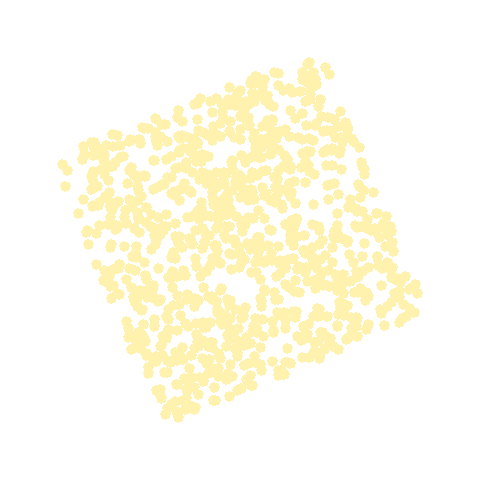} &
      \includegraphics[scale = 0.75]{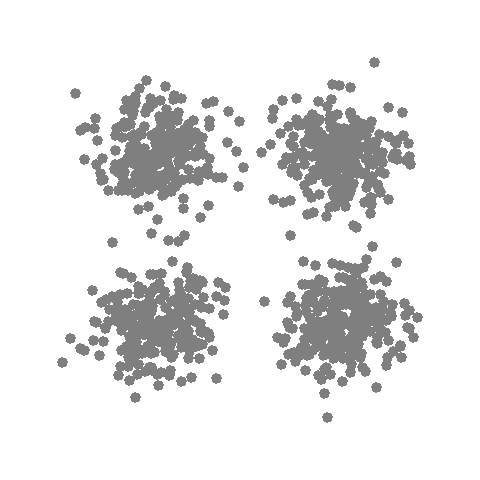} \\
      \includegraphics[scale = 0.75]{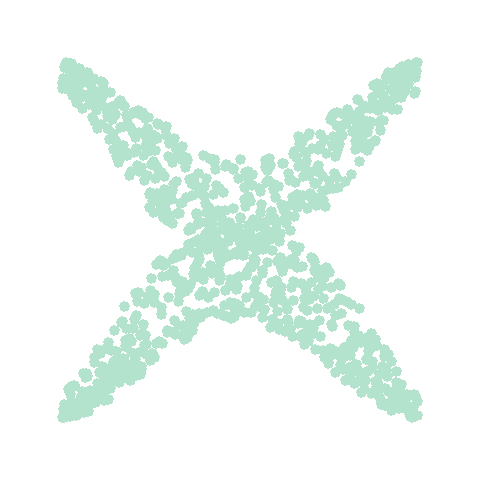} & 
      \includegraphics[scale = 0.75]{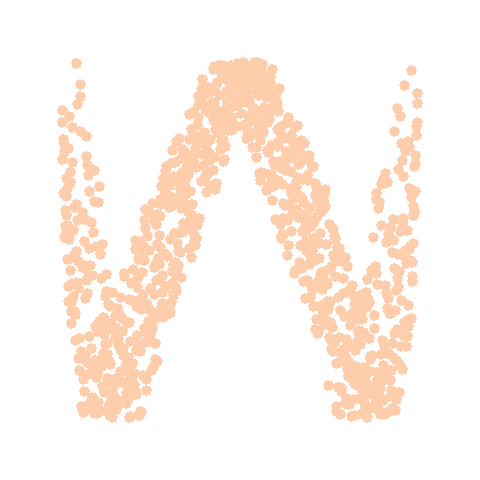} &
      \includegraphics[scale = 0.75]{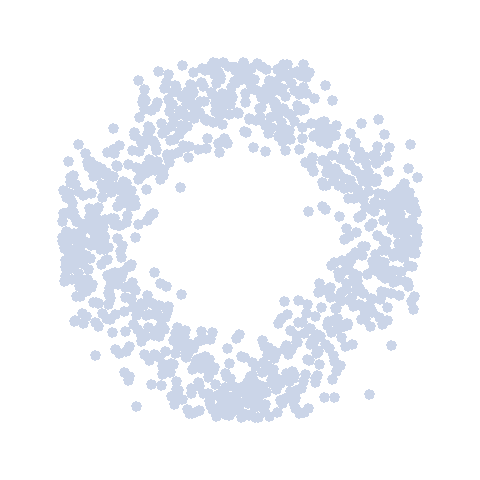} & 
      \includegraphics[scale = 0.75]{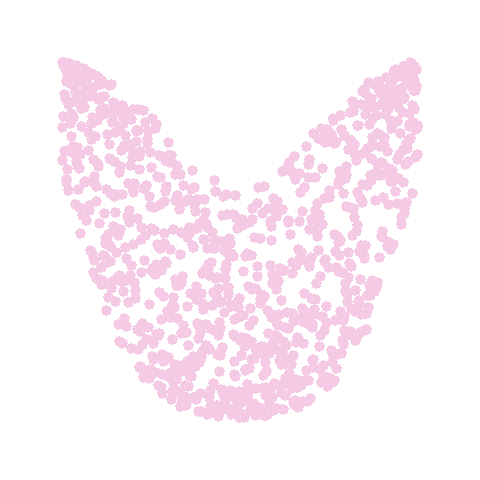} &
      \includegraphics[scale = 0.75]{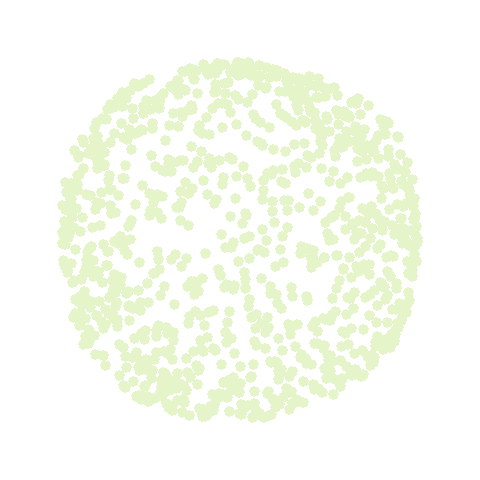} &
      \includegraphics[scale = 0.75]{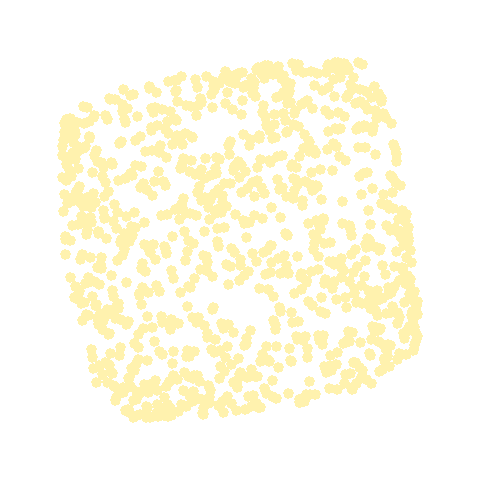} &
      \includegraphics[scale = 0.75]{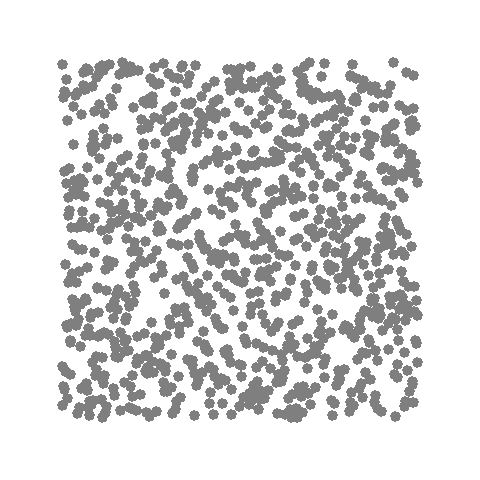}
    \end{tabular}
  \caption{{\footnotesize Six dependent data patterns: cross, wave, ring, saddle, circle, rotated square, and noise (an independent pattern of four clusters), as in  \cite{newton2009introducing}.  
 First row shows $(x, y)$;  second, the ranks $(s, t)$.  }}
  \label{fig:simulatedPatterns}
\end{center}
\end{figure}

To investigate random recursive binning, the algorithm is applied to simulated pairwise data having  $(x, y)$ configurations as shown
in the top row of Figure \ref{fig:simulatedPatterns}
with corresponding $(s, t)$  rank patterns in the bottom row (all configurations  have zero correlation).  
The independent case appears as uniform scatter in the rightmost rank configuration.

\begin{figure}[!h]
\begin{center}
\includegraphics[scale = 0.7]{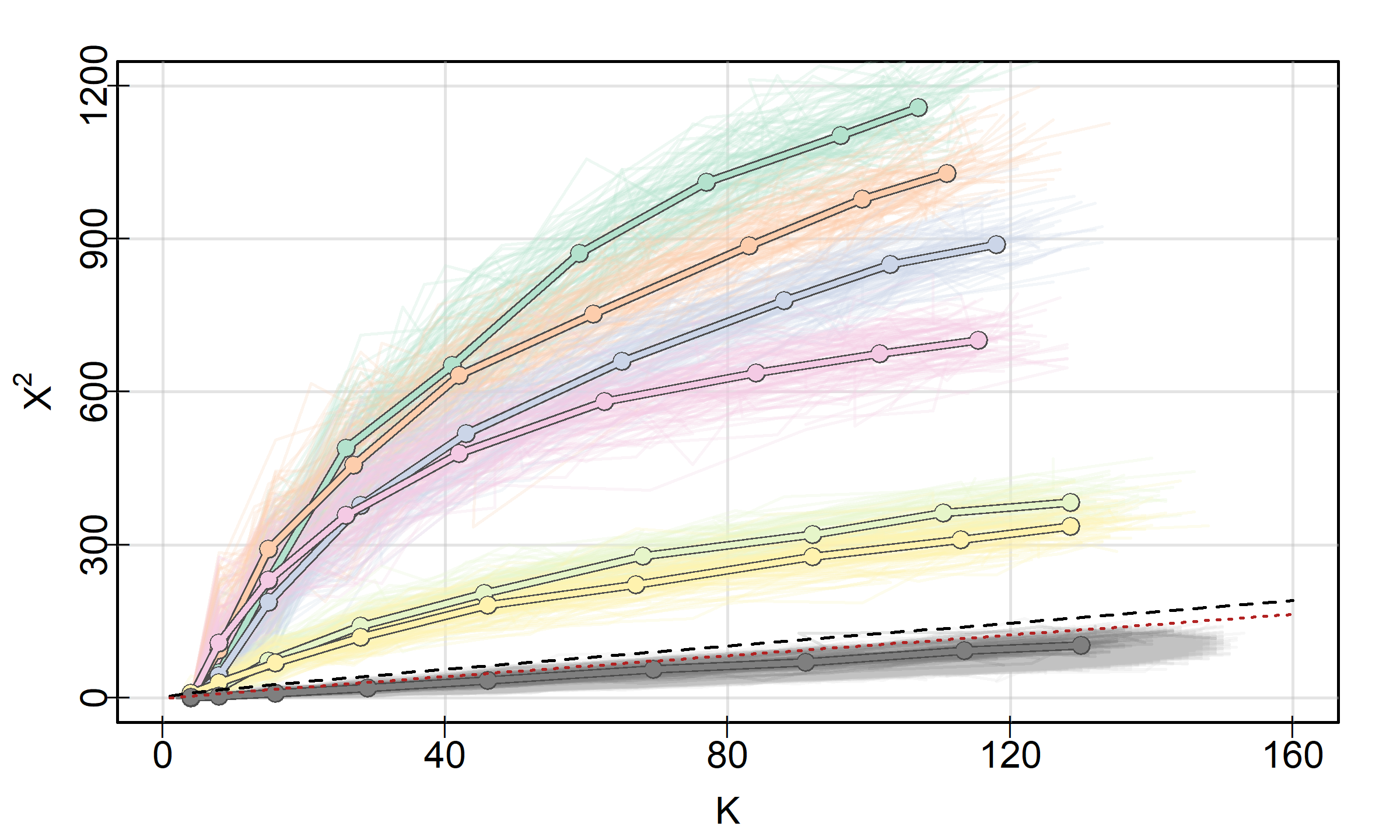} 
\caption{{\footnotesize Recursive random binning of exemplars.  Shown colour coded are 100 replications as paths for each of the seven exemplar data patterns.  The median path of each exemplar is plotted with a thicker line width and, top to bottom, match the left to right order of  Figure \ref{fig:simulatedPatterns}.
Dashed and dotted curves delineate the 0.95 quantile of a $\chi^2_{K-1}$ and of a 
 $\chi^2_{(\sqrt{K} -1)^2}$.}}
\label{fig:simDataBinAllRandPaths}
\end{center}
\end{figure}

For each configuration, 100 samples of $N =$ 1,000 observations were generated.  Recursive random binning was applied to each sample for maximal depths $d \in \{2, 3, \ldots, 10\}$ with minimal expected number  $e_k \ge 5$ for all bins and empty bins left unsplit. 
A new binning was made for every pattern, sample, and depth.
Consequently, no sample and depth pair have the same binning.
Each binning results in a value of $\bigChi$ for its $K \leq 2^d$ bins and the pairs $(K, \bigChi)$ can be connected to form a single path for each sample from any configuration's pattern.
These are plotted, coloured by configuration, for each of that pattern's 100 samples in
Figure \ref{fig:simDataBinAllRandPaths}. 

Overlaid on those 100 curves is a thicker line connecting the coordinates of the median $K$ and median $\bigChi$ values over each configuration and $d$ (shown as filled circles).
As with the paths for each sample, these curves of median pairs are coloured to match their configuration in Figure \ref{fig:simulatedPatterns}.
The 100 individual paths for each configuration are noisy due to random sampling and random binning but the smoother curves given by the paths of the median $(K, \bigChi)$ coordinates provide more easily interpreted summaries.

How large a $K$ can be achieved is peculiar to each configuration, with the null configuration tending to have the greatest number of bins.
This holds even when, as here,  the depths and minimal expected counts are identical across configurations. 
The difference arises because of the choice to leave empty bins whole.
Configurations with more contiguous empty space have a greater chance of having an empty bin created, thereby ending the splitting of that bin at that depth. 
The result is fewer bins for the same maximal depth.  Comparing the rank configuration patterns of Figure \ref{fig:simulatedPatterns} to their median curves in Figure \ref{fig:simDataBinAllRandPaths} bears this out.

The non-null configurations clearly separate from the null (bottom-most) configuration for depths as low as 4, and increasingly do so as depth and number of bins, $K$, increase.  Configurations of similar pattern in Figure \ref{fig:simulatedPatterns} have curves closer together and those patterns which are most obvious have curves farthest from the null configuration.  

Dashed and dotted curves mark the 0.95 quantile of two different $\chi^2$ approximations based on $(K-1)$ and $(\sqrt{K}-1)^2$ degrees of freedom, respectively.
As the curves from the null configuration mostly fall below these, neither approximation would produce strong evidence against $\Hyp_0$ when it is true.  
Moreover, the non-null configurations all fall above these curves indicating small $p$-values which become exceedingly small as the number of bins grows. 

\begin{figure}[!ht]
	\begin{center}
		\begin{tabular}{ccc}
			\includegraphics[scale = 1]{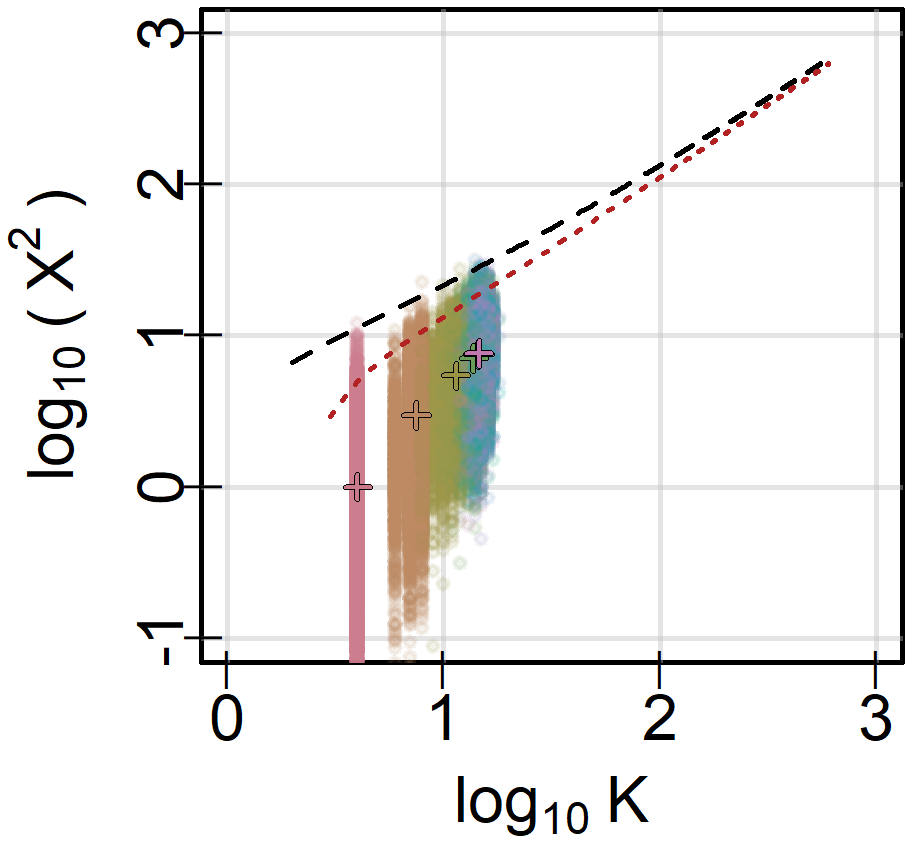} & \includegraphics[scale = 1]{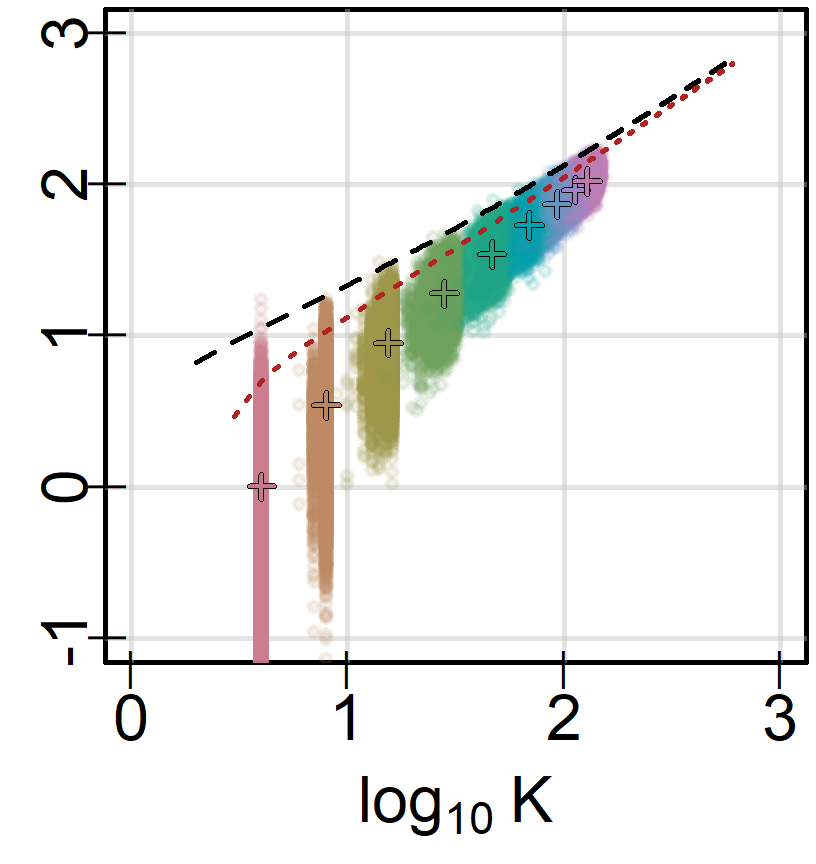} & \includegraphics[scale = 1]{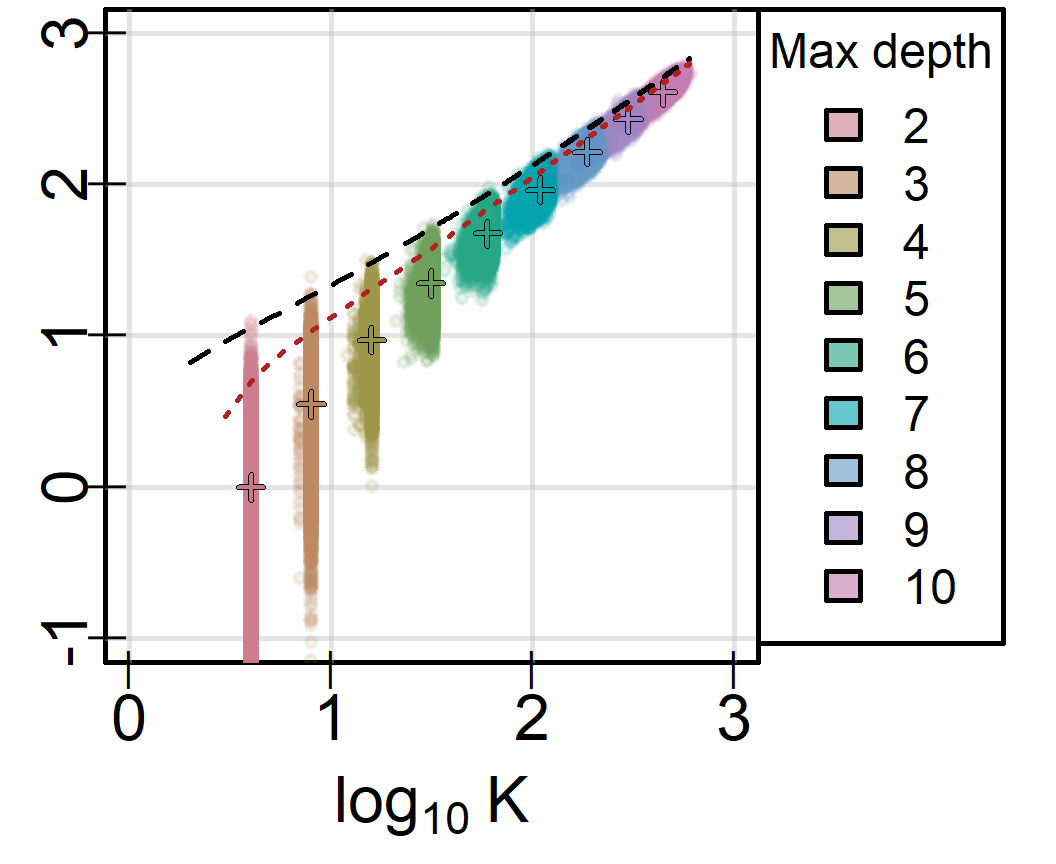} \\
			{\footnotesize (a) $N =$100} & {\footnotesize (b) $N =$1,000} & {\footnotesize (c) $N =$10,000} 
		\end{tabular}
	\caption{{\footnotesize Recursive random binning under $\Hyp_0$. 
	For each depth/colour, 10,000 independent replications are plotted.
	Crosses identify average $K$ and $\bigChi$ for each depth/colour. Dashed  and dotted curves follows the 0.99 quantile for $\chi^2_{K-1}$ and $\chi^2_{(\sqrt{K} -1 )^2}$, respectively.}}
	\label{fig:depthBySampleSizeRand}
	\end{center}
\end{figure}

The two different $\chi^2$ approximations are compared under the null hypothesis in 
Figure \ref{fig:depthBySampleSizeRand}.
There $\bigChi$ values from recursive random binning are plotted against the number of bins for different sample sizes and maximal depths. The total number of bins $K$ will differ within depth due to our binning parameters ($e_k \geq 5$, stop splitting when $n_k = 0$).
Each colour therefore shows the empirical joint density of $(K, \bigChi)$ for samples of size $N$ (on a log-log scale) using recursive random binning under $\Hyp_0$ at that maximal depth.
As can be seen, the null distribution of $\bigChi$ changes with the number of bins for each sample size $N$. 
For reference, the black dashed curve tracks the 0.99 quantile of the $\chi^2_{K-1}$ distribution
and the red dotted curve the same for a  $\chi^2_{(\sqrt{K}-1)^2}$ approximation. 
As can be seen in Figure \ref{fig:depthBySampleSizeRand}, the $\chi^2_{K-1}$ appears to be a poorer approximation than $\chi^2_{(\sqrt{K}-1)^2}$  --  for all sample sizes, all depths, and
all numbers of bins.   

Judging by Figure \ref{fig:depthBySampleSizeRand}, testing independence with random binning at a fixed  level $\alpha$ depending only on the number of bins seems promising.
Critical values from $\chi^2_{(\sqrt{K} -1 )^2}$ appear to be fairly accurate while using the more naive $\chi^2_{K-1}$ critical values would at best be a conservative test of size $\alpha$ under $H_0$.  Section \ref{sec:null_dist} discusses approximation of the null distribution of $\bigChi$ in more detail for recursive random binning (see also Appendix \ref{app:mle_approx}).
 
\subsection{Departure displays}
\label{sec:graphic}
To investigate where, if anywhere, the data do not support the hypothesis of independence, the bins are plotted on the unit square with each one coloured according to its standardized Pearson 
residual, red for positive and blue  for negative, with a saturation increasing in the absolute magnitude \cite[e.g., similar to that proposed by][for mosaic plots of cross-classified data]{FriendlyMosaic94}. 
To emphasize the patterns, colouring begins only when the standardized residuals are large, say greater than 2 in absolute value.  The colouring used here is described in Appendix \ref{app:colouring}.
The resulting display reveals the data's pattern of greatest departures from the null hypothesis.  These departure displays appear best when bins are ``squarified'' as very narrow bins are difficult to discern.

\begin{figure}[!h]
	\begin{center}
		\begin{tabular}{c}
			\includegraphics[scale = 0.8]{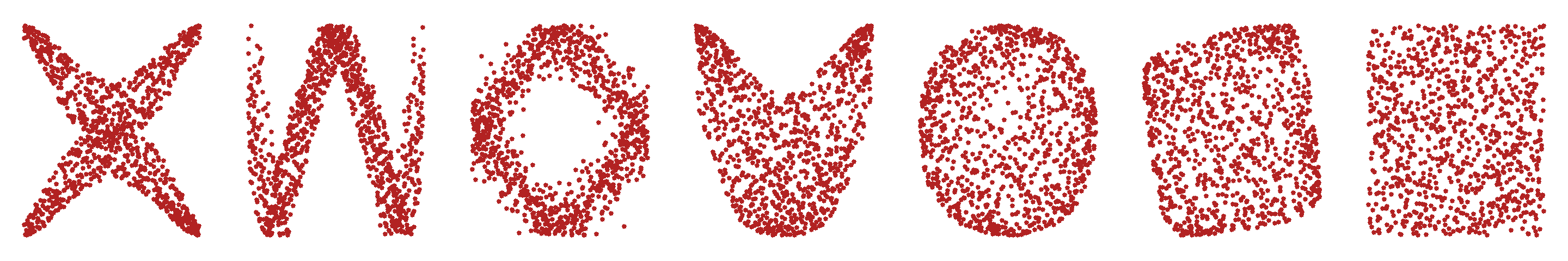} \\
			\includegraphics[scale = 0.8]{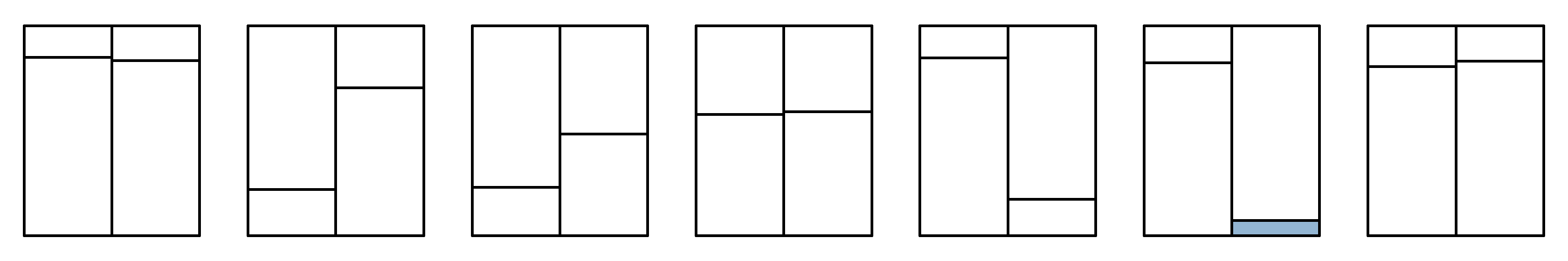} \\
			\includegraphics[scale = 0.8]{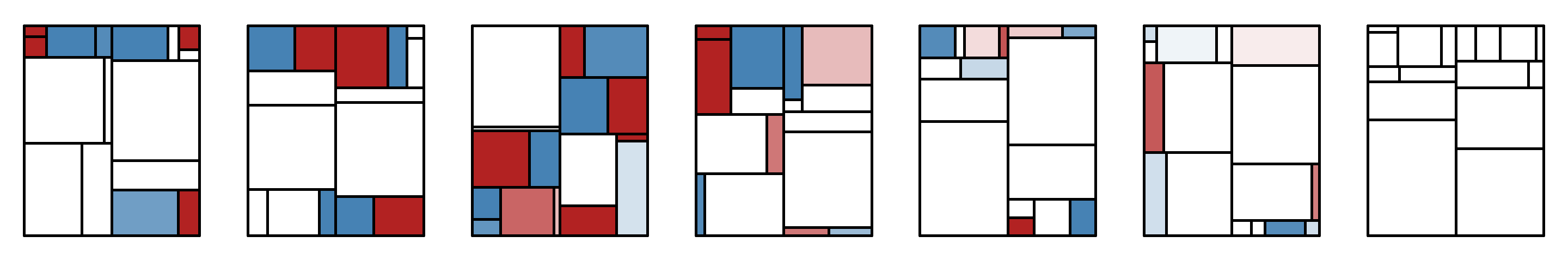} \\
			\includegraphics[scale = 0.8]{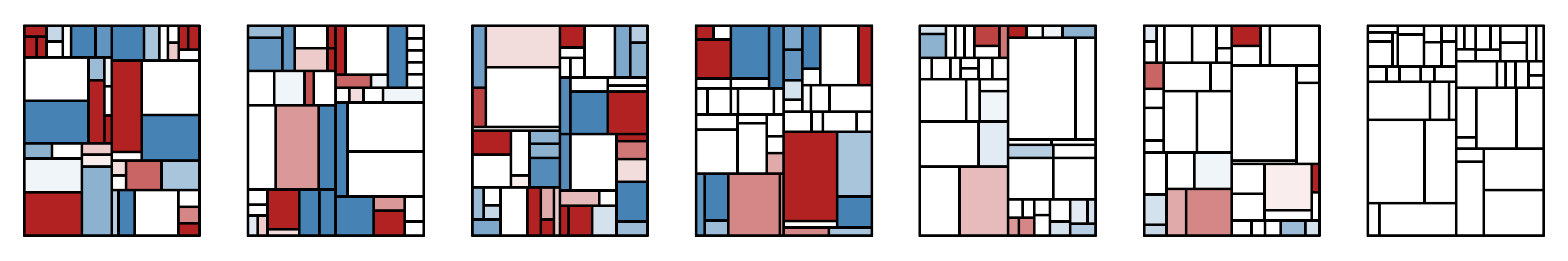} \\
			\includegraphics[scale = 0.8]{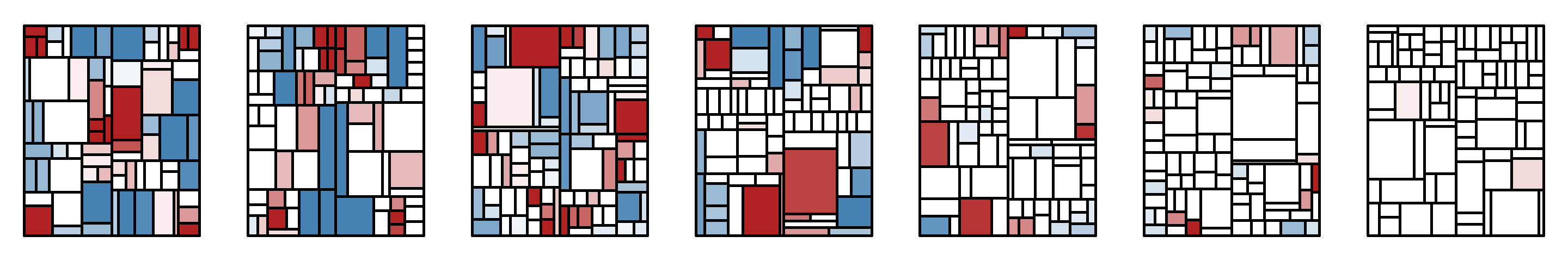} \\
			\includegraphics[scale = 0.8]{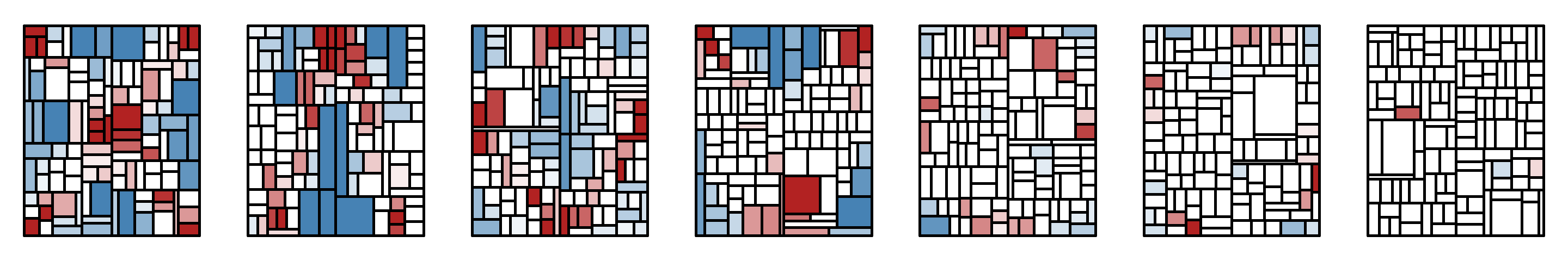} \\
			\includegraphics[scale = 0.8]{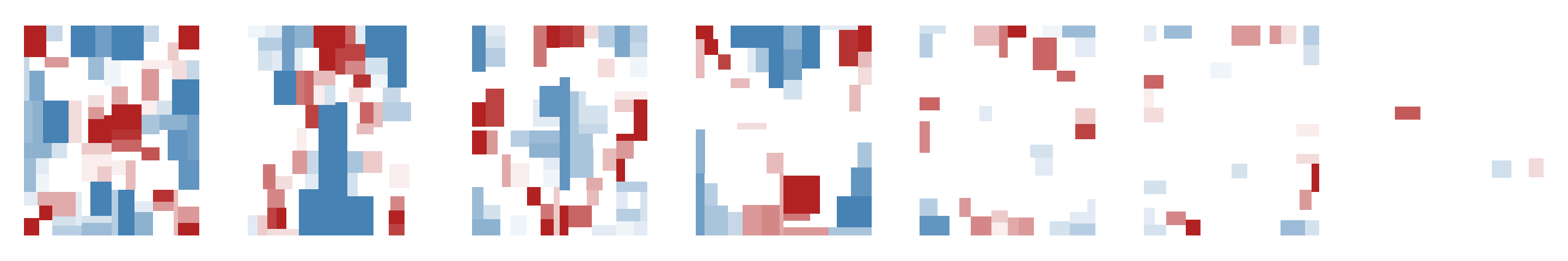} \\
		\end{tabular}
	\end{center}
	\caption{{\footnotesize Departure displays from recursive random binning.  Rank patterns are shown in the first row. 
	Binning depths of 2, 4, 6, 8, and 10 appear within each column for that configuration; the last row is again depth 10 but without bin boundaries.}}
	\label{fig:simDataRandEvol}
\end{figure}

Figure \ref{fig:simDataRandEvol} 
shows departure displays for the exemplar data configurations by column and increasing binning depths by row for one recursive random binning on each configuration.  
As the depth increases, the departures are better detected and the non-null configurations appear -- unusual data concentration as red and paucity as blue.  
For large numbers of bins, marking the boundaries can obscure the pattern; these are removed in the last row for $d = 10$.  Note that the null configuration in the rightmost column shows little departure even when $d = 10$.

\section{On the null distribution of $\bigChi$ when binning ranks}
\label{sec:null_dist}
The bins and counts used to form the $\bigChi$ statistic are based on the paired ranks $(s, t)$ of the data, or, equivalently, on the pseudo-observation pairs $(\frac{s}{N+1}, \frac{t}{N+1})$.  If these could be treated as pairs of independent uniforms (on their corresponding ranges), then under $\Hyp_0: X \indep Y$,  the goodness of fit statistic $\bigChi$ would indeed be approximately $\chi^2_{K-1}$ for $K$ mutually exclusive bins.
However, the ranks are heavily constrained and this approximation is no longer justified.

The null probability for the occurrence of any rank pair, as well as that of the count of rank pairs appearing in any rectangular bin, is derived in detail in Appendix \ref{app:paired_ranks}.  Unfortunately, derivation of the null distribution of $\bigChi$ is challenging
and approximations must be considered.
The following subsections explore a few alternative methods.

\subsection{Permutation distribution}
\label{sec:permutationTest}
The first, and most obvious, is to determine the \textit{exact} permutation distribution for $\bigChi$ -- this is the ``gold standard''.
Under $\Hyp_0$, the vectors $\ve{s} = (s_1, \dots, s_N)$ and $\ve{t} = (t_1, \dots, t_N)$ will be realizations of independent random permutations of $(1, \dots, N)$.
Each of the $N!$ possible rank configurations $(s_1, t_1), \dots, (s_N, t_N)$ gives a value of $\bigChi$ and this empirical distribution is then used to determine a $p$-value for the observed $\bigChi$.  
Unfortunately, this is only practicable for very small $N$.  

Instead, the distribution is estimated by equiprobable sampling of $n$ permutations to produce the $\bigChi$ values.  
To construct each sample, pairs $(x_i, y_i)$ can be taken to be realizations of $X_i  \overset{\text{iid}}{\follows} F_X(x)$ and  $Y_i  \overset{\text{iid}}{\follows} F_Y(y)$ for any continuous distribution functions $F_X(x)$ and $F_Y(y)$ with $X \indep Y$ -- a simple choice is $F_X(x) = F_Y(y) = U(0,1)$.  Rank pairs and then $\bigChi$ are determined for each sample and quantiles of this empirical distribution of $\bigChi$ used to estimate the $p$-value.   

This works well for $n$ large enough (though much less than $N!$); for example, $n =$ 10,000 is used for this purpose in Appendix \ref{app:mle_approx}. 
Even so, when both $n$ and $N$ are large, or when a great many pairs of variates are being investigated at once, the computation can be burdensome (especially when the purpose is that of simple screening).
Direct calculation without such simulation would be of value.
The next few subsections consider some possibilities.

\subsection{The inverse probability integral transform}
\label{sec:pit}
Since the problem is that the rank pairs $(s_i, t_i)$ are not really pairs of uniform random variates, one solution would be to transform them into pairs $(u_i, v_i)$ that are, and  then calculate $\bigChi$ based on these pairs.
This suggests the probability integral transform 
$u_i = F_X(x_i)$  and its inverse $x_i = F^{-1}_X(u_i)$; and similarly for  $v_i$ and $y_i$.
Had $F_X(x)$ and $F_Y(y)$ been available, the pairs $(u_i, v_i)$ would be used and the approximation $\bigChi \follows \chi^2_{K-1}$ would be valid.  
Replacing the distribution functions by their empirical versions led to considering the rank pairs and the inverse transform would lead back to the original $(x_i, y_i)$ pairs which, unfortunately, are not necessarily uniform.

It is well known  that the dependence structure between two random variates is captured by their copula \cite[e.g., see][Chapter 6]{copulaBook}. The original marginal distributions $F_X(x)$ and $F_Y(y)$ could be any continuous distribution functions and are  replaced by marginal uniforms in the copula via the probability integral transform.  
It follows that the observed rank pairs (which determine the empirical copula) could have been generated by realizations of marginally uniform random variates without any change to the underlying dependence structure. 

This suggests transforming the paired ranks to paired uniforms.   
Independently generate $2N$ uniform values $u_1, \ldots, u_N$ and $v_1, \ldots, v_N$ and determine their order statistics
$u_{(1)} \leq u_{(2)} \leq \cdots \leq u_{(N)}$ and $v_{(1)} \leq v_{(2)} \leq \cdots \leq v_{(N)}$.  The pairs 
\[(u_{(s_i)}, v_{(t_i)}) ~~ i = 1, \ldots, N\]
are realizations from a joint distribution whose marginals are uniform.  Moreover, they produce the same rank pairs $(s_i, t_i)$ as the original $(x_i, y_i)$ pairs and hence the same empirical copula -- which is all we have to assess independence.

The pairs $(u_{(s_i)}, v_{(t_i)})$ are now used to produce frequency counts within the bins -- the expected frequencies are unchanged -- and to calculate the $\bigChi$ value.  The null distribution of this $\bigChi$ statistic will be approximately $\chi^2_{K-1}$ as desired.   

At the cost of a single simulation of $N$ pairs,  an approximate $p$-value can be calculated for the data. 
We call this the \textit{pit1} estimate of the $p$-value.

\subsection{A simple asymptotic approach}
\label{sec:asymp}
A more conventional approach to finding an approximate $p$-values would be to determine the distribution of the observed frequencies and use this to determine the distribution of the quadratic form given by $\bigChi$ \cite[e.g., see][Section 6b]{RaoBook73}.
Typically, the vector of (Poisson) standardized frequencies has approximate (asymptotic) multivariate normal distribution with zero mean and a variance-covariance matrix proportional to an orthogonal projection matrix.  The squared length of this vector is equivalent to $\bigChi$ and has approximate (asymptotic) distribution of $\chi^2_d$ where $d$ is the rank of the variance-covariance matrix.

In Appendix \ref{app:paired_ranks}, the count of paired ranks in any $R \times C$ bin of the $N \times N$ lattice on which paired ranks can appear is shown to be a hypergeometric random variate.
As such,  the standard  $N(0,1)$  approximation to each individual Poisson standardized count follows.  
It turns out that the variance-covariance matrix of the vector of observed counts, derived in Appendix \ref{app:paired_ranks}, is not, in this instance,  
an orthogonal projection matrix and so will not have eigen-values that are only either zero or one.

Assuming that the joint distribution of the vector of standardized counts can be approximated by a multivariate normal distribution, its squared length, the statistic $\bigChi$,  will still be a linear combination of $d$ independent $\chi^2_1$ random variates
with coefficients equal to the $d$ non-zero eigen-values of the appropriate variance-covariance matrix (see Appendix \ref{app:simple_approx} for details).   
The distribution is more complicated now because these eigen-values are not all equal to one.

Fortunately, this is not an uncommon expression for goodness of fit statistics \cite[e.g., see][]{UnifiedGeneralChiSq1975} and exact methods for calculating its tail probabilities have long been known  \cite[e.g.,][]{Imhof1961, Davies_1980, Farebrother1984, Farebrother1990}.
Unfortunately, as shown in Appendix \ref{app:simple_approx}, in the present case the values produced are rarely close to the correct values derived  by simulation (with $n=$ 10,000).   
Moreover, common approximations \cite[e.g., ][]{Kuonen_Saddlepoint_1999, LiuEtAl2009}  fare no better.

It would appear that in the present case, the usual multivariate normal approximation is not fit for purpose and will not be considered further.

\subsection{$\chi^2$ approximations}
\label{sec:chi_approx}
It remains true that the $\bigChi$ statistic is the sum of $K$ terms, each of which has distribution that is (asymptotically) approximated by a $\chi^2_1$ distribution.  They are not, however,  independent nor does the vector of their signed square roots  seem to be well approximated by a multivariate normal distribution.
Even so, one would like to imagine that a $\chi^2$ approximation would still be available.

To this end, in Appendix \ref{app:mle_approx}, the null distribution for $\bigChi$ was simulated ($n =$ 10,000) for a variety of different conditions (bin patterns, number of bins, varying sample sizes $N$).  
Examination of the resulting sample histograms suggested that a $\chi^2$ approximation might be achievable in each case.
If not $\chi^2_{df}$ for some degrees of freedom $df$ (not necessarily integer), then $\Gamma(\alpha, \theta)$ for some shape and scale parameters $\alpha$ and $\theta$.

For each combination, the maximum likelihood estimates of the degrees of freedom, $\widehat{df}$, assuming a $\chi^2$ distribution, and the maximum likelihood estimates of the shape and scale parameters, $\widehat{\alpha}$ and $\widehat{\theta}$ assuming a Gamma distribution, were fitted to that empirical distribution.
These were then considered as a function of the number of bins, or blocks $K$, used in the determination of $\bigChi$.
Appendix \ref{app:mle_approx} gives the details.

Gratifyingly, some simple $\chi^2$ approximations turned out to be serviceable.  
Most interesting was that a simple and easy to remember pattern for the approximate degrees of freedom worked surprisingly well. 
Depending on the types of the pairs of variates, the \textit{simple degrees of freedom} for this approximate $\chi^2$ distribution had the following intuitive forms:
\begin{enumerate}
	\item Both $Y$ and $X$ are categorical, resulting in a contingency table of $R$ rows and $C$ columns.
	         The simple degrees of freedom are the classic  
	         \[ df_{simple} = (R - 1)(C - 1)\]
	\item Both $Y$ and $X$ are continuous and the rank pair lattice is tessellated into $K$ rectangular bins.
	         The simple degrees of freedom now are  
	         \[ df_{simple} = (\sqrt{K} -1)^2 = (\widehat{R} -1)(\widehat{C} -1)\]
	         as if the lattice were tessellated into $K$ rectangles perfectly aligned in $\widehat{R} = \sqrt{K}$ rows and $\widehat{C} = \sqrt{K}$ columns.
	\item $Y$ is categorical with $R$ categories, $X$ is continuous and tessellation occurs within rows giving a total of $K$ rectangular bins. The simple degrees of freedom estimate is
	        \[df_{simple} =  \left (R - 1 \right )( \widebar{C}- 1) \]
	        where $\widebar{C} = \frac{K}{R}$ is the average number of bins per row.
	\item $Y$ is continuous but  $X$ is categorical with $C$ categories and tessellation occurs within columns giving a total of $K$ rectangular bins. The simple degrees of freedom estimate is
	        \[df_{simple} =  \left (\widebar{R} - 1 \right )(C - 1) \]
	        where $\widebar{R} = \frac{K}{C}$ is the average number of bins per column.
\end{enumerate}
The latter two are the same case, simply reversing the roles of $X$ and $Y$.
Only in the first case must these simple degrees of freedom be integer.  
They are all analogous to the classic contingency case and so easy to remember.  
For any such binning of the ranks,  an approximate $p$-value is calculated as
\[ p_{simple} = Pr\left(\chi^2_{df_{simple}} \geq \bigChi \right). \]
Though imperfect, the ease with which this calculation is remembered and, so, applied, makes it
eminently useful in practice, particularly for screening many pairs of variates of different types.

The quality of this simple $\chi^2$ approximation is discussed in some detail in Appendix \ref{app:mle_approx}.
It was arrived at by first fitting the mles of the degrees of freedom to the number of bins, or bins and columns (or rows)
over a large variety of conditions.   The fits were surprisingly tight, yielding multiple correlation coefficients of $R^2$ around $99.97 $ and $99.98\%$.  These \textit{fitted} degrees of freedom are denoted $df_{fitted}$ and are close to $df_{simple}$.  
Similarly, the mles $\widehat{\alpha}$ and $\widehat{\theta}$ were modelled as functions of $df_{simple}$.
In the case of one continuous and one categorical variate the Gamma approximations produced null tail probabilities (size of test) more reliably than either $\chi^2$ approximation.

In what follows, these three different $p$-values tests will be denoted as
\[ p_{simple} = Pr\left(\chi^2_{df_{simple}} \geq \bigChi \right),
 ~~~~ p_{fitted} = Pr\left(\chi^2_{df_{fitted}} \geq \bigChi \right),\]
 \[   \text{ and }~~~~
 p_{gamma} = Pr\left(\Gamma(\alpha_{fitted}, \theta_{fitted}) \geq \bigChi \right).
 \]
 The functions giving these fitted values can be found in Appendix \ref{app:mle_approx}.

\subsection{Recursive maximal score binning}
\label{sec:maximal}
Among grid-based measures of dependence, a common approach to improve sensitivity is to select a grid maximizing the measure (e.g., \citealp{reshef2011MIC, helleretal2016consistent, caoetal2021improved}).
The same tactic could be applied here -- at each depth, rather than a random split, the bin is split at whatever rank gives the greatest value of $\bigChi$.
The splitting algorithm \ref{algo:maxSplit} of Appendix \ref{app:algorithms} accommodates this possibility in addition to random splitting.

Generally, the result will be a more sensitive test and a departure display that more readily reveals non-uniform pattern but at the cost of false positives were the maximization ignored and any of the preceding approximations used
(for comparisons see  
Appendix \ref{app:RecursiveMaxScoreBinning}).
Though promising, without further investigation to properly calibrate both test and display, recursive maximal binning cannot yet be recommended in practice.

\section{Quasi power}
\label{sec:power}
\begin{figure}[!ht]
\begin{center}
\begin{tabular}{cccccc}
\includegraphics[width= 0.12\textwidth]{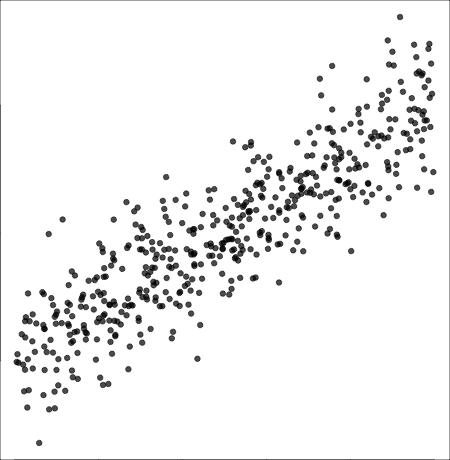} &
\includegraphics[width= 0.12\textwidth]{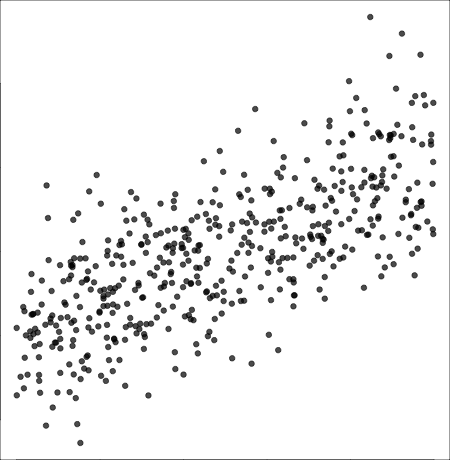} &
\includegraphics[width= 0.12\textwidth]{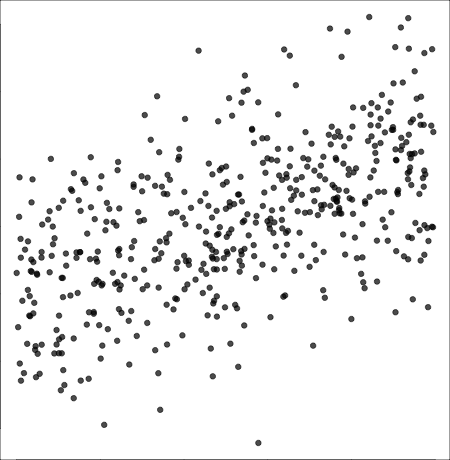} &
\includegraphics[width= 0.12\textwidth]{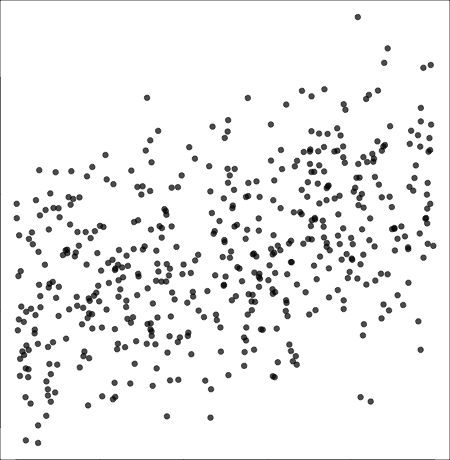} &
\includegraphics[width= 0.12\textwidth]{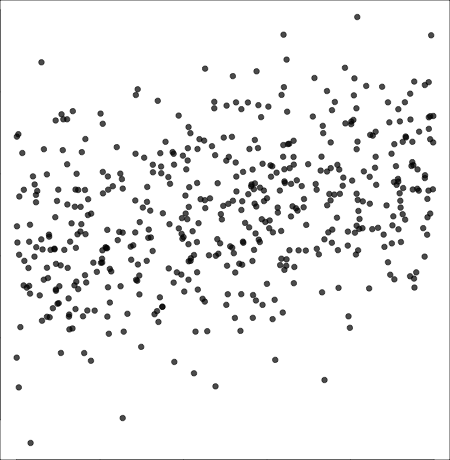} &
\includegraphics[width= 0.12\textwidth]{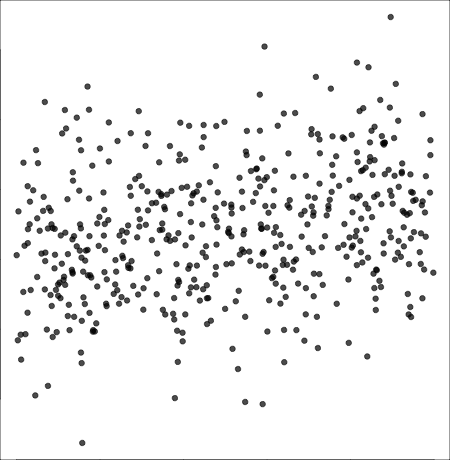} 
\\
\includegraphics[width= 0.12\textwidth]{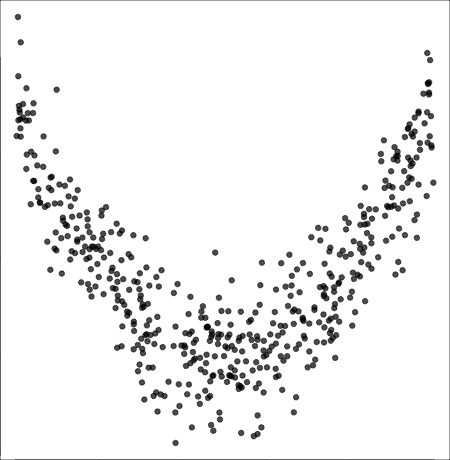} &
\includegraphics[width= 0.12\textwidth]{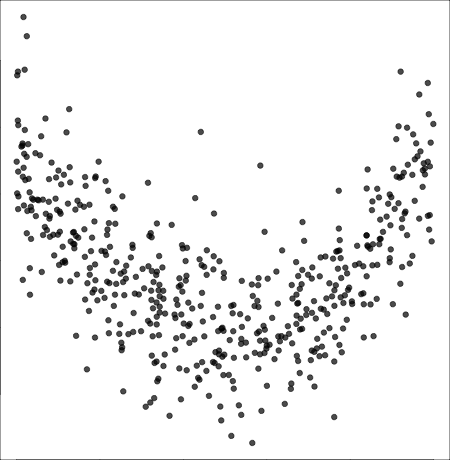} &
\includegraphics[width= 0.12\textwidth]{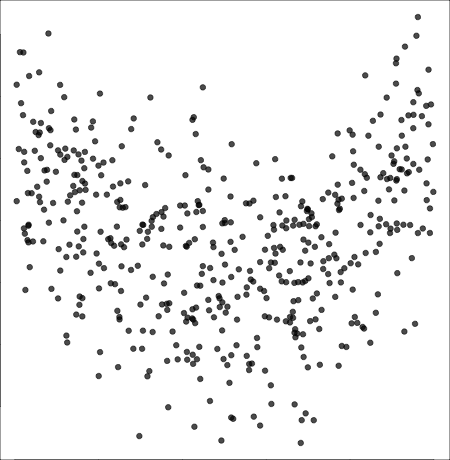} &
\includegraphics[width= 0.12\textwidth]{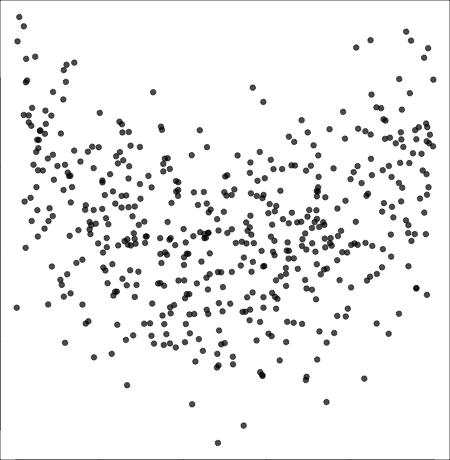} &
\includegraphics[width= 0.12\textwidth]{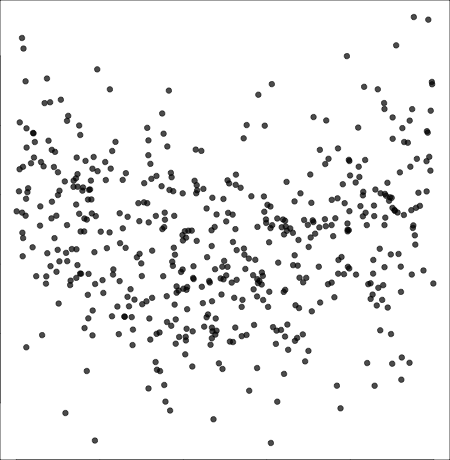} &
\includegraphics[width= 0.12\textwidth]{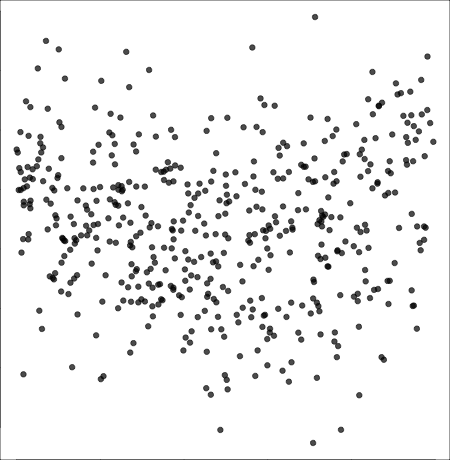} 
\\
\includegraphics[width= 0.12\textwidth]{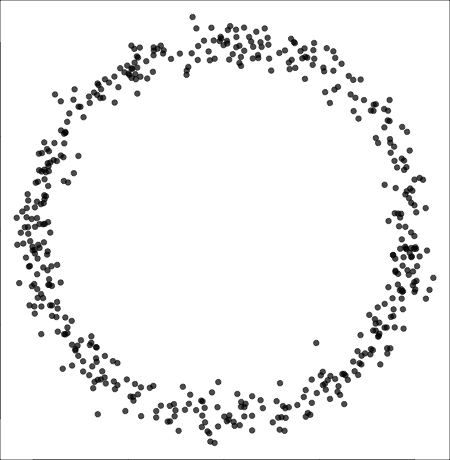} &
\includegraphics[width= 0.12\textwidth]{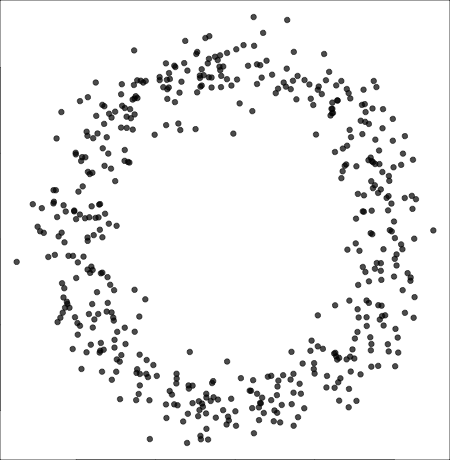} &
\includegraphics[width= 0.12\textwidth]{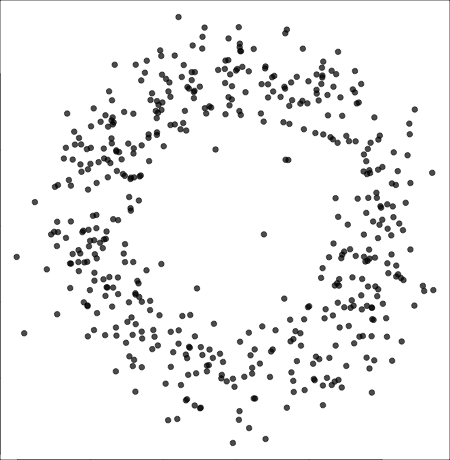} &
\includegraphics[width= 0.12\textwidth]{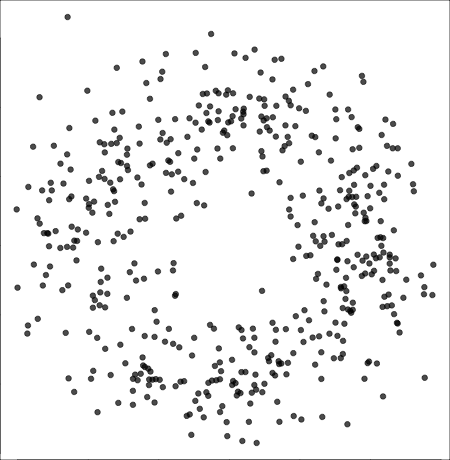} &
\includegraphics[width= 0.12\textwidth]{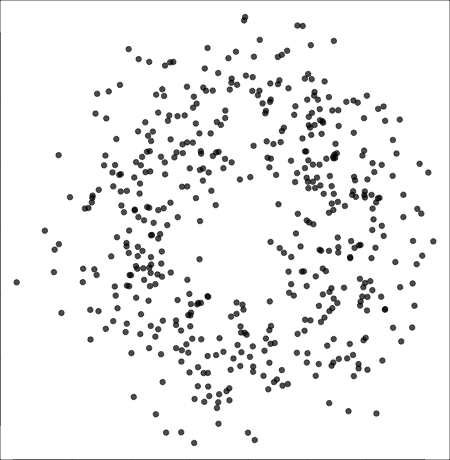} &
\includegraphics[width= 0.12\textwidth]{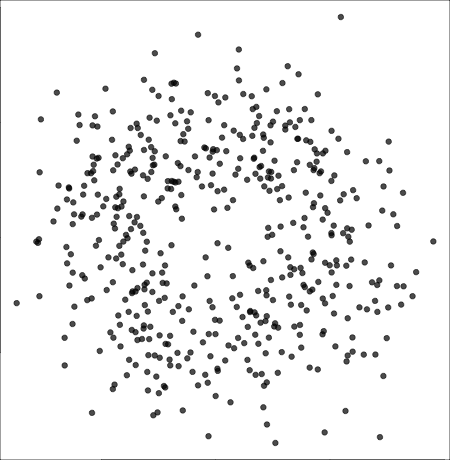} 
\\
\includegraphics[width= 0.12\textwidth]{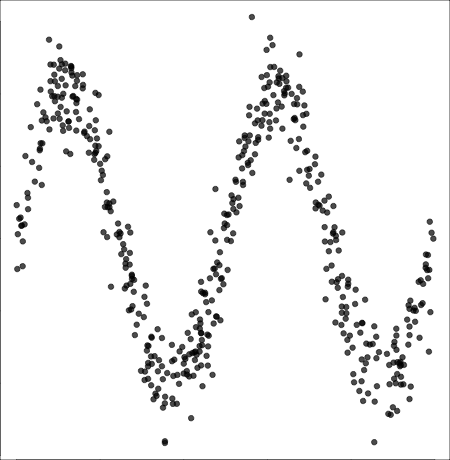} &
\includegraphics[width= 0.12\textwidth]{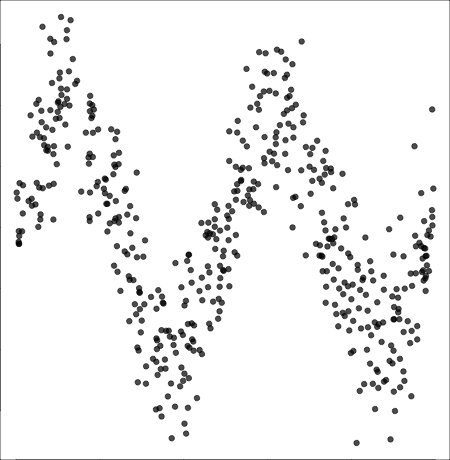} &
\includegraphics[width= 0.12\textwidth]{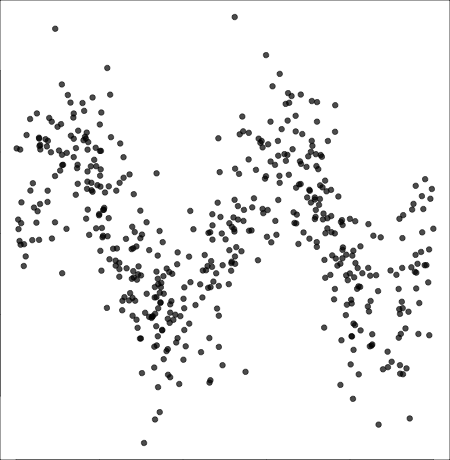} &
\includegraphics[width= 0.12\textwidth]{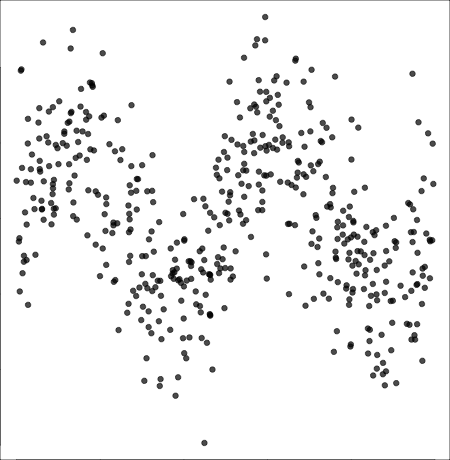} &
\includegraphics[width= 0.12\textwidth]{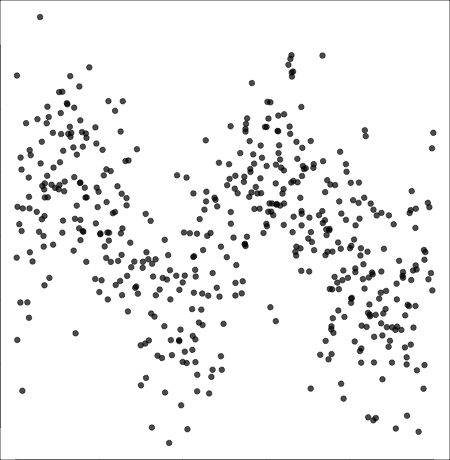} &
\includegraphics[width= 0.12\textwidth]{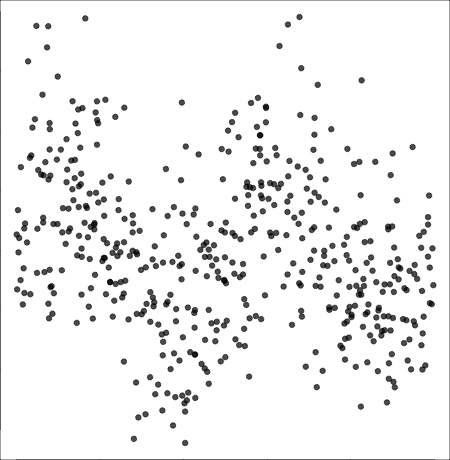} 
\\
\includegraphics[width= 0.12\textwidth]{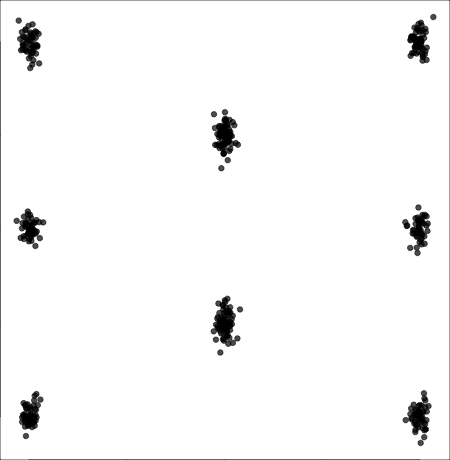} &
\includegraphics[width= 0.12\textwidth]{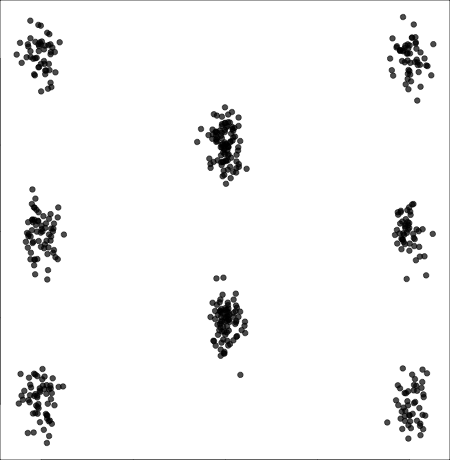} &
\includegraphics[width= 0.12\textwidth]{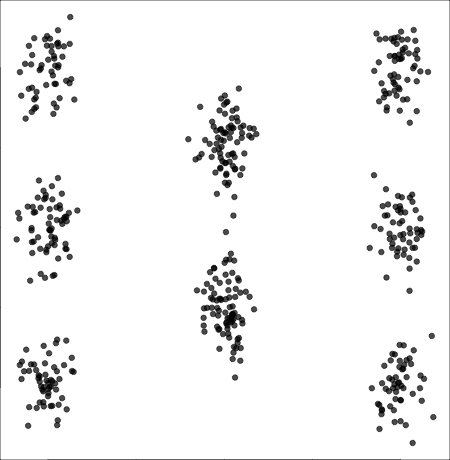} &
\includegraphics[width= 0.12\textwidth]{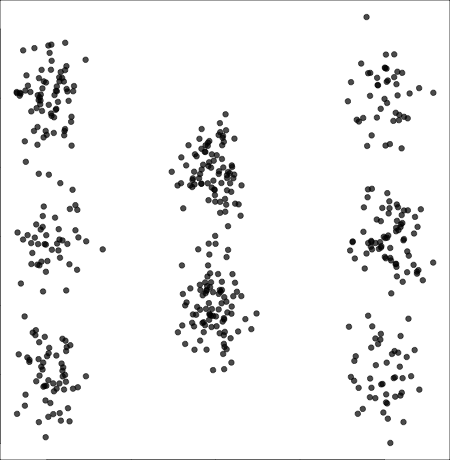} &
\includegraphics[width= 0.12\textwidth]{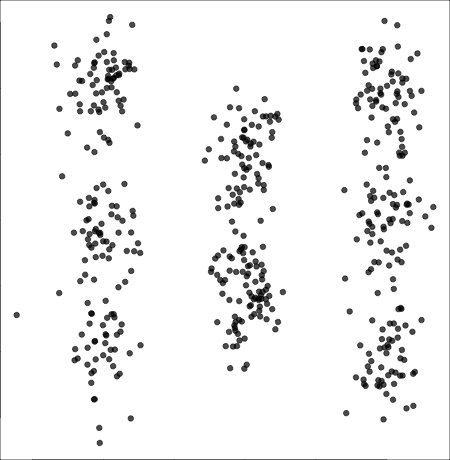} &
\includegraphics[width= 0.12\textwidth]{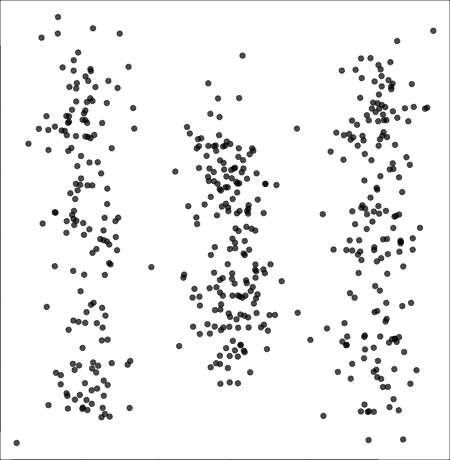} 
\\
\includegraphics[width= 0.12\textwidth]{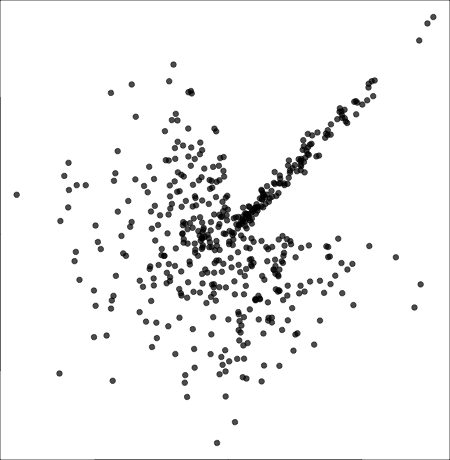} &
\includegraphics[width= 0.12\textwidth]{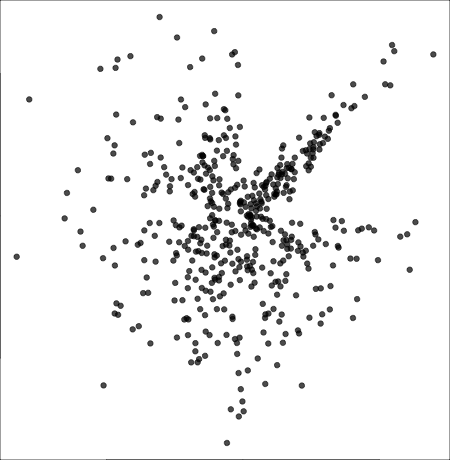} &
\includegraphics[width= 0.12\textwidth]{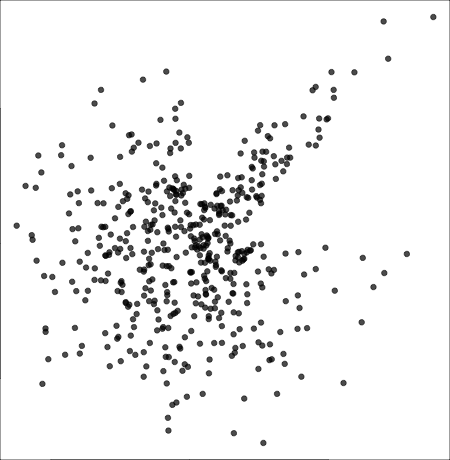} &
\includegraphics[width= 0.12\textwidth]{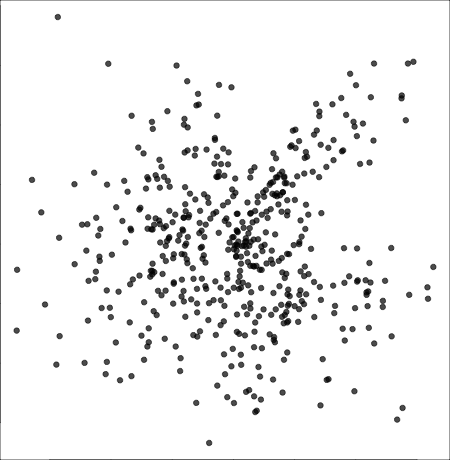} &
\includegraphics[width= 0.12\textwidth]{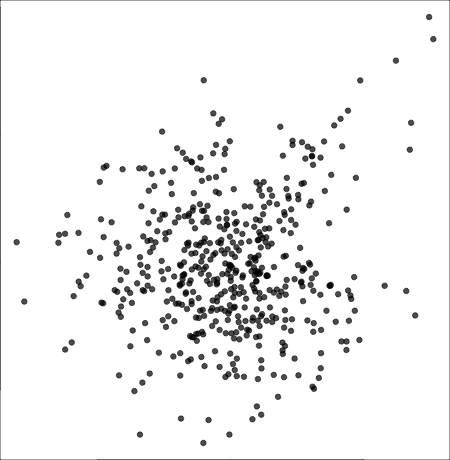} &
\includegraphics[width= 0.12\textwidth]{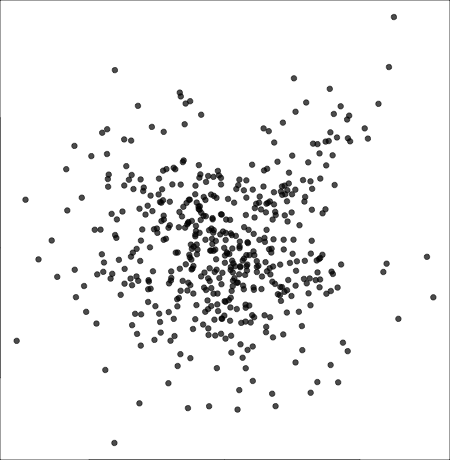} 
\end{tabular}
\caption{The noise patterns for $N = 500$ and left to right noise levels 1,2, \ldots, 6.
Top to bottom the patterns are ``linear'',  ``parabolic'',    ``circular'', ``sine'', ``checkerboard'', and ``local''.}
\label{fig:noise_patterns}
\end{center}
\end{figure}

The four $p$-value calculations for recursive random binning are now tested against several different point configurations generated for dependent $X$ and $Y$.  
The simulation patterns are those first appearing in \cite{FES_paper} and used
again, though with slightly modified parameters, in \cite{zhang2019BET}. 
The latter set of parameters are used here \cite[see][for details]{zhang2019BET}.
For each pattern, there is a noise level that can be increased to obscure the dependency between $X$ and $Y$ and
bring the pattern closer to a null configuration.
In this way, each test's ability to detect the dependency can be assessed as a function of the noise level.
Examples of the noise patterns for thirty-six samples of size 500 are shown in Figure \ref{fig:noise_patterns}
for noise levels 1 to 6.

In addition to the four possible $p$-values for recursive binning, the comparison includes two grid-based binning methods, namely Fisher exact scanning (FES) of \cite{FES_paper}  and binary expansion testing (BET)  of \cite{zhang2019BET}.
Both are multi-scale binning methods designed to test independence of rank pairs over ever finer divisions of the unit square. 
Each performs many tests of independence for carefully constructed $2 \by 2$ tables and applies  an adjustment for multiple testing to arrive at a final $p$-value.

With FES, the unit square is recursively divided into a grid having up to $2^{k_1}$ equi-sized divisions on one margin and $2^{k_2}$ on the other (typically, $k_1 = k_2$).  Roughly, at each recursion, a new $2 \by 2$ table appears within each cell of the coarser grid and Fisher's exact test of independence is applied.  The $p$-values at all recursions are combined and adjusted for multiple testing.  In the present simulation, the \code{fisherScan()} function from \cite{FES_software} is used with  $k_1 = k_2 = 4$ (yielding a total of 256 possible cells) and \code{fixed.M = TRUE}; the final $p$-value reported is the thrice corrected ``Sidek'' value.  

In the case of BET, the grid is similarly defined at the $(k_1, k_2)$ level but now the $2^{k_1} \by 2^{k_2}$ contingency table likelihood 
is the basis for inference.  \cite{zhang2019BET} effects a re-parameterization by introducing  binary ``interaction'' variates constructed from the coefficients of the finite binary expansion (up to $k_1$ terms in the horizontal direction and $k_2$ in the vertical) which locate each cell in the unit square.  Each of these $2^{k_1 + k_2} -1$ binary variates divides the unit square into two equal area regions.  The difference in counts of these regions measure their balance, or symmetry, and provide sufficient statistics for the likelihood.  \cite{zhang2019BET} shows that the null hypothesis of independence for this contingency table is equivalent to the expectation of the symmetry measure being zero for all the $2^{k_1 + k_2} -1$ binary variates not defined solely in terms of one margin.  The problem is turned into one of multiple testing (for which a Bonferroni adjustment is made) by determining the $p$-value via the maximal absolute symmetry.  This is the ``max BET'' procedure implemented as  \code{maxBET()}  \R package called \pkg{BET} \cite{BET_pkg}. 
 As with FES, the values $k_1 = k_2 = 4$ were selected as the depth  in the simulation.

A total of six different tests were then applied to simulated data for the null distribution and the six patterns of Figure \ref{fig:noise_patterns} for noise levels from 1 to 10.
Four tests were based on the different $p$-values for random recursive binning using a depth of $d = 8$ to match the $k_1 = k_2 = 4$ depths of FES and BET.  For each replication, a random set of (squarified) bins were created each having expected frequency no less than 5; this meant that no replication for random recursive binning approached the maximum possible number possible of 256 bins (see top left histogram of Figure \ref{fig:quasi_power_rBEX}). 
Then a sample was generated for each pattern (with varying noise) and the six different $p$-values determined for each sample.
A total of 200 independent replications were produced.

\begin{figure}[!ht]
\begin{center}
\begin{tabular}{ccc}
\includegraphics[width=0.45\textwidth]{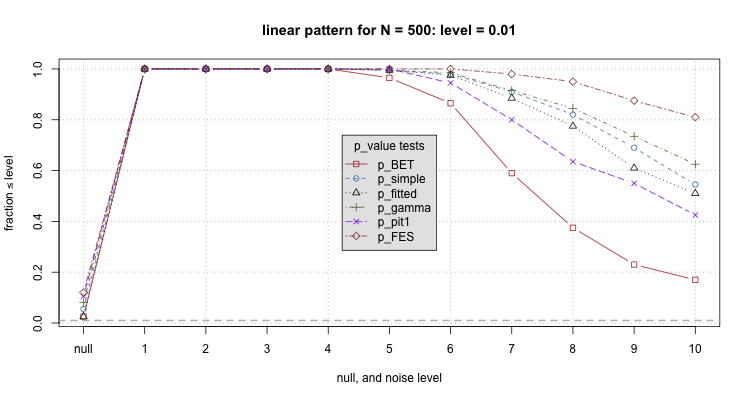}
& $~~~$ &
\includegraphics[width=0.45\textwidth]{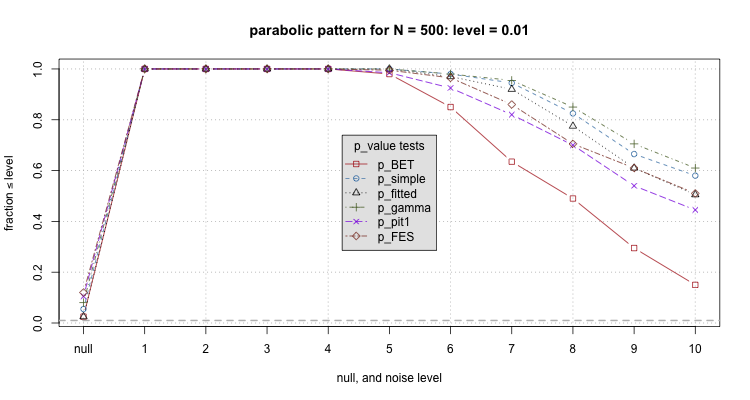}\\

\includegraphics[width=0.45\textwidth]{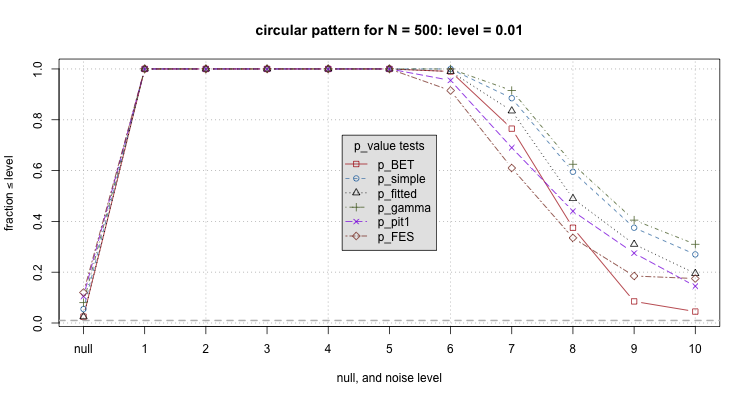}
& $~~~$ &
\includegraphics[width=0.45\textwidth]{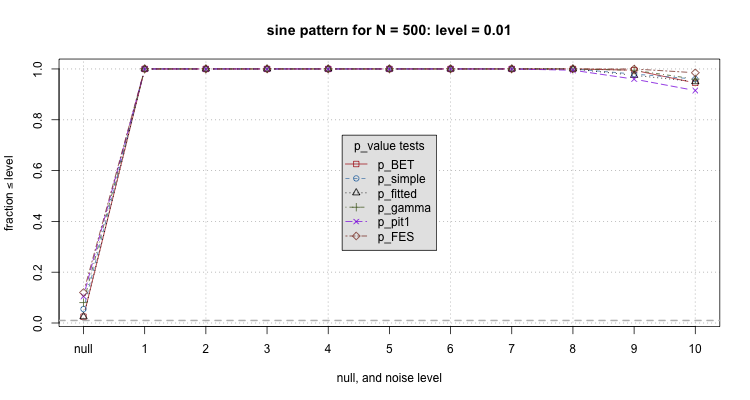}\\
\includegraphics[width=0.45\textwidth]{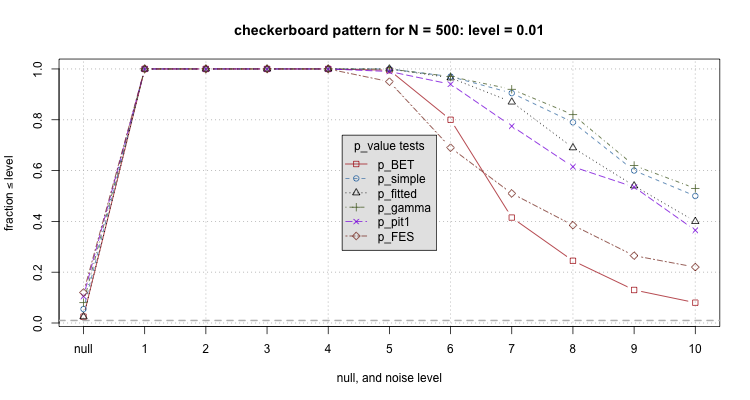}
& $~~~$ &
\includegraphics[width=0.45\textwidth]{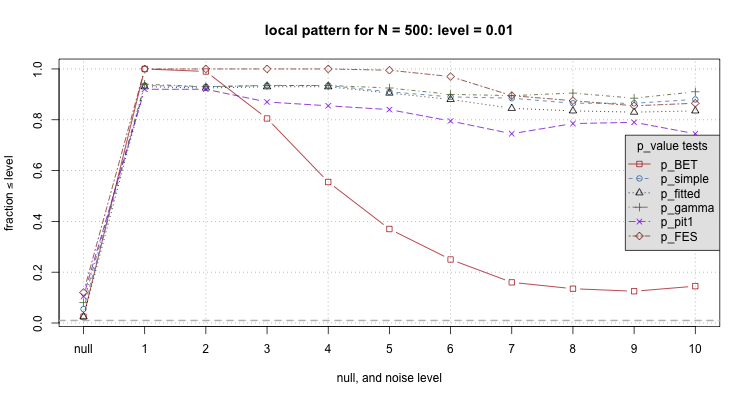}\\
\end{tabular}
\caption{Quasi-power curves for the different patterns as a function of increasing noise.}
\label{fig:quasi_power_curves}
\end{center}
\end{figure}

Figure \ref{fig:quasi_power_curves}
shows quasi-power curves which track, for each test, the proportion of $p$-values observed to be less than some critical value, here 0.01.  The null case is included to show how well each test achieves the putative size shown as a horizontal dashed line.

As can be seen, all four recursive binning methods are competitive and often dominate with the $p_{gamma}$ method being the best, followed closely by the $p_{simple}$ method.  Even the $p_{pit1}$ method, based on a single generation of independent uniform pairs can compete with FES and  BET and sometimes outperform them, especially BET.

In the ``linear'' case, and for lower noise in the ``local'' case, the FES method dominates.
FES fares poorly in the checkboard case, suggesting perhaps that it is the linear structure apparent in the low noise ``local'' case being picked up by it.  Looking at the null condition (the same in all six plots), FES has the highest proportion of $p$-values being observed -- at least an order of magnitude larger than the putative size of the test. 
This suggests that the power observed for the noise patterns might be inflated for FES.

The poorest performer overall is BET, performing particularly poorly in the case of ``local'' structure -- \cite{zhang2019BET} give some theoretical support as to why this might be the case.  In comparing BET with FES, Figure \ref{fig:quasi_power_curves} suggests either neither dramatically dominates the other or FES dominates BET.  In contrast, for sample size 128 and critical level 0.1, 
\cite{zhang2019BET} found that BET dominated FES for the ``circular'', ``sine'' and ``checkerboard'' patterns.
Note however that the fraction observed under null conditions is very near the critical value for BET, where FES is a little over 10 times larger, suggesting that BET's power may not be as inflated as that of FES.

\begin{figure}[!ht]
\begin{center}
\begin{tabular}{cccccc}

\includegraphics[width= 0.12\textwidth]{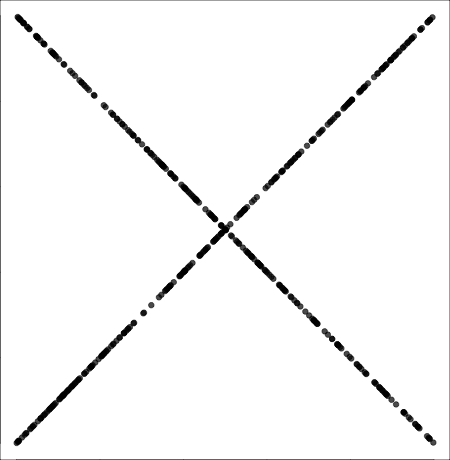} &
\includegraphics[width= 0.12\textwidth]{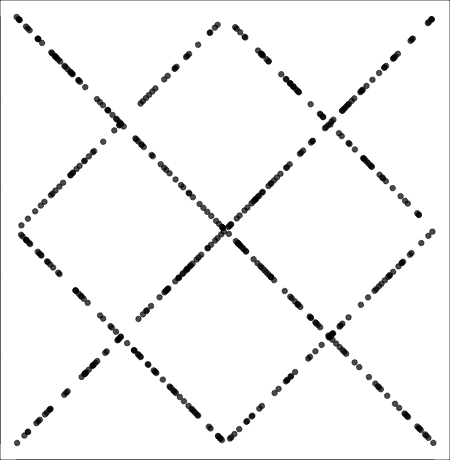} &
\includegraphics[width= 0.12\textwidth]{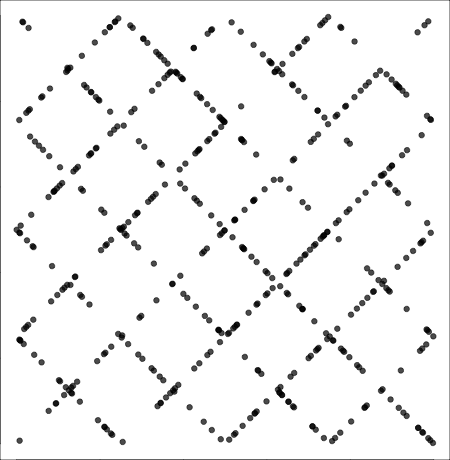} &
\includegraphics[width= 0.12\textwidth]{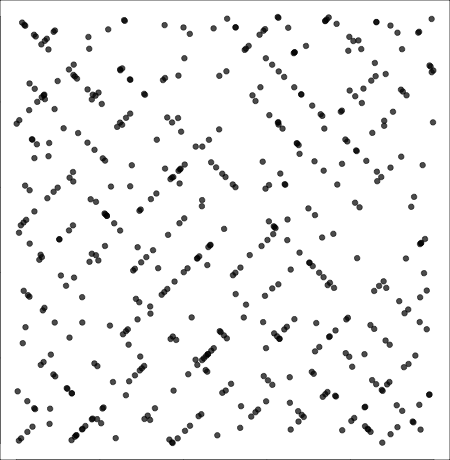} &
\includegraphics[width= 0.12\textwidth]{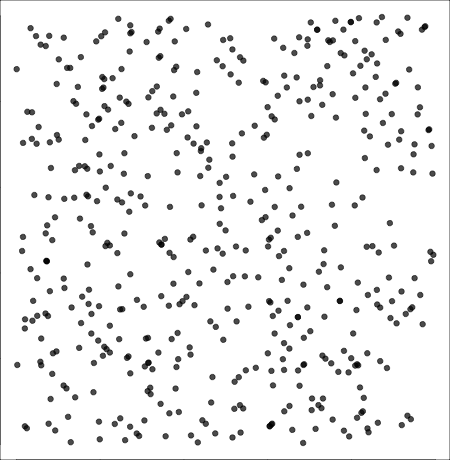} &
\includegraphics[width= 0.12\textwidth]{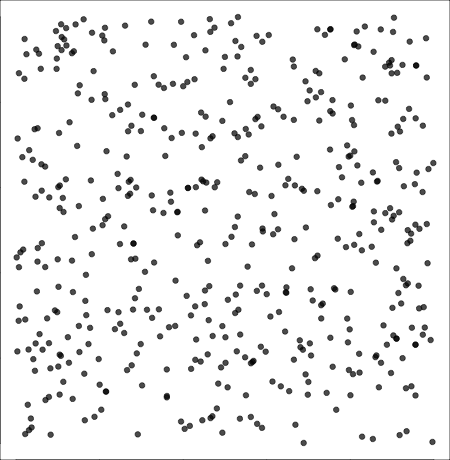} 
\\
\includegraphics[width= 0.12\textwidth]{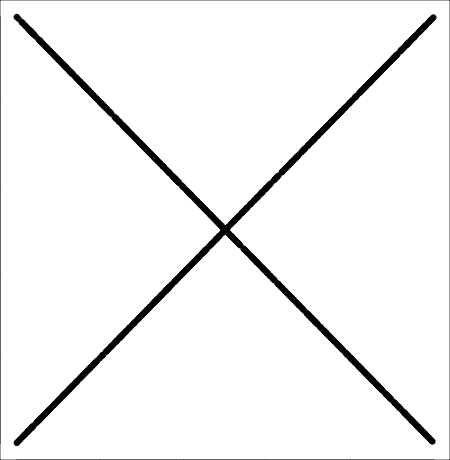} &
\includegraphics[width= 0.12\textwidth]{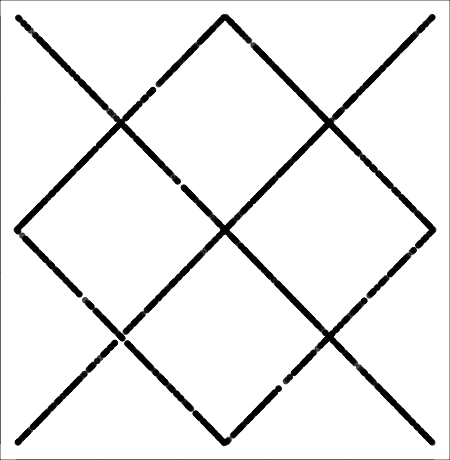} &
\includegraphics[width= 0.12\textwidth]{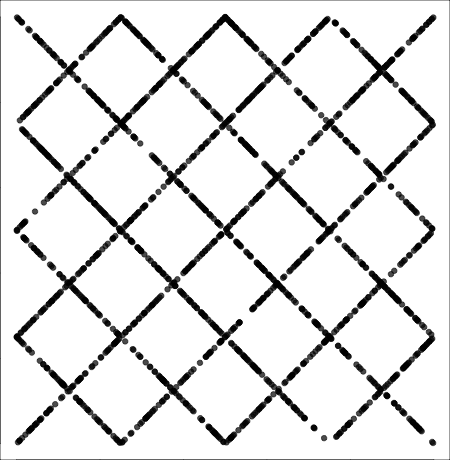} &
\includegraphics[width= 0.12\textwidth]{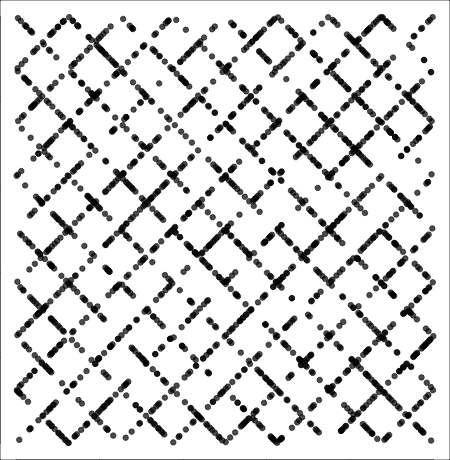} &
\includegraphics[width= 0.12\textwidth]{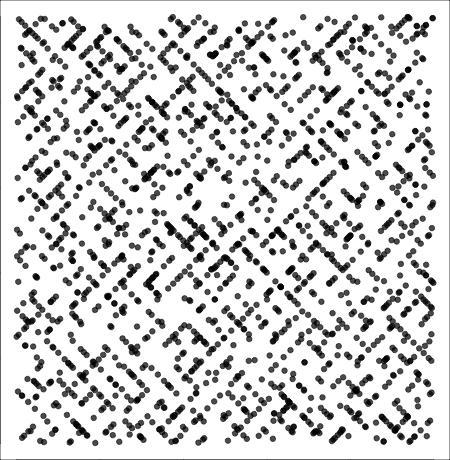} &
\includegraphics[width= 0.12\textwidth]{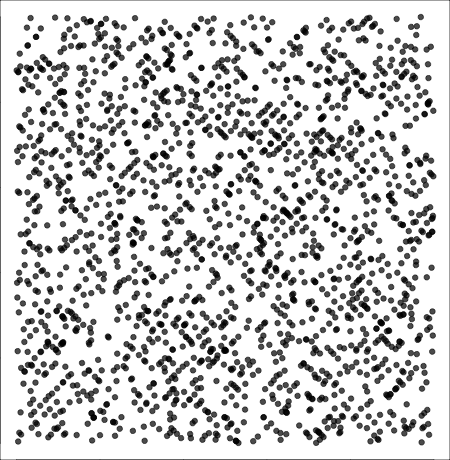} 
\\
\end{tabular}
\caption{rBEX patterns for resolutions 1, 2, \ldots, 6.  \\Randomly drawn samples of size $N = 500$ in the first row and $N = 2000$ in the second.}
\label{fig:rBEX_patterns}
\end{center}
\end{figure}

\begin{figure}[!ht]
\begin{center}
\begin{tabular}{ccc}
\includegraphics[height= 0.175\textheight]{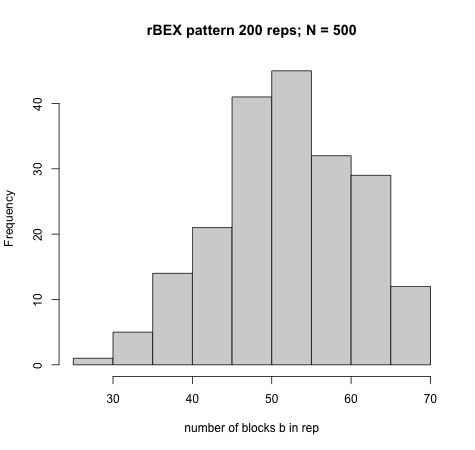}
& $~~~$ &
\includegraphics[height= 0.175\textheight]{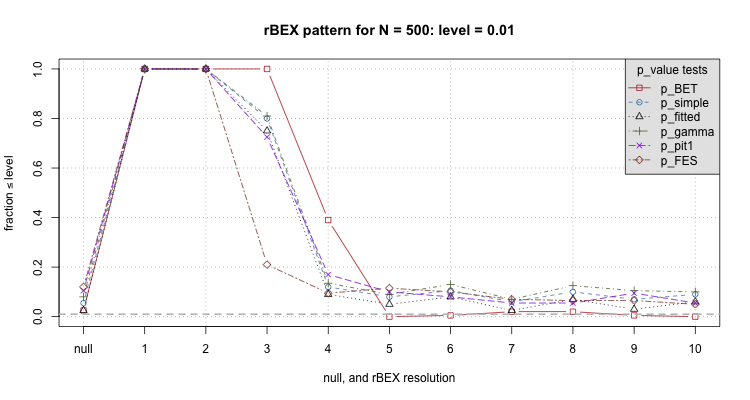}\\
\includegraphics[height= 0.175\textheight]{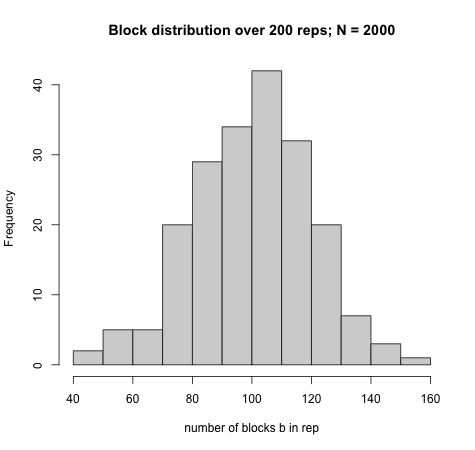}
& $~~~$ &
\includegraphics[height= 0.175\textheight]{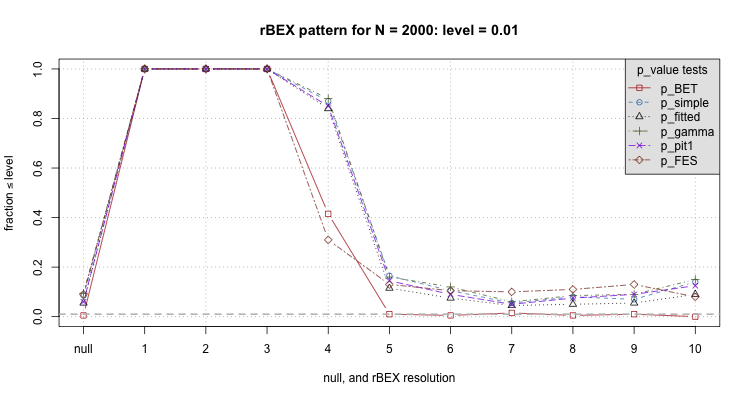}\\
\end{tabular}
\caption{Quasi-power curves for the rBEX patterns and two different sample sizes.}
\label{fig:quasi_power_rBEX}
\end{center}
\end{figure}

In motivating BET, \cite{zhang2019BET} provide the example of random variates uniformly distributed on a bisection expanding cross manifold (BEX), essentially a 45\degree{} rotated grid.  How fine this grid will be is determined by a resolution parameter akin to that of BET's depth.  Figure \ref{fig:rBEX_patterns} 
random samples of size 500 and 2,000 for resolutions 1 to 6; we call these rBEX patterns to distinguish the random sample from its support.
\cite{zhang2019BET} note that the marginal distributions will be uniform and that the joint distribution, though degenerate, becomes arbitrarily close to that of product of independent uniforms as the resolution increases. 
\cite{zhang2019BET} state that the ``power loss due to nonuniform consistency can be severe'' and that ``simulations \ldots show that many CDF and kernel based tests are powerless in detecting BEX at [resolution] 4 even when the sample size is as high as 20,000.'' Though visually detectable for such large sample sizes, ``\ldots  many existing tests
cannot distinguish it from independence.''
In contrast, BET should  be able to detect BEX patterns whose resolution is less than or equal to BET's depth  (i.e., $k_1 = k_2$) as its grid of squares will isolate each  cross in the pattern \cite[see][Figure 3]{zhang2019BET}.

The first row of Figure \ref{fig:quasi_power_rBEX}
shows the results of the six tests on the rBEX pattern for sample size 500.
As the quasi-power curves show,  BET (with depth 4 and its 256 grid squares of the right size) picks up the rBEX patterns for resolutions less than or equal to BET's depth, though for this sample size resolution 4 is much more difficult.
For resolutions beyond its depth, BET fails to detect a signal, recording proportions close to that of independence.
In contrast, beyond resolution 4, all other methods are able to detect a signal on occasion.

The four tests with recursive binning tests dominate FES  for resolution 3 and below and have some power in detecting resolutions 4 and beyond.  
The FES performance resolution 3 and 4 poorer performance is a little surprising since it can make use of the same
256 cells, though in groups of four as $2 \by 2$ tables, as does BET.

The second row of Figure \ref{fig:quasi_power_rBEX} repeats the analysis but for sample size 2,000.  
As expected, the power of all methods increases.
Surprisingly, however, every recursive binning method now dominates both FES and BET for resolution 4.
Beyond resolution 4, BET continues to be unable to distinguish the rBEX patterns from uniform independence, whereas the power curves for the other methods appear to increase with the sample size as expected.
With larger sample sizes, the recursive binning tests have access to far more bins as the histogram shows.

Finally, note that for the recursive random binning methods, every binning was used across all configurations to reduce experimental variation when comparing configurations.  
Since this binning was done independent of all configurations (stopping depending only on depth and minimum expected count), it might have introduced an apparent splitting of empty bins in non-null configurations having a lot of empty space.  This would not change $\bigChi$ but would increase the number of bins, and hence the degrees of freedom in any of the $\chi^2$ approximations, the result being that the $p$-value reported would be larger than that which would be used in practice.  Consolidating any recursively split empty bins would better represent practice; this would have reduced the $p$-value and increased the power of the test for non-null configurations.

\subsection{$p$-value distributions}
The quasi-power curves tell a bit of the story but focus entirely on a fixed size test.
While often mathematically convenient, it is not necessary computationally to reduce the entire collection of
$p$-values to a single statistic at each alternative.  More can be gleaned from examination of the entire sample of $p$-values.

\begin{figure}[!ht]
\begin{center}
\begin{tabular}{c}
\includegraphics[width=0.75\textwidth]{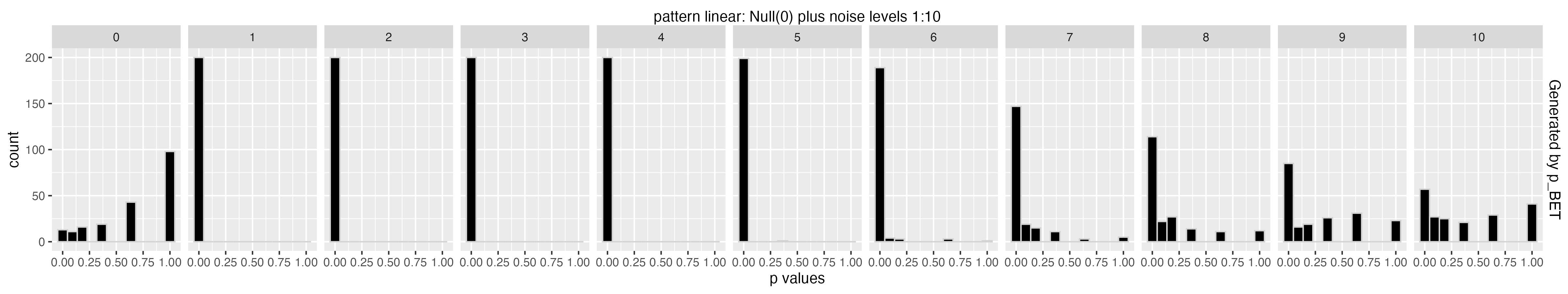}\\
\includegraphics[width=0.75\textwidth]{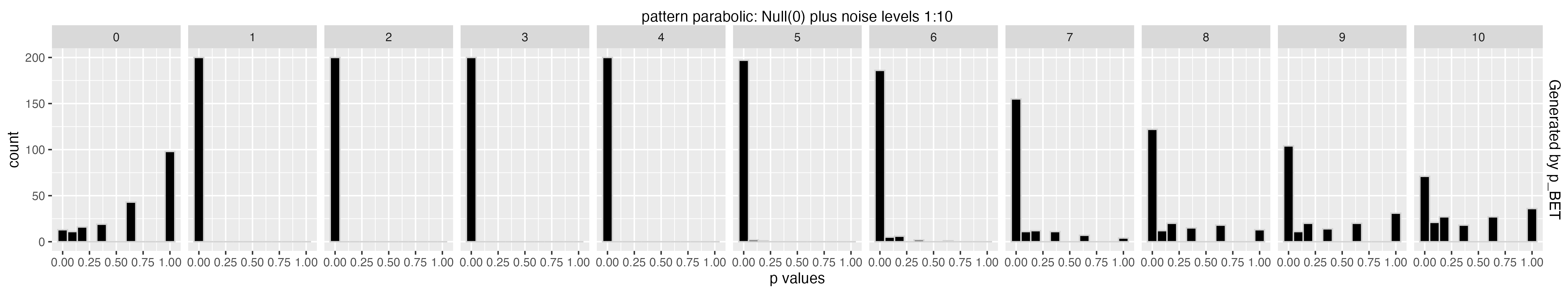}\\
\includegraphics[width=0.75\textwidth]{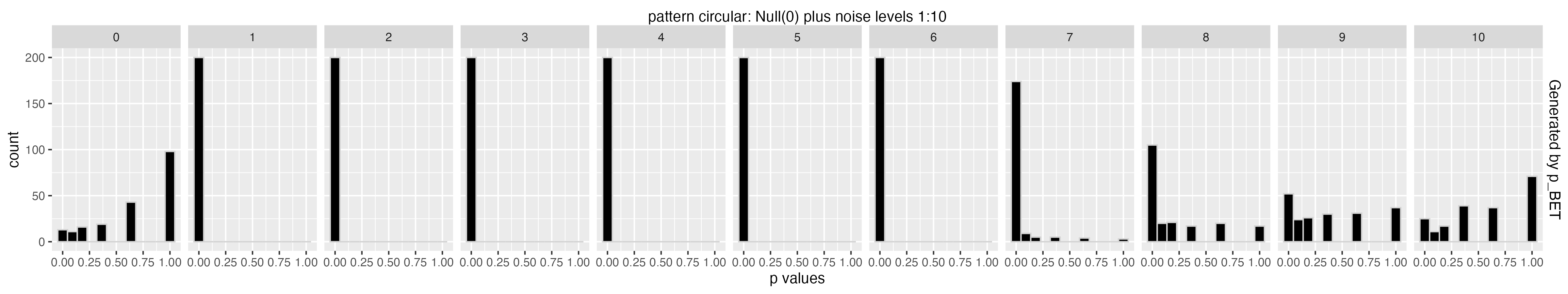}\\
\includegraphics[width=0.75\textwidth]{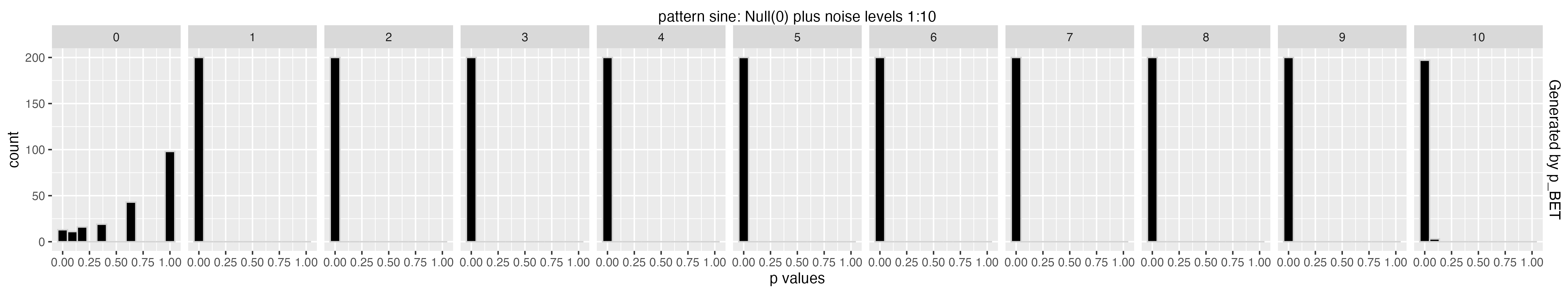}\\
\includegraphics[width=0.75\textwidth]{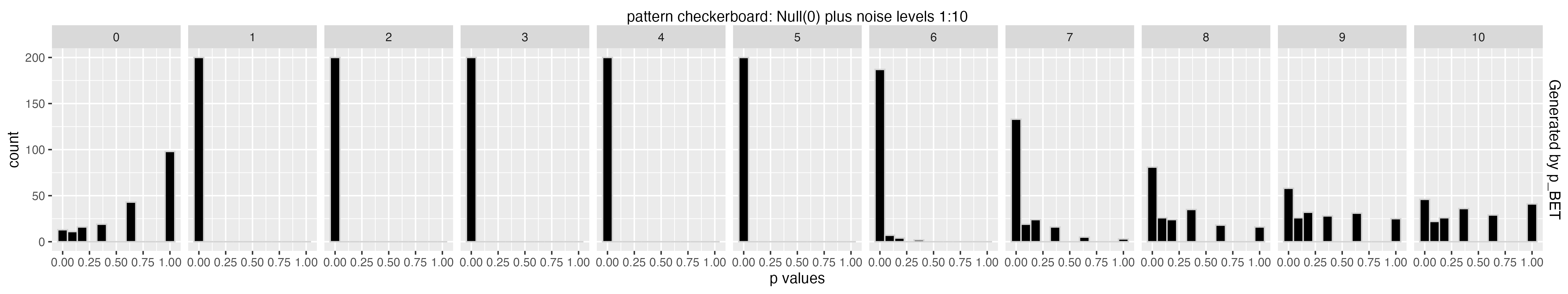}\\
\includegraphics[width=0.75\textwidth]{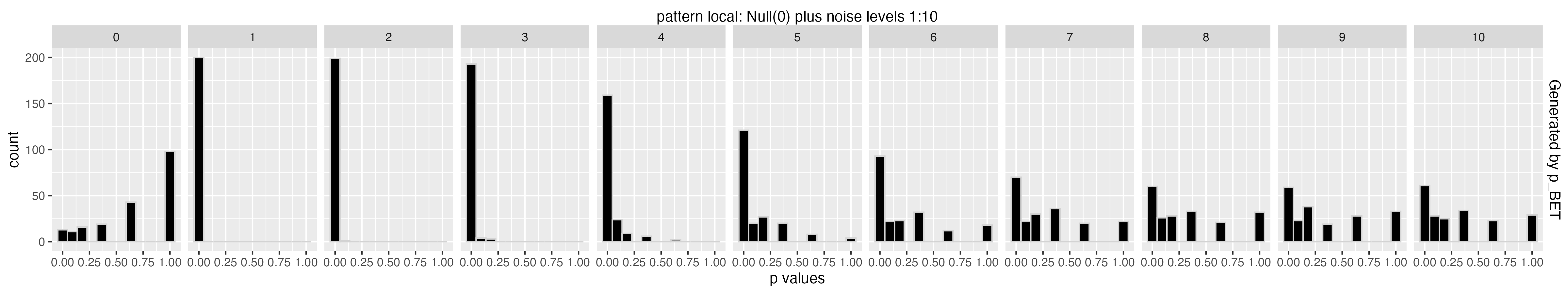}\\
\end{tabular}
\caption{Histograms of $p$-values using BET.  Top to bottom patterns are ``linear'', ``parabolic'', ``circular'', ``sine'', ``checkerboard'', and ``local'';  Left most is the ``null''.  Left to right, levels from 1 to 10.}
\label{fig:hist_pBET}
\end{center}
\end{figure}

For example, the 200 $p$-values generated by BET for the six patterns and all noise levels are shown 
in Figure \ref{fig:hist_pBET}.
The leftmost histogram represents the $p$-value distribution under the null hypothesis of independence and is the same in all rows. 
Clearly, the $p$-value distribution under independence is not uniform, as one might expect.
Indeed, the null distribution for BET favours high values near one.  Depending on where one fixed the critical value, it might or might not reflect the size of the test.  
Moving left to right, for most small critical values, the height of the leftmost histogram bar is about proportional to the quasi-power of the test -- each row's left bar height follows the corresponding ``p\_BET`` curves from each display in  Figure \ref{fig:quasi_power_curves}. 

\begin{figure}[!hbt]
\begin{center}
\begin{tabular}{c}
\includegraphics[width=0.75\textwidth]{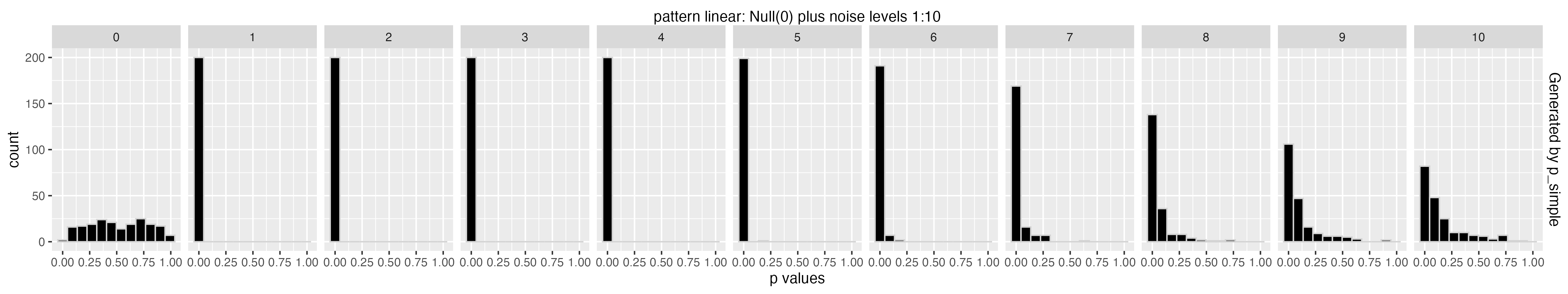}\\
\includegraphics[width=0.75\textwidth]{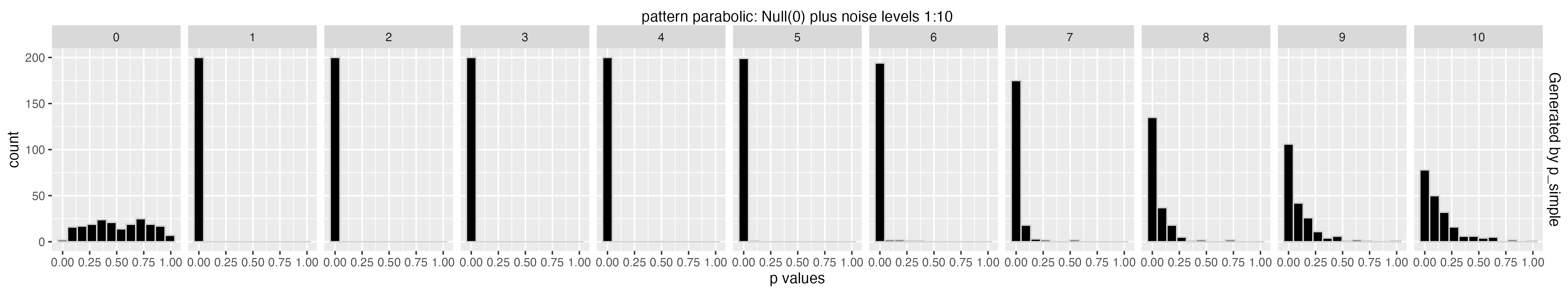}\\
\includegraphics[width=0.75\textwidth]{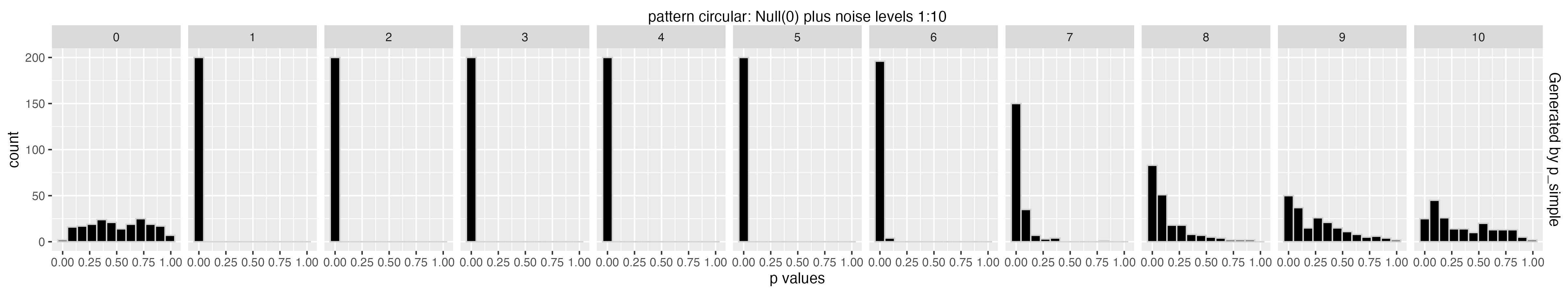}\\
\includegraphics[width=0.75\textwidth]{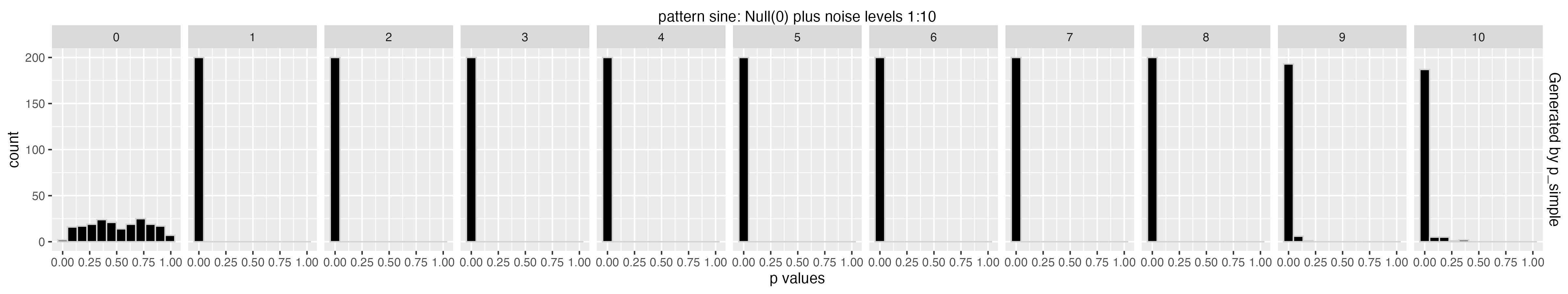}\\
\includegraphics[width=0.75\textwidth]{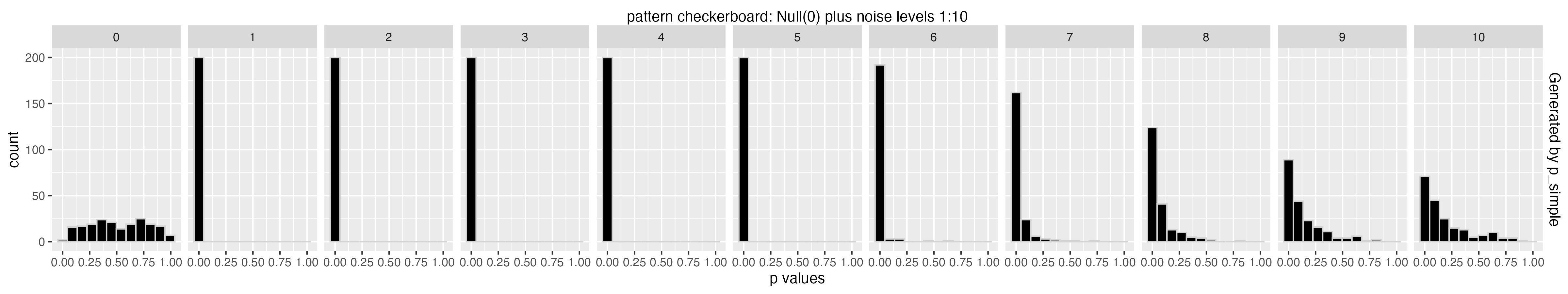}\\
\includegraphics[width=0.75\textwidth]{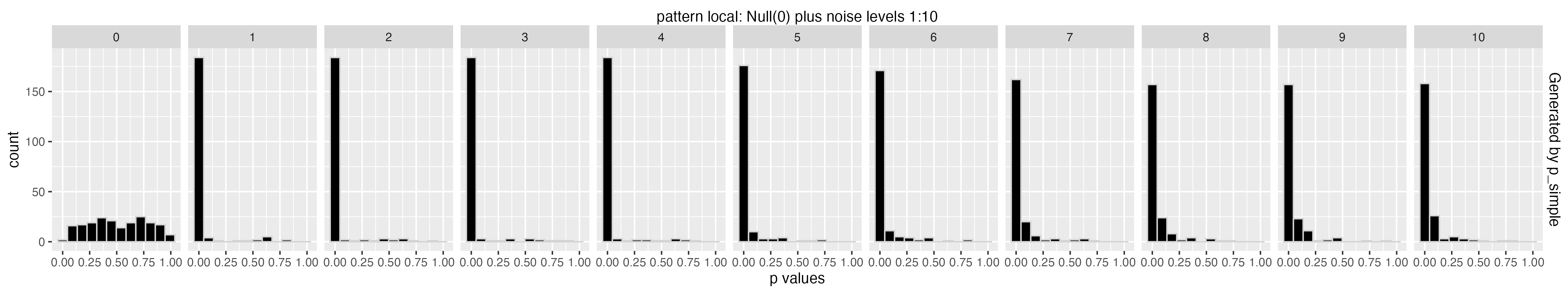}\\
\end{tabular}
\caption{Histograms of $p$-values using recursive random binning and $\chi^2$ on $df_{simple} =(\sqrt{K}-1)^2$ degrees of freedom.  Top to bottom patterns are ``linear'', ``parabolic'', ``circular'', ``sine'', ``checkerboard'', and ``local'';  Left most is the ``null''.  Left to right, levels from 1 to 10.}
\label{fig:hist_p_simple}
\end{center}
\end{figure}

In contrast, the histograms of $p$-values generated using $p_{simple}$ on the same data are shown to be much
better behaved in shape in Figure \ref{fig:hist_p_simple}.
The null distribution is as it should be, approximately uniformly distributed. 
Whatever critical value might be chosen, it can represent the size of the test.
Under the noisy alternatives, the histograms for $p_{simple}$ also have the desired shape -- right-skewed so that the $p$-values concentrate on low values.  
This is not the case for BET.

Two tests may be compared for almost any critical value by marking the critical value on the horizontal axis of each histogram.
The weight of the histogram to the left of that mark will be proportional to the power on each alternative.   
The more heavily skewed right is the histogram, the greater will be its power for any critical value.
For the small critical values typically entertained, it will be enough to compare the leftmost bar heights.
For example, the superior performance of $p_{simple}$ over BET on the ``local'' pattern is easily seen by comparing the last rows of 
Figures  \ref{fig:hist_pBET} and \ref{fig:hist_p_simple}.

All methods but BET produced histograms of the desired shape under null and alternative hypotheses for all patterns including rBEX.
When the noise was too great to discern dependence, the histograms of all methods (but BET) looked uniform as under the null.

\section{Example: dependence in wine data}
\label{sec:wineEx}

The wine data of \cite{wineData,wine_quality_186} consists of $N =$ 6,497 Portuguese \textit{vinho verde} wines on which thirteen variates were recorded:  eleven measuring different physical and chemical properties (including percent alcohol), each wine's type (red or white), and a subjective quality rating  (scaled 1 to 10).  
Because there were so few having the lowest or highest qualities, we reduced the  wine quality variate to have five ordered categories: ``$\leq 4$'', ``5'', ``6'', ``7'', and ``$\geq 8$''.
For illustration,  we transformed a continuous variate,  the percent alcohol content,   to a categorical one,  alcohol content, having possible values of  ``low'', ``medium'', or ``high''.  
As a check, two continuous variates, $U$ and $V$, that are independently generated from a $U(0,1)$ distribution are added to the data.
This gives a final data set of size $N =$ 6,497  having  three categorical and ten continuous variates from real measurements as well as two 
artificial continuous variates generated independently of one another and of all other variates.

There are a total of 105 pairs to investigate, 78 of which are real data pairs, one is the artificial pair, and twenty-six a combination of artificial and real.
While these are few enough that, in principle at least, all could be examined at once in, say, a scatterplot matrix, we use the 105 pairs to illustrate how recursive random binning would be applied (here with depth of 8 and minimum expected value of 10) to identify the most and least dependent pairs of variates. 
The departure displays from the selected pairs can then be examined to identify patterns of dependence.

\begin{figure}[!hbt]
  \begin{center}
  \begin{tabular}{ccc}
  \includegraphics[width = 0.4 \textwidth]{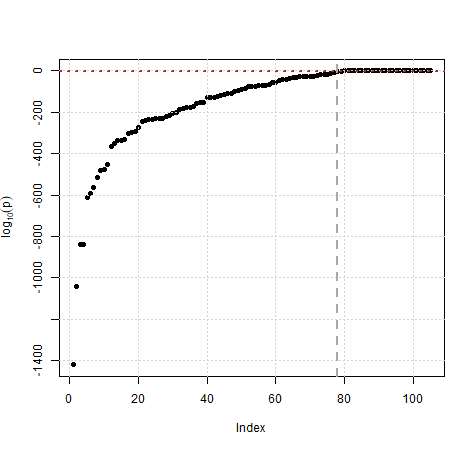}
  &$~~~$ &
  \includegraphics[width = 0.4 \textwidth]{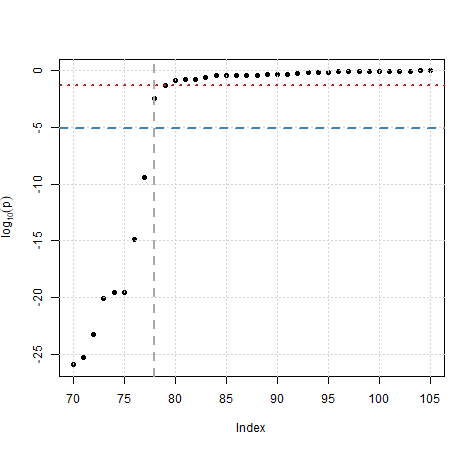} \\
  {\footnotesize (a) All 105 pairs} &&
  {\footnotesize (b) Least significant pairs}\\
  \end{tabular}
  \caption{Wine data: Ordered  $\log_{10} p_{simple}$ values. Horizontal (red) dotted line at the nominal $p = 0.01$.  Horizontal (blue) dashed at right is the same but Bonferroni corrected.  Vertical (grey) dashed line at pair index 78.} 
  \label{fig:wine_pvals}
  \end{center}
\end{figure}

Figure \ref{fig:wine_pvals}
plots the ordered $\log_{10} p$  for the $\chi^2$ approximation having $df_{simple}$ as its degrees of freedom.  Of the 105 assessed, 27 appear above the $p = 0.01$ line in Figure \ref{fig:wine_pvals}(b) (also have $p > 0.05$; not marked) -- all involve at least one  of $U$ or $V$.  Applying the conservative Bonferroni correction for multiple testing, the pair of variates ``pH'' versus ``quality'' (index 78) is judged to be non-significant at the 1\% level (also, though not shown, at the Bonferroni corrected 5\% level as well).  All other pairs have Bonferroni corrected $p << 0.01$.  The uncorrected $p$-value, from small to large, orders the pairs in decreasing strength of the dependency observed between them.

\begin{figure}[!ht]
  \begin{center}
  \begin{tabular}{cc}
  \includegraphics[width = 0.45\textwidth]{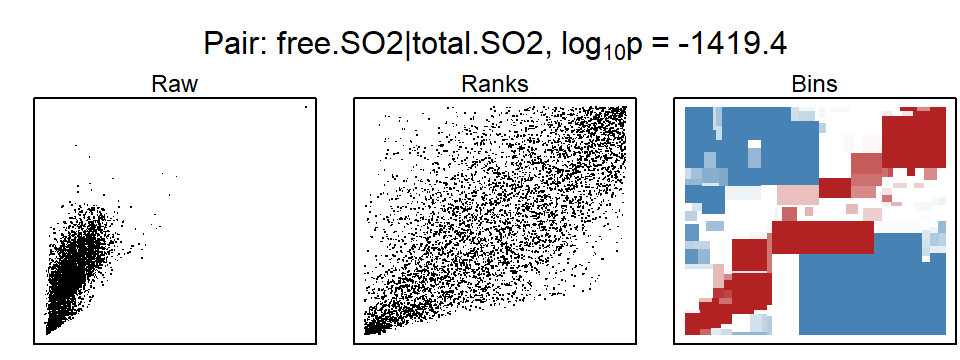}&
  \includegraphics[width = 0.45\textwidth]{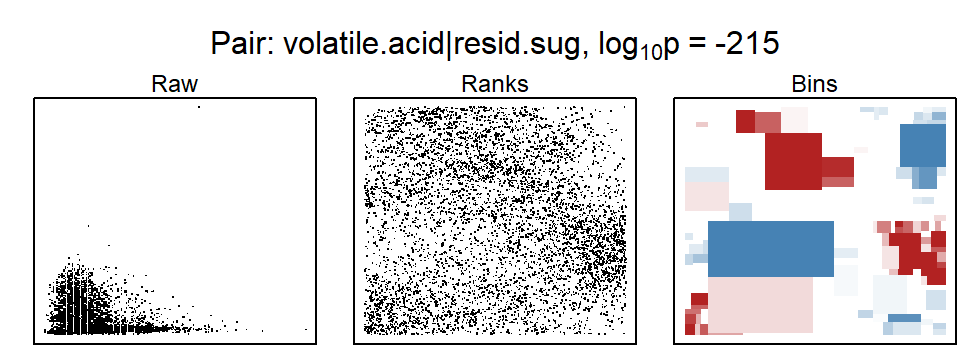}\\
  \includegraphics[width = 0.45\textwidth]{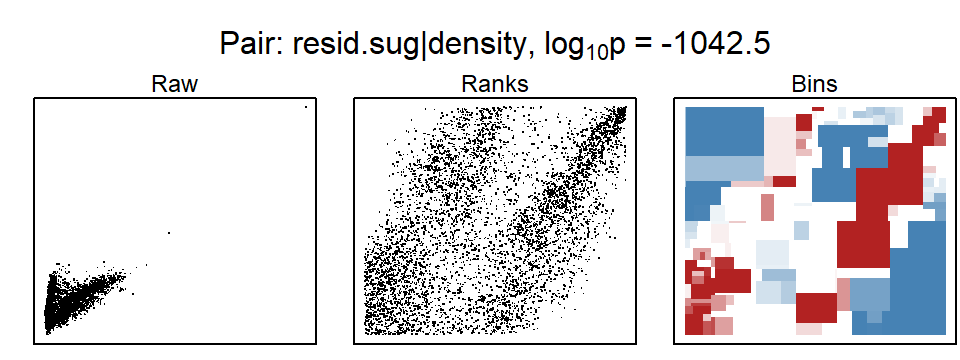}&
  \includegraphics[width = 0.45\textwidth]{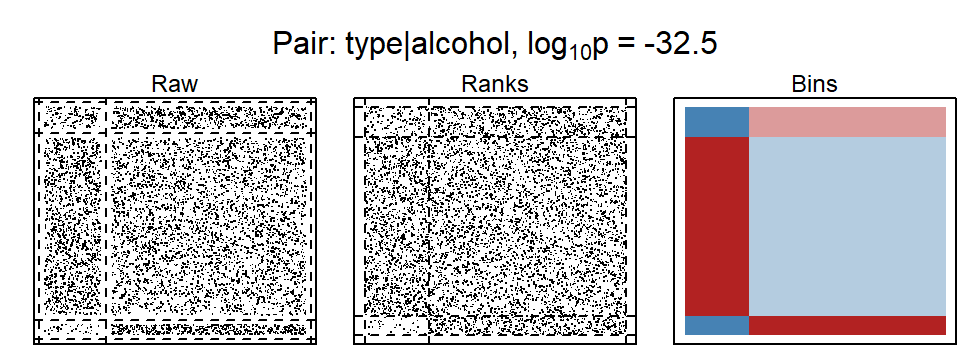}\\
  \includegraphics[width = 0.45\textwidth]{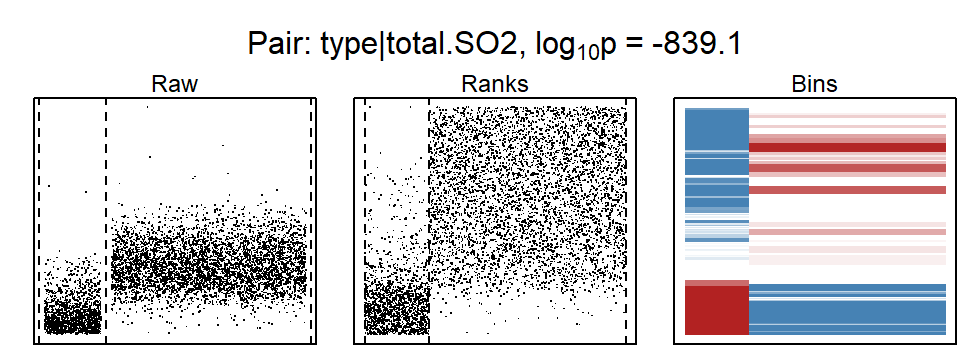}&
  \includegraphics[width = 0.45\textwidth]{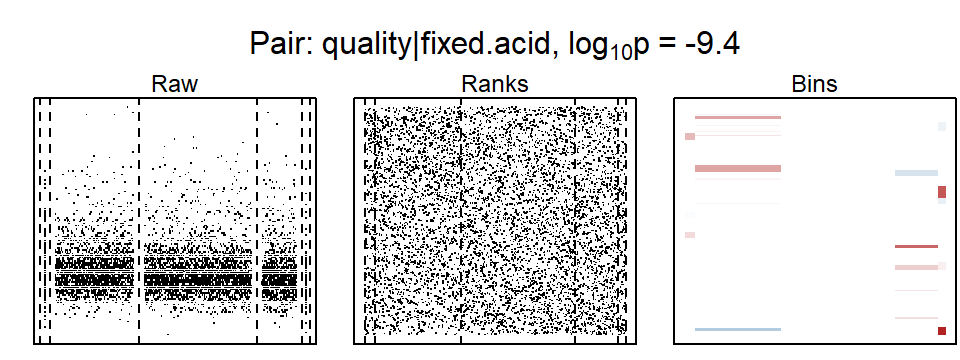}\\
  \includegraphics[width = 0.45\textwidth]{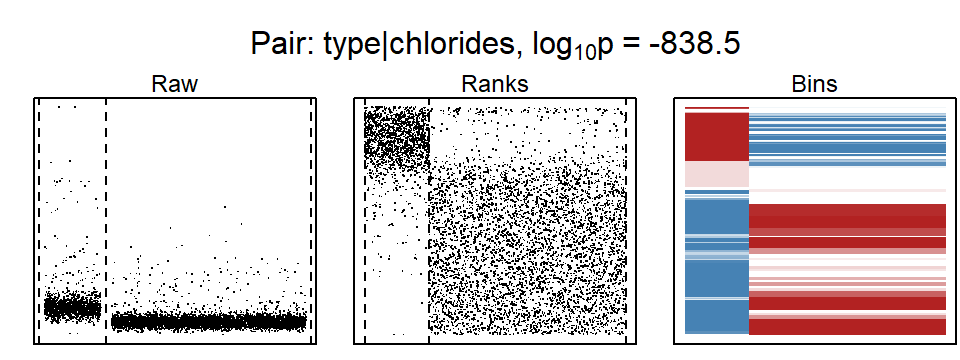}&
  \includegraphics[width = 0.45\textwidth]{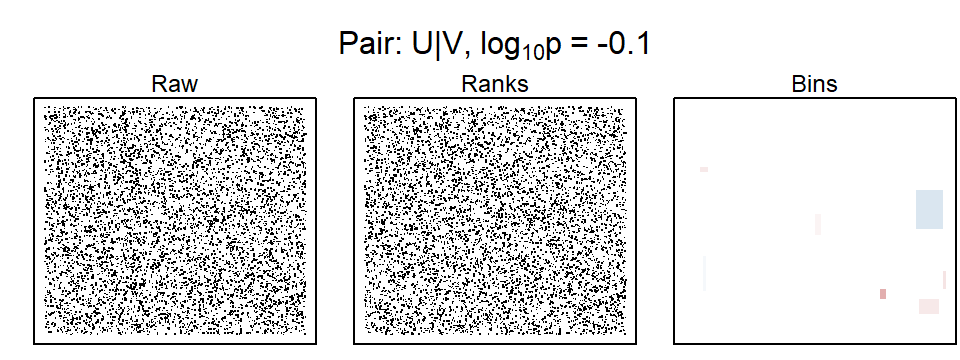}
  \\
  {\footnotesize (a) Four most dependent pairs} &
  {\footnotesize (b) Pairs indexed  29, 65, 77, and 99}
  \end{tabular}
  \caption{The pairs with $p_{simple}$-values of rank 1, 2, 3, 4, 29, 65, 77, and 99 in the wine data.}
  \label{fig:wineExample}
  \end{center}
\end{figure}

The dependencies of eight different variate pairs are shown in Figure \ref{fig:wineExample}
as a triple of displays for each pair.  The first of each triple is a scatterplot of the raw data (uniformly distributed within category for a categorical variate), the second the corresponding plot of ranks, and the third the departure display from recursive random binning.  Note that all values are slightly inset from the edges of each display and that categories are separated by dashed lines.  The colouring of bins in the departure displays mark discrepant bins (having absolute large standardized Pearson residuals) with deep saturation indicating the strength of the discrepancy and colour indicating sign (red is positive, blue negative; see Appendix \ref{app:colouring} for details). 

The four pairs of variates having smallest $p_{simple}$ values are arranged from smallest to largest down the left column of triples in Figure \ref{fig:wineExample}.  Down the right column, again in order of decreasing significance, are four more pairs selected at roughly equally spaced intervals from the remaining ordered $p$-values.

The most significant pair has ($x$, $y$) = (``free $SO_2$'' , ``total $SO_2$''), displayed in the top left triple.  The  saturated  colours clearly show strong lower and upper tail dependencies; the broadly linear pattern in the departure display suggests a monotonic relation between the original variates (with variation).  Next down is  ``density'' versus ``residual sugar'', which shows two strong monotonic relations, one having a higher upper tail dependency and the other a stronger lower tail dependency.
The next two down the left column show continuous versus categorical variates:  first ``total $SO_2$'' and then ``chlorides''  versus ``type'' of wine (left red, right white).   Low values of sulphur are associated with red and high values with white; chlorides show the opposite association.
Note that categorical boundaries determine the bins on that variate (width proportional to observed count); recursive random binning is only applied to continuous variates.

Moving through the pairs to the 29th most significant yields a pair of continuous variate, ``residual sugar'' versus ``volatile acidity'', shown at top right in Figure \ref{fig:wineExample}.  Here the dependence is strong but difficult to describe -- certainly a functional relation does not seem warranted.
Simply, the deeply saturated bins show where there are far more (red) or far fewer (blue) observations than expected under independence.  

Next down shows  ``alcohol content'' versus ``type''  having three and two categories respectively.
The strong independence here is easy to describe with red wines an unexpectedly low number having either high or low alcohol content as well as an unexpectedly high number with the middle level of alcohol content.  For white wine something like the opposite occurs -- there are definitely far more white wines than expected with low alcohol content, somewhat more with high content, and somewhat fewer with the middle level of content.  

The next pair shows a continuous (``fixed acidity'')  versus a five category variate (``quality'').
Though the $p$-value (even Bonferroni corrected) is still very small,  its departures appear to be isolated in small regions within essentially a single category,  the small dark red regions of the rightmost category.  
It is not hard to imagine that, while statistically significant, this departure may not be of any practical significance -- the quality of wine might well be treated in practice as independent of its fixed acidity (the exceptions may be too few to matter). 

The last pair shows the two artificially generated uniform variates and, as expected, no evidence appears against the hypothesis of independence and a very few bins are faintly coloured.

\section{Discussion} \label{sec:discussion}

Recursive random binning is applicable to assess the hypothesis of independence between any two variates.  When both are categorical,  no random binning occurs and it defaults to the usual two-way $R \by C$ contingency table test of independence, comparing Pearson's $\bigChi$ to a $\chi^2$ distribution with $(R-1)(C-1)$ degrees of freedom.  When both are continuous, the rank pairs are recursively binned at random to produce $K$ bins in total; the null distribution of the $\bigChi$ can now be usefully approximated by a $\chi^2$ with $(\sqrt{K} - 1)^2$ degrees of freedom.  If only one is categorical, say having $C$ categories, then the ranks of the continuous variate are recursively randomly binned within category and the null distribution of $\bigChi$ can be usefully approximated by a $\chi^2$ with $(\widebar{R} -1)(C-1)$ degrees of freedom, where $\widebar{R} = K/C$ is the average number of random bins per category.
Other approximations are also available, as discussed in Section \ref{sec:null_dist} and in more detail in Appendix \ref{app:mle_approx}.  

All approximations were shown to be competitive with, and sometimes superior to,  the recent grid-based methods FES and BET of \cite{FES_paper} and \cite{zhang2019BET}, respectively.
Being an irregular, random, and hence less predictable, tessellation of the unit square, recursive random binning is not \textit{a priori} tuned to any particular type of departure from independence.
It can uncover departures that occur for example from a noisy functional relation, a non-uniform but regular pattern, dependent patterns appearing only locally, simple tail dependence,  and even difficult to describe dependencies such as some arising in the wine data example of Section \ref{sec:wineEx}.  Moreover, the method is not constrained to only pairs of continuous variates.
As such, recursive random binning is well suited to screening pairs of variates of all types and ordering them according to the strength of the observed dependence.

Recursive random binning works hand in glove with the departure displays, where colouring and its saturation depend on the value of the standardized Pearson residuals in each bin.  The display shows the pattern of dependence when dependence is found, and shows little if any when no dependence exists.
Moreover, even, as in the ``fixed acid'' versus ``quality'' case observed in the wine data, the dependence information provided by the departure display allows the analyst to better assess the important scientific difference between statistical and practical significance.
As with the method on which it is based, the departure display is not tuned to any pattern of departure and provides a universal consistent display to interpret, regardless of whether the variates are continuous or categorical.  This again makes it well suited to screening pairs as only a single display type need be interpreted.

The method and display are widely applicable, consistent in construction and presentation, and
are not tuned to detect any particular non-null structure.  All of which makes them well suited to
data exploration where discovery of unanticipated patterns may be more important than those expected in advance.   Of course, should such \textit{a priori} patterns be important as well, and for which powerful detection methods are available, all could be combined in some ensemble method such as the recent BERET method of \cite{BERET2023} designed to supplement BET with complementary methods powerful where BET is not.

Another possibility which, like FES, BET, and BERET, would introduce the problem of multiple testing, would be to run some number of recursive random binnings on the same data and combine the $p$-values in some fashion \cite[e.g, see][for a number of possibilities]{SalahubMultipleTesting2023}.  
This too would take some study -- the binnings are independent but the data are not.

Recursive random binning is not in principle restricted to assessing only pairwise independence of variates.  
Just as with multi-way tables, the methodology naturally extends to assessing the joint independence of any number of variates at once.  Of course, the open question is whether the approximations to the null distribution of $\bigChi$ also extend just as naturally.
This would require further investigation but should not be out of reach for at least a small number of variates at once.
A simpler question would be whether one group of variates is independent of another group of variates.
Recursive random binning and departure displays could be applied to this problem by first transforming each set of variates (multivariate $X$ and multivariate $Y$) via a collapsing function (e.g., an inter-point distance function) to new univariate $X$ and $Y$  \cite[e.g., see][for details]{hofertOldfordEtAl_JMA}.
It is not hard to imagine taking this approach to sets and subsets of variates (of all types) in a large exploratory analysis.

Of course recursive random binning need not be constrained to use of $\bigChi$ as the measure of goodness of fit.  Any score
\[D = \sum_{k = 1}^K d(o_k, e_k)\]
could be computed over the final $K$ bins and used to sort variate pairs.
For example, the log likelihood ratio statistic \cite[e.g., see][]{BFH_loglinear}
\[ 
\bigGstat = 2 \sum_{k = 1}^K  o_k \log \left ( \frac{o_k}{e_k} \right ),
\]
or the mutual information  \cite[e.g.,][]{reshef2011MIC}
\[ 
\mutInf = \sum_{k = 1}^K  \frac{o_k}{N} \log \left ( \frac{o_k}{e_k} \right ) = \frac{\bigGstat}{2N},
\]
might be calculated instead.  
Any change is easily accommodated by the software available in the package \pkg{AssocBin}.
Again, approximations to the null distribution of either would need to be investigated though that of $\bigGstat$ might be expected to be close to that of $\bigChi$ (as it is for multiway tables).
Similarly, deviance residuals might be expected to be exchanged with the Pearson residuals in the departure displays.

Finally, the detection and display of dependence is an important part of any data analysis. Being pairwise, it can be considered as one more scagnostic \cite[e.g.,][]{TukeyGraphicsEDA85, wilkinson2005graph} that should be applied to all pairs of variates in an exploratory data analysis.  As such, it and the departure display could be incorporated into existing visualization methods including scatterplot matrices, or for large numbers of pairs into navigation graphs \citep{navGraphs2011} or zenplots \citep{hofertoldford2020zigzag}.  
If incorporated as interactive \cite[e.g., using \pkg{loon}][]{loonR} the display of ordered $p$ values and the departure displays could be made even more powerful  for data exploration.
It is hard to imagine a data analysis that would not be better informed by first searching over all pairs for their dependence structure.  

\section*{Software}
All methods and displays are implemented in the \pkg{AssocBin} \R{} package available on CRAN (\citealp{AssocBin, Rlang}; see Appendix \ref{app:AssocBin} for construction of plots in this example).
\bibliographystyle{chicago}

\bibliography{./fullbib}

\newpage
\appendix

\section{On the null distribution of paired ranks}
\label{app:paired_ranks}
\begin{figure}[!ht]
	\begin{center}
		\begin{tabular}{ccc}
			\includegraphics[width= 0.2\textwidth]{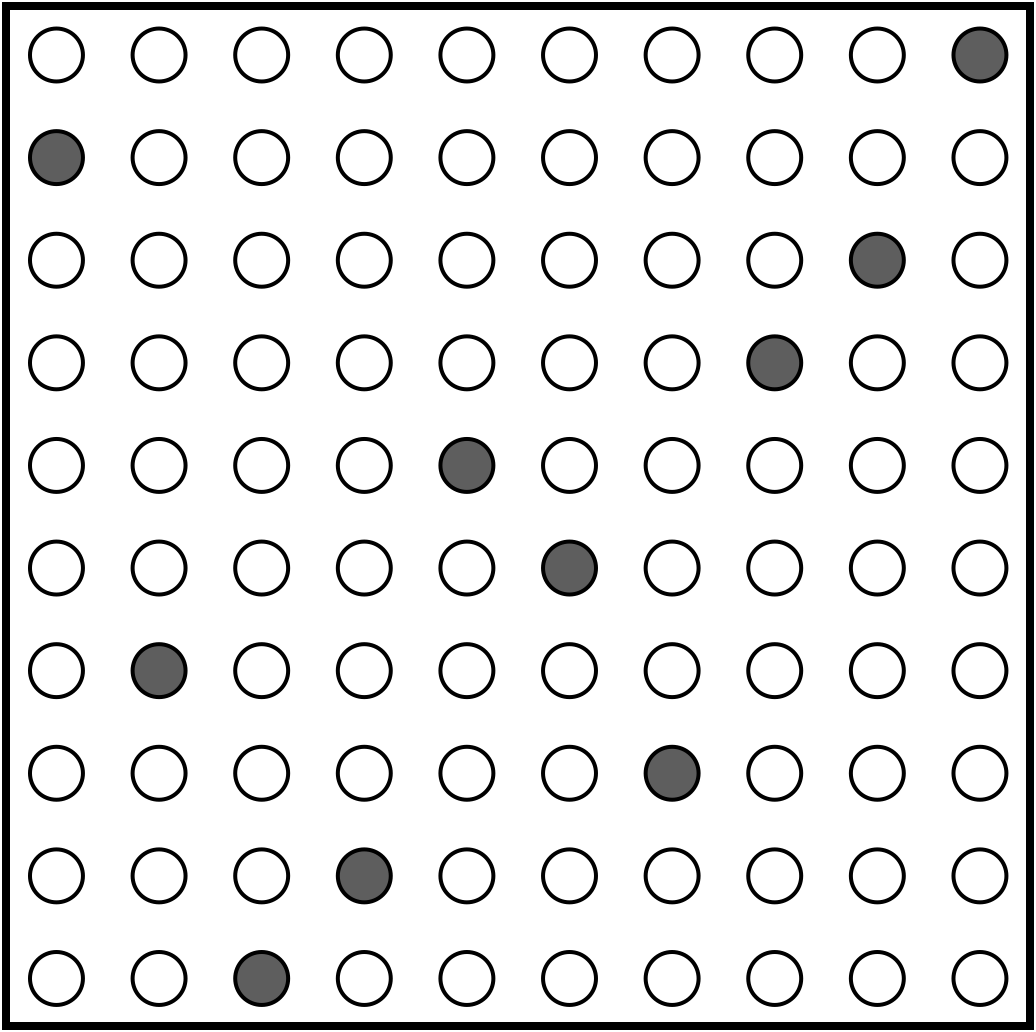} & 
			$~~~~~~~~~~~$ &
			\includegraphics[width= 0.2\textwidth]{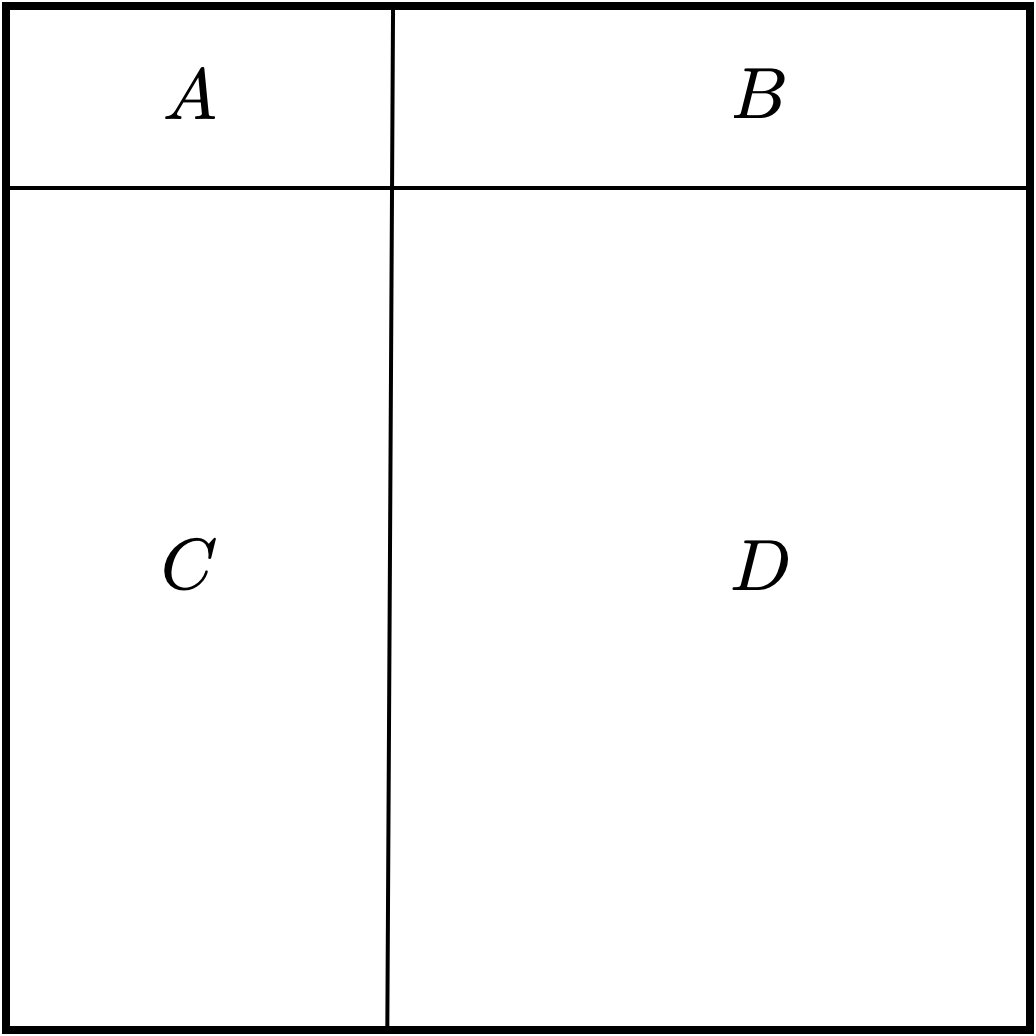}  \\
			{\footnotesize (a) An $\samp{S}$ for $N = 10$} & 
			 &
			{\footnotesize (b) Block labels } 
		\end{tabular}
		\caption{Under $\Hyp_0$: (a) Filled circles identify a generated configuration for $N = 10$; (b) Determining the probability of observed counts in an $r \times c$ rectangle $A$.}
		\label{fig:rank_lattice}
	\end{center}
\end{figure}

Consider the set of rank pairs $(i, j) \in \{1, 2, \ldots, N\} \times \{1, 2, \ldots, N\} $  as points on an $N \times N$ lattice and let 
$\samp{S}$ be a sample of $N$ rank pairs generated under the null hypothesis $\Hyp_0$ of independence.
Further, let
$Z _{ij}$ be a Bernoulli random variable taking the value one if the rank pair $(i,j) \in \samp{S}$ and zero otherwise.
Assuming that no ties are possible in the ranks, the $Z_{ij}$ are subject to the following constraints:
\[ \sum_{i=1}^N Z_{ij} = 1 ~\text{ for all } ~ j ~ \text{ and }  \sum_{j=1}^N Z_{ij} = 1  ~\text{ for all } ~ i .\]
That is, exactly one location in each row and exactly one location in each column of the lattice can be occupied in any given sample (e.g., see Figure \ref{fig:rank_lattice}(a) for a sample $\samp{S}$ when $N = 10$).

It follows, for example, that 
\[ Z_{i_1j_1} \times Z_{i_2j_2} = 0  ~~ \text{ whenever } i_1 = i_2 ~~\text{ or }  j_1 = j_2.\]
The total number of possible samples is easily seen to be $N!$, found by choosing from the first column one row to be occupied from the $N$ available, followed by one to be occupied from the remaining $N-1$ rows available in the second column, and so on.

Under the null hypothesis of independence $\Hyp_0$, each such sample of $N$ paired ranks, $\samp{S}$, has the same probability of being realized, namely 
\[ Pr(\samp{S}) = \frac{1}{N!} .\]

\subsection{Probabilities of $Z_{ij}$s}
From this, we can calculate the probability that $Z_{ij} = 1$ for any specified rank pair $(i, j)$. 
This is
\[Pr(Z_{ij} = 1) = Pr(\samp{S} \ni (i,j)) = \frac{\text{number of samples containing } (i, j) }{ \text{total number of samples}}
= \frac{(N-1)!}{N!} = \frac{1}{N}.\]
Clearly, $Pr(Z_{ij} = 0) \rightarrow 1$ and $Pr(Z_{ij} = 1) \rightarrow 0$  as $N \rightarrow\infty$.

Expectations follow immediately. For any pair of ranks $(i, j)$,
\[ E\left(Z_{ij}\right) = 1 \times Pr(Z_{ij} = 1) + 0 \times Pr(Z_{ij} = 0) = \frac{1}{N}\]
and
\[Var(Z_{ij}) = E(Z_{ij}^2) - \left(E(Z_{ij})\right)^2 = \frac{1}{N} - \left( \frac{1}{N}\right)^2 = \frac{N-1}{N^2}.\]

Similarly, for $i_1 \neq i_2$ and $j_1 \neq j_2$, we can determine
\[ Pr \left(\samp{S} \supset \{(i_1, j_1), (i_2, j_2) \} \given i_1 \neq i_2 \text{ and } j_1 \neq j_2 \right)  = Pr \left( Z_{i_1j_1} \times Z_{i_2j_2} = 1 \right)  = \frac{1}{N(N-1)}.\]
And, since each row and column must sum to one,  it must be that
\[  Pr \left( Z_{ij_1} \times Z_{ij_2} = 1 \right) = Pr \left( Z_{i_1j} \times Z_{i_2j} = 1 \right) = 0.\]
More generally, for any specified set $\samp{S}_k$ of $k$ specified legal positions in $\samp{S}$, the probability is
\[  Pr \left(\samp{S} \supset \samp{S}_k \right) = Pr \left(\prod_{(i,j) \in \samp{S}_k}  Z_{ij} = 1 \right) 
= \frac{1}{N^{[k]}}
 = \frac{1}{N(N-1) \cdots (N - k + 1)} 
= \frac{(N-k)!}{N!}.\]
The following conditional probabilities trivially hold based on the constraints:
\[ Pr \left( Z_{ij_1} = 1 \given Z_{ij_2} = 1 \right) = 0  ~~\text{ and } ~~ Pr \left( Z_{ij_1} = 0 \given Z_{ij_2} = 1 \right) = 1\]
\[ Pr \left( Z_{i_1j} = 1 \given Z_{i_2j} = 1 \right) = 0  ~~\text{ and } ~~ Pr \left( Z_{i_1j} = 0 \given Z_{i_2j} = 1 \right) = 1\]
\[ Pr \left( Z_{ij_1} = 1 \given Z_{ij_2} = 0 \right) = \frac{1}{N-1}  ~~\text{ and } ~~ Pr \left( Z_{ij_1} = 0 \given Z_{ij_2} = 0 \right) = \frac{N-2}{N-1}\]
\[ Pr \left( Z_{i_1j} = 1 \given Z_{i_2j} = 0 \right) = \frac{1}{N-1}  ~~\text{ and } ~~ Pr \left( Z_{i_1j} = 0 \given Z_{i_2j} = 0 \right) = \frac{N-2}{N-1}\]
For those in different rows and columns, we have
\[ Pr( Z_{i_1j_1} = 1 \given  Z_{i_2j_2} = 1) 
= \frac{Pr( Z_{i_1j_1} = 1, Z_{i_2j_2} = 1)}{Pr( Z_{i_2j_2} = 1)}
= \frac{ Pr \left( Z_{i_1j_1} \times Z_{i_2j_2} = 1 \right) }{Pr( Z_{i_2j_2} = 1)}
= \frac{1}{N-1}
\]
and so
\[ Pr( Z_{i_1j_1} = 0 \given  Z_{i_2j_2} = 1) 
= \frac{N-2}{N-1}
\]
Similarly,  it can be determined that
\[\begin{array}{rcl}

Pr(Z_{i_1j_1} = 1,   Z_{i_2j_2} = 0) &=& \frac{\text{number of samples containing}~ (i_1,j_1)~ \text{but }\textbf{not} ~ (i_2,j_2)}{N!}\\
&&\\
&=& \frac{ 1 \times (N-2) \times (N- 2)!}{N!}\\
&&\\
&=& \frac{  (N-2)}{N(N-1)} = Pr(Z_{i_1j_1} = 0,   Z_{i_2j_2} = 1).
\end{array}
\]
From which it follows that
\[ Pr(Z_{i_1j_1} = 1 \given  Z_{i_2j_2} = 0) 
= \frac{ \frac{  (N-2)}{N(N-1)}}{\frac{N-1}{N}} 
= \frac{N-2}{(N-1)^2}.
\]
It now must be that
\[\begin{array}{rcl}
Pr(Z_{i_1j_1} = 0,   Z_{i_2j_2} = 0) &=& 1 - Pr(Z_{i_1j_1} = 1,   Z_{i_2j_2} = 1) - 2 \times Pr(Z_{i_1j_1} = 1,   Z_{i_2j_2} = 0)\\
&&\\
&=& 1 - \frac{1}{N(N-1)} -  2 \times \frac{  (N-2)}{N(N-1)}.\\
&&\\
&=& \frac{ N(N-1) - 1 - 2(N-2)}{N(N-1)}\\
&&\\
&=& \frac{ N^2 - 3N +3}{N(N-1)}.
\end{array}
\]
And so,
\[\begin{array}{rcl}
Pr(Z_{i_1j_1} = 0 \given Z_{i_2j_2} = 0) 
&=&  \frac{\frac{ N^2 - 3N +3}{N(N-1)}}{\frac{N-1}{N}} \\
&&\\
&=& \frac{ N^2 - 3N +3}{(N-1)^2} = 1 - Pr(Z_{i_1j_1} = 1 \given  Z_{i_2j_2} = 0).
\end{array}
\]

While the random variates $Z_{ij}$ are identically distributed under $\Hyp_0$, they are clearly not distributed independently of one another.  Indeed, no pair of $Z_{ij}$s are independently distributed, whatever their indices.

\subsection{Covariances}
Since they are not independent, covariances between pairs of $Z_{ij}$s will also depend upon the indices.
As before, take $i_1 \neq i_2$ and $j_1 \neq  j_2$.

Now
\[ 
E(Z_{i_1j_1} \times Z_{i_2j_2} ) = Pr(Z_{i_1j_1} \times Z_{i_2j_2} = 1) = \frac{1}{N(N-1)}
\]
gives covariance
\[
Cov(Z_{i_1j_1},  Z_{i_2j_2}) = \frac{1}{N(N-1)} - \left(\frac{1}{N}\right) \times \left(\frac{1}{N}\right)
= \frac{1}{N^2(N-1)}
\]
and correlation
\[
Cor(Z_{i_1j_1},  Z_{i_2j_2}) = \frac{ \frac{1}{N^2(N-1)}}{\frac{N-1}{N^2}} = \frac{1}{(N-1)^2}.
\]
Those in the same row (or the same column) have
\[
E(Z_{ij_1} \times Z_{ij_2} ) = Pr(Z_{ij_1} \times Z_{ij_2} = 1) = 0
\]
giving covariance

\[
Cov(Z_{ij_1},  Z_{ij_2}) = - \left(\frac{1}{N^2}\right) 
\]
and correlation
\[
Cor(Z_{ij_1},  Z_{ij_2}) = - \frac{1}{N-1}.
\]

The $N^2 \by N^2$ variance-covariance matrix will have rank $(N-1)^2$ since there are $2N-1$ independent linear
constraints on the $Z_{ij}$s.

\subsection{On expected counts}
Consequently, for any set $\set{S}_m$  of $m$ pairs on the $N \times N$ lattice, 
\[ E\left( \sum_{(i, j) \in \set{S}_m} Z_{ij}\right)  =   \sum_{(i, j) \in \set{S}_m}  E\left(Z_{ij}\right) = \frac{m}{N}.\]
In particular, when $\set{S}_m$ is a rectangular block in the array having $r$ rows and $c$ columns, $m= rc$ and the
expected number of occupied locations in the rectangular block will be
\[ \left(\begin{array}{c}\text{expected count in} \\ r \times c  ~\text{block} \end{array} \right) = \frac{rc}{N} = N \times \frac{rc}{N^2} 
=  \left(\text{sample size} \right) \times ~\left(\begin{array}{c} \text{proportion of lattice} \\ \text{within that block} \end{array} \right).\]

For convenience,  suppose that $\samp{S}_m$ is an $r$ by $c$ rectangular block in the position of $A$ in Figure \ref{fig:rank_lattice}.  Then $(i, j) \in \samp{S}_m$  has $i = 1, \ldots, r$ and $j = 1, \ldots, c$ with $m = rc$.
%
%

\subsection{On observed counts}
Let $A$ be defined by $r$ rows and $c$ columns of the lattice of possible rank pairs. 
Without any loss of generality under $\Hyp_0$,  $A$ can be a contiguous $r \times c$ rectangular block located in the top left corner as shown in Figure \ref{fig:rank_lattice}(b).
The remaining blocks labelled $B$, $C$, and $D$ in Figure \ref{fig:rank_lattice}(b) follow from the definition of $A$ and consist of 
sub-lattices of dimensions $r \times (N-c)$, $(N-r) \times c$, and $(N-r) \times (N-c)$, respectively.
Of interest is the probability that $A$ contains exactly $k$ elements of a sample configuration $\samp{S}$ generated when  $\Hyp_0$ holds.  

That is, we want the probabilities
\[ p_k = Pr\left( \sum_{(i, j) \in A} Z_{ij} = k \biggiven \Hyp_0 ~\text{holds}\right) ~~\text{for any} ~~k \in \{0, 1, 2, \ldots, \min(r, c)\}\]
Since all possible samples $\samp{S}$ are equiprobable, $p_k$ will simply be the proportion of samples for which 
$\sum_{(i, j) \in A} Z_{ij} = k$.  

We need only determine the number of such samples and divide by $N!$ which we do by counting 
how many arrangements are possible for each of the blocks $A$, $B$, $C$, and $D$ when exactly $k$ lattice points in $A$
must appear in $\samp{S}$. 

To simplify, suppose $r \leq c$ so that $k \in \{0, 1, 2, \ldots, r\}$.  

First for block $A$.  If there are $k$ pairs from $\samp{S}$ in $A$, they can be located within $A$ in exactly
\[ \binom{r}{k} \times c^{[k]} =  \binom{r}{k} \times \frac{c!}{(c-k)!}\]
ways by first choosing which rows contain a point in the configuration and then selecting which column.

For block $B$, $k$ rows may be removed according to which rows of $A$ contain a configuration.  
Block $B$ is thus reduced to be of size $(r-k) \times (N-c)$ and exactly $(r-k)$ of these lattice points must be in $\samp{S}$.
For this to be possible, it must be that $N-c \geq r-k$.
It is now simply a matter of choosing a different column from each of the $(r-k)$ in succession 
which can be done in
\[(N-c)^{[(r-k)]} = \frac{(N-c)!}{(N-c-r+k)!} \]
ways.

For block $C$, $k$ columns can similarly be removed.
This leaves a block of size $(N-r) \times (c-k)$ available for points to appear from the configuration  $\samp{S}$ not already 
assigned to blocks $A$ or $B$.
Similarly, the block $D$ can be reduced by those columns of $B$ containing points from $\samp{S}$.  
There were $(r-k)$ such columns, making the reduced $D$ a lattice of dimension $(N-r) \times (N-c -(r-k))$.
Prepending to this the unassigned columns of $C$ gives a lattice of size $(N-r) \times (N-r)$; 
the remaining $N-r$ points of $\samp{S}$ are now to be assigned (one per row and one per column) to this new  $(N-r) \times (N-r)$ lattice. 
This can be done in exactly
\[ (N-r)! \]
ways.

Putting the three pieces together and dividing by $N!$, the total number of samples possible, gives
\[
\begin{array}{rcl}
p_k &=& \left( \binom{r}{k} \times \frac{c!}{(c-k)!}\right) \times \left( \frac{(N-c)!}{(N-c-r+k)!}\right) \times \left((N-r)!\right) \times \left(\frac{1}{N!}\right) \\
&& \\
&=&
  \frac{r!c! (N-c)!(N-r)!}{k!(r-k)!(c-k)!(N-c-r+k)!N!}  \\
&& \\
&=&
  \frac{\binom{c}{k}
          \binom{N-c}{r-k}} 
          {\binom{N}{r}}
 \end{array}
 \]
giving $p_k = Pr(K=k)$ for a hypergeometric random variable $K \follows Hypergeometric(N, r, c)$.

This is the probability when $r \leq c$ and $N - c - r + k \geq 0$.   
If $r > c$ then the symmetry of the problem allows the same calculation, and the correct formula is had by swapping $r$ and $c$ in the above formula.  But, as the second line shows (and as is well known for the hypergeometric), the formula is interchangeable in $r$ and $c$.  Indeed, the following are equivalent:
\[
\begin{array}{rcl}
p_k &=&  \frac{\binom{c}{k}
          \binom{N-c}{r-k}} 
          {\binom{N}{r}}  \\
&& \\
&=&
  \frac{\binom{r}{k}
          \binom{N-r}{c-k}} 
          {\binom{N}{c}} \\
&& \\
&=&
  \frac{\binom{r}{k}}{\binom{N}{r}} \times
  \frac{\binom{c}{k}}{\binom{N}{c}} \times
  \frac{\binom{r+c}{r} \binom{N+k}{r+c}}
         {\binom{N+k}{k}}
 \\
\end{array}
\]
with the last expression emphasizing the symmetry of $r$ and $c$.  This holds for all $k$ from
$\max(0, c + r -N)$ to $\min(r, c)$.

Being hypergeometric, numerous results are available, including that
\[
E(K) = \frac{rc}{N} ~~~ \text{as found before, and } ~~~ Var(K) = \frac{rc}{N}~~\frac{(N-r)(N-c)}{N(N-1)}.
\]
\subsection{Covariance of counts}
Any $r \by c$ rectangular block $B$ can be defined by two $N \by 1$ vectors of indicators: the first $\ve{r} = \tr{(r_1, \ldots, r_N)}$ identifies which rows are partially covered by $B$, and $\ve{c} = \tr{(c_1, \ldots, c_N)}$ identifies which columns are.  Each coordinate is an indicator
\[ r_i = \left\{
             \begin{array}{ll}
               1 & B ~\text{covers row }~i  \\
                0 &\text{otherwise} 
                \end{array}
            \right. 
    ~~~ \text{ and } ~~~
     c_j = \left\{
     \begin{array}{ll} 
     1 & B ~\text{covers column }~j \\
     0  & \text{otherwise.}
      \end{array}
       \right.
 \]
 The count of paired ranks appearing in $B$ can now be written as
 \[K = \sum_{(i,j)\in B}Z_{ij} = \tr{\ve{r}}\m{Z}\ve{c} \]
 where $\m{Z} = [Z_{ij}]$ is the $N \by N$ matrix of the $Z_{ij}$s.
 Applying the standard $vec$ operator to $\m{Z}$ stacks its column vectors as
 \[vec(\m{Z}) = \tr{(Z_{11}, Z_{21}, \ldots, Z_{N1}, Z_{12}, \ldots, Z_{N2}, Z_{13}, \ldots, Z_{NN})}\]
 and provides another expression for $K$, namely,
\[K = \tr{\ve{a}} vec(\m{Z})\]
where $\ve{a}$ is an $N^2 \by 1$ vector of indicators selecting the appropriate elements of $vec(\m{Z})$.  The
 $\ve{a}$ can be written as
\[\ve{a} = \ve{c} \otimes \ve{r}\]
where $\otimes$ is the Kronecker product.

This is notationally convenient for any two rectangular blocks, say $B_1$ and $B_2$ on the lattice and their corresponding sums $K_1 = \tr{\ve{a}_1} vec(\m{Z})$ and  $K_2 = \tr{\ve{a}_2} vec(\m{Z})$.  The covariance of the two counts is simply 
\[Cov(K_1, K_2) = \tr{\ve{a}_1} \sm{\Sigma}_{\m{Z}} \ve{a}_2 \]
where $\ve{a}_i = \ve{c}_i \otimes \ve{r}_i$, for $i = 1, 2$, and $\sm{\Sigma}_{\m{Z}}$ is the $N^2 \by N^2$ 
variance covariance matrix of $vec(\m{Z})$, the elements of which were determined earlier.
Each $\ve{a}_i = \ve{c}_i \otimes \ve{r}_i$ is constructed from the indicators of the rows and columns of $B_i$.

Indeed, if there are $b$ such blocks, $B_1, \ldots, B_b$ (e.g., possibly tessellating the lattice as in recursive binning), with counts
in each of $K_1, \ldots, K_b$, then the variance covariance matrix of the $b \by 1$ vector of random counts $\ve{K} = \tr{(K_1, \ldots, K_b)}$ will be the $b \times b$ matrix
\[ \sm{\Sigma}_{\ve{K}} = Var(\ve{K}) = \tr{\m{A}} \sm{\Sigma}_{\m{Z}}\m{A}\]
where $\m{A} = [\ve{a}_1, \ldots, \ve{a}_b]$ and each $\ve{a}_i = \ve{c}_i \otimes \ve{r}_i$ is determined by those rows and columns of the lattice which determine $B_i$.

\subsection{Covariance of a pair of counts}
\begin{figure}[!ht]
	\begin{center}
			\includegraphics[width= 0.4\textwidth]{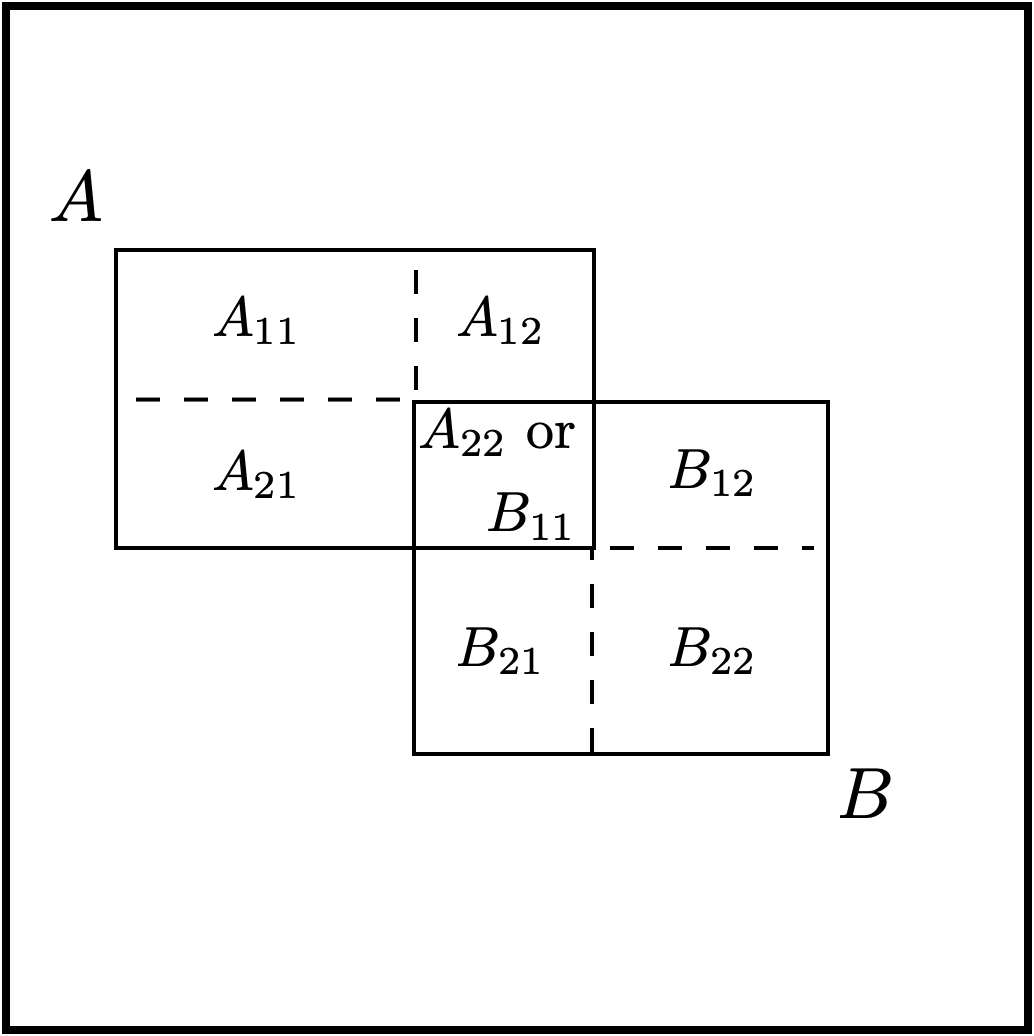}		
			\caption{Intersecting blocks: $A$ contains an $r_A \times c_A$  rectangle of the rank lattice, $B$ contains an $r_B \by c_B$ rectangle.  Rectangular sub-blocks are $A_{ij}$ and $B_{ij}$ for $i, j = 1,2$ with $A_{22} = B_{11} = A\intersect B$ of dimension $r_I \by c_I$.}
		\label{fig:intersecting_blocks}
	\end{center}
\end{figure}

Each entry of $Var(\ve{K})$ will be the covariance of the counts in two different blocks on the lattice. 
To be completely general, suppose $A$ and $B$ are two (possibly intersecting) blocks as shown in Figure \ref{fig:intersecting_blocks}.
Of interest is the covariance $Cov(K_A, K_B)$ of the counts $K_A$ and $K_B$ in $A$ and $B$.  This will be the sum of the covariances, $Cov(Z_{ij}, Z_{kl})$, for all pairs $(i,j) \in A$ and pairs $(k, \ell) \in B$. 
To determine this sum we break it into pieces, by considering the $Z_{ij}$s in each of $A_{11}, A_{12}, A_{21}, A_{22}$ with the $Z_{ij}$s in each sub-region of $B$.  For each comparison, the number of $Z$ pairs having the same row and column, the same row but different columns, the same column but different rows, and different rows and columns needs to be determined.

The results are as follows:
\begin{enumerate}
\item $A_{11}$ with each of $B_{ij}$s.  There is no common row or column with any of the $B_{ij}$s.  Total contribution to the covariance is
therefore 
\[(r_A - r_I)(c_A - c_I) \times r_B c_B  \times \left( \frac{1}{N^2(N-1)}\right) \]

\item $A_{12}$ with each of the $B_{ij}$s. 
     \begin{itemize}
     \item $A_{11} \times (B_{12} \union B_{22})$. No rows, no columns in common.  Contributions to the covariance:
     \[(r_A - r_I)c_I \times r_B (c_B - c_I) \times \left( \frac{1}{N^2(N-1)}\right) \]
     \item $A_{12} \times(B_{11} \union B_{21})$. Two conditions hold: those $Z_{ij} \in A_{12}$ and $Z_{k\ell} \in (B_{11} \union B_{21})$ that 
     \begin{enumerate}
          \item share columns ($j = \ell$), but not rows ($ i \neq k$)  contribute
          \[  c_I \times (r_A - r_I) \times r_B \times \left(-\frac{1}{N^2} \right)\]
          \item share neither ($i \neq k$ and $j \neq \ell$)  contribute
          \[c_I \times (r_A - r_I) \times (c_I -1)r_B \times \left( \frac{1}{N^2(N-1)}\right) \]
     \end{enumerate}
     \end{itemize}
\item $A_{21}$ with each of the $B_{ij}$s. 
     \begin{itemize}
     \item $A_{21} \times (B_{21} \union B_{22})$. No rows, no columns in common.  Contribution to the covariance:
     \[r_I (c_A - c_I) \times (r_B - r_I) c_B \times \left( \frac{1}{N^2(N-1)}\right) \]
     \item $A_{21} \times(B_{11} \union B_{12})$. Two conditions hold: those $Z_{ij} \in A_{21}$ and $Z_{k\ell} \in (B_{11} \union B_{12})$ that 
     \begin{enumerate}
          \item share rows ($i = k$), but not columns ($j \neq \ell$) which contribute
          \[  r_I \times (c_A - c_I)\times   c_B \times \left(-\frac{1}{N^2} \right)\]
          \item share neither ($i \neq k$ and $j \neq \ell$) which contribute
          \[r_I \times (c_A - c_I) \times (r_I -1) c_B \times \left( \frac{1}{N^2(N-1)}\right) \]
     \end{enumerate}
     \end{itemize}
\item $A_{22}$ with each of the $B_{ij}$s. 
     \begin{itemize}
     \item $A_{22} \times B_{11}$. Four conditions hold:  those $Z_{ij} \in A_{22}$ and $Z_{k\ell} \in B_{11}$
     \begin{enumerate}
          \item share both rows and columns ($i = k$ and $j = \ell$) which contribute
          \[ r_I \times c_I \times \left( \frac{N-1}{N^2}\right)\]
          \item share rows ($i = k$), but not columns ($j \neq \ell$) which contribute
          \[  r_I \times (c_I - 1) \times c_I \left(-\frac{1}{N^2} \right)\]
          \item share columns ($j = \ell$), but not rows ($ i \neq k$) which contribute
          \[  c_I \times (r_I - 1) \times r_I \left(-\frac{1}{N^2} \right)\]
          \item share neither row nor column  ($i \neq k$ and $j \neq \ell$) which contribute
          \[ r_I \times c_I \times (r_I -1)(c_I -1)\left( \frac{1}{N^2(N-1)}\right) \]
     \end{enumerate}
     \item $A_{22} \times B_{12}$. Two conditions hold: those $Z_{ij} \in A_{22}$ and $Z_{k\ell} \in B_{12}$ that 
     \begin{enumerate}
          \item share rows ($i = k$), but not columns ($j \neq \ell$) which contribute
          \[  r_I \times c_I \times (c_B - c_I) \left(-\frac{1}{N^2} \right)\]
          \item share neither row nor column  ($i \neq k$ and $j \neq \ell$) which contribute
          \[ r_I \times c_I \times (r_I -1)(c_B - c_I)\left( \frac{1}{N^2(N-1)}\right) \]
     \end{enumerate}
     
     \item $A_{22} \times B_{21}$. Two conditions hold: those $Z_{ij} \in A_{22}$ and $Z_{k\ell} \in B_{21}$ that 
     \begin{enumerate}
          \item share columns ($j = \ell$), but not rows $i \neq k$) which contribute
          \[  c_I \times r_I \times (r_B - r_I) \left(-\frac{1}{N^2} \right)\]
          \item share neither row nor column  ($i \neq k$ and $j \neq \ell$) which contribute
          \[ c_I \times r_I \times (c_I -1)(r_B - r_I)\left( \frac{1}{N^2(N-1)}\right) \]
     \end{enumerate}
     \item $A_{22} \times B_{22}$. All $Z_{ij} \in A_{22}$ and $Z_{k\ell} \in B_{22}$ share neither row nor column ($i \neq k$ and $j \neq \ell$). They contribute 
          \[ c_I \times r_I \times (c_B - c_I)(r_B - r_I)\left( \frac{1}{N^2(N-1)}\right)
           \]

     \end{itemize}

\end{enumerate}

The covariance $Cov(K_A, K_B)$ will be the sum of all of these contributions.  Namely, 
\[
\begin{array}{rcl}
Cov(K_A, K_B) & = &\left(\frac{1}{N^2 (N-1)}\right)\\
&&~~~ \times 
 \left[ (r_A - r_I)(c_A - c_I)r_Bc_B 
              + (r_A - r_I)c_Ir_B(c_B - 1) \right. \\
        
              && ~~~~~~~~~~~~
  		+ ~~~~r_I(c_A -c_I)(r_B-1)c_B 
   		+ r_Ic_I(r_B -1)(c_B -1) \\
		&&\\
		&& ~~~~~~~~~~~~ 
		-(N-1) \left[(r_A - r_I)c_Ir_B +
		                        r_I(c_A - c_I) c_B 
		                        \right. \\ 
		&&\left. ~~~~~~~~~~~~~~~~~~~~~~ ~~ ~~~~~~~~~~+ ~~  r_Ic_I (c_B -1) + r_I c_I (r_B - 1)                                                 
 \right]\\ 
                 &&\\
		&& \left.  ~~~~~~~~~~~~  + (N-1)^2  r_Ic_I   ~~~ \right]                        
 \end{array}
 \]

When $A$ and $B$ do not overlap, as is the case in a tessellation, either $r_I = 0$ or $c_I = 0$, or both are zero.
For tessellation the following cases matter
\begin{enumerate}
\item If $r_I = c_I = 0 $ then 
\[
Cov(K_A, K_B) = \frac{r_Ac_Ar_Bc_B}{N^2(N-1)} = \frac{1}{N-1} \left(\frac{r_Ac_A}{N}\right) \left(\frac{r_Bc_B}{N}\right)
 \]
\item If $r_I = 0$, $c_I \neq 0 $ then 
\[
Cov(K_A, K_B) = \frac{r_Ar_B(c_Ac_B -Nc_I)}{N^2(N-1)} = \frac{1}{N-1} \left(\frac{r_Ac_A}{N}\right)\left[\left(\frac{r_Bc_B}{N}\right)
                                                                                                                          -  \left(\frac{c_I}{c_A}\right) r_B\right]
 \]
\item If $r_I \neq 0$, $c_I = 0 $ then 
\[
Cov(K_A, K_B) = \frac{c_Ac_B(r_Ar_B -Nr_I)}{N^2(N-1)} = \frac{1}{N-1}\left(\frac{r_Ac_A}{N}\right)\left[ \left(\frac{r_Bc_B}{N}\right)
                                                                                                                          -  \left(\frac{r_I}{r_A}\right)c_B\right]
 \]
\item If $r_I = r_A$, then 
\[
\begin{array}{rcl}
Cov(K_A, K_B) & = &\left(\frac{r_A}{N^2 (N-1)}\right)\\
&&~~~ \times 
 \left[ (c_A -c_I)(r_B-1)c_B 
   		+ c_I(r_B -1)(c_B -1)   \right.\\
		&&\\
		&& ~~~~~~~~~~~~ 
		-(N-1) \left[
		                        (c_A - c_I) c_B 
		                        +  c_I (c_B -1) + c_I (r_B - 1)                                                 
 \right]\\ 
                 &&\\
		&& \left.  ~~~~~~~~~~~~  + (N-1)^2  c_I   ~~~ \right]                        
 \end{array}
 \]
 N.B. If $r_A = r_I = r_B$ they are two blocks occupying identical rows.
 
\item If $c_I = c_A$, then 
\[
\begin{array}{rcl}
Cov(K_A, K_B) & = &\left(\frac{c_A}{N^2 (N-1)}\right)\\
&&~~~ \times 
 \left[ (r_A - r_I)r_B(c_B - 1) 
   		+ r_I(r_B -1)(c_B -1) \right. \\
		&&\\
		&& ~~~~~~~~~~~~ 
		-(N-1) \left[(r_A - r_I)r_B + r_I(c_B -1) + r_I (r_B - 1)                                                 
 \right]\\ 
                 &&\\
		&& \left.  ~~~~~~~~~~~~  + (N-1)^2  r_I  ~~~ \right]                        
 \end{array}
 \]
 N.B. If $c_A = c_I = c_B$ they are two blocks occupying identical columns.

\item If $r_A = r_I$ and $c_I = c_A$, then $A \subset B$ and
\[
\begin{array}{rcl}
Cov(K_A, K_B) & = &\left(\frac{r_Ac_A}{N^2 (N-1)}\right)
 \left[  (r_B -1)(c_B -1) \right. \\
		&&\\
		&& ~~~~~~~~~~~~ ~~~~~~~~~~~~ 
		-(N-1) \left[ (c_B -1) + (r_B - 1)                                                 
 \right]\\ 
                 &&\\
		&& \left.  ~~~~~~~~~~~~ ~~~~~~~~~~~~  + (N-1)^2   ~~~ \right]      \\
		&&\\
&=&\left(\frac{r_Ac_A}{N^2 (N-1)}\right)(N - r_B)(N - c_B)                   
 \end{array}
 \]
 N.B. If, further, $r_A = r_I = r_B$ and $c_A = c_I = c_B$ then $A = B$ and the above expression is the variance of its count. 
 This will be the same as from the hypergeometric earlier, namely,
\[ Var(K_A) = Cov(K_A, K_A) =  \frac{r_Ac_A}{N}~~\frac{(N-r_A)(N-c_A)}{N(N-1)}.\]

\end{enumerate}

\subsection{As $N$ grows}
Suppose that the lattice is suitably scaled to fit within the unit square (e.g., divide each rank by $(N+1)$) and that a bin is defined to be that part of the lattice within a fixed rectangle of the unit square.
As $N$ increases, the number of lattice points inside the rectangle, and hence the size $(r \by c)$ of the bin, will increase accordingly. 
Similarly, the expected counts, the variances, and covariances will all increase with $N$.

To see this, take a block $A$ to be that part of the lattice within some $p_r \by p_c$ rectangle within $(0,1)  \by  (0, 1)$.
Then 
$r_A \approx N p_r$ and $c_A \approx N p_c$ gives, for example, 
\[E(K_A) \approx  Np_rp_c, ~~\text{ and } ~~ Var(K_A) \approx Np_rp_c (1-p_r)(1-p_c) \frac{N}{N-1} \approx Np_rp_c (1-p_r)(1-p_c).\]
Similar calculations could be done for the covariance of two blocks.

\subsection{When one variable is categorical}
\label{sec:case_categorical}
Suppose that $X$ is categorical with $C$ categories, indexed $c \in \{ 1, \ldots, C\} = \Nset{C}$.   Category totals, $n_c$, sum to $N$.
The continuous variate $Y$ has $N$ values, each rank $i \in \{1, \ldots, N \} = \Nset{N}$ being associated with a value $c$ of $X$ subject to a total of $n_c$ values of $c$.  A simple way to generate data under the null hypothesis is to create an $N$ vector 
$\ve{c} = (c_1, \ldots, c_N)$ as a random
permutation of a vector containing $n_1$ repetitions of $1$, $n_2$ of 2, and so on.  The pairs $(c_i, i)$ for $i = 1, \ldots, N$ then constitute a sample of categories of $X$ and ranks of $Y$ under the null hypothesis.

This is simply a constrained version of the continuous by continuous case.
Without loss of generality, the categories can be mapped onto the $N \by N$ lattice of rank positions as contiguous columns in the lattice.
That is, each category, $c$, is taken to be $n_c$  side by side columns arranged with category $c = 1$ being the leftmost $n_1$ columns of the $N \by N$ lattice, category $c = 2$ the next $n_2$, and so on.  Each category, then, has its own set of column indices, say $\set{J}_c$, of the $N \by N$ lattice of size $\cardinality{\set{J}_c} = n_c$ which defines that category.  

The random variates $Z_{ij}$ are now collapsed within each category ($j \in \set{J}_c$) as
\[ W_{ic} = \sum_{j \in \set{J}_c} Z_{ij} ~~~ \text{ for } ~ i \in  \Nset{N}, ~~ \text{ and } ~ c \in  \Nset{C}. \]
It follows that $\sum_{i = 1}^N W_{ic} = n_c.$
Each $W_{ic}$ is an indicator random variate taking the value one whenever the continuous variate having rank $i \in \Nset{N}$ is paired with the categorical variate in category $c$, and zero otherwise.  

A reduced $N \times C$ lattice
for the continuous $\by$ categorical  is thus constructed from the $N \times N$ lattice.   Data under the null hypothesis can be simulated as before.  The recursive binning would occur within each category and the probability results would transfer from the $Z_{ij}$s to the $W_{ic}$s as expected.

Conversely, the $N \times N$ lattice of $Z_{ij}$ could be generated from the $N \times C$ one of $W_{ij}$.
Within category $c$, create $n_c$ column vectors $\ve{Z}_j = \ve{0}$ for $j \in \set{J}_c$ and then set $Z_{ij} = 1$ by distributing the values $W_{ic} = 1$ across the $n_c$ columns, one value to each column such that $\sum_i Z_{ij} =1$ for all $j$.  Finally, randomly permute the columns $\ve{Z}_j$ within each category.

The categorical and continuous merely represent a particular form of binning where the data are divided into columns along one axis.
\newpage
\section{A simple approach to the asymptotic distribution}
\label{app:simple_approx}
Since the count $K \follows Hypergeometric(N, r, c)$ for any rectangular region $A$ containing an $r \by c$ portion of the lattice,   the standard asymptotic approximations will follow.
Namely, for large $N$, the probability $p_k = Pr (K = k)$ can be approximated 
by a Poisson($\lambda$) random variable $K$ having $\lambda = rc/N$.  That is,
\[Pr(K = k) \approx \frac{\lambda^k e^{-\lambda}}{k!}. \]
From there, asymptotic approximations follow as
\[ \frac{K - \lambda}{\sqrt{\lambda}} \follows N(0,1) ~~~\text{and, so, } ~~~ \frac{(K-\lambda)^2}{\lambda} \follows \chi^2_1.\]
The latter is the contribution the observations found in any $r \times c$ rectangle $A$ make to the $\bigChi$ statistic.
As $N$ increases, any fixed rectangular region $A$ will encompass an ever finer $r \by c$ lattice as $r$ and $c$ each
grow with $N$ to fix the size of the rectangle.

For $i = 1, \ldots, b$ such blocks, with $K_i$ the random variate representing the count in block $i$ (having $r_i$ rows and $c_i$ columns), the Pearson goodness of fit statistic will be
\[ \bigChi = \sum_{i=1}^b \left(\frac{K_i - \lambda_i}{\sqrt{\lambda_i}}\right)^2  \]
a sum of $b$ (approximately) $\chi^2_1$ random variables where $\lambda_i = r_ic_i/N$.
Unfortunately, these are not independently distributed.

Moreover, their sum is not immediately expressible as the sum of some number, say $d < b$ of independent $\chi^2_1$ random variates.

\subsection{Approximation by a multivariate normal}
\label{sec:mvn}
To see this, suppose that the multivariate distribution of the vector of standardized counts 
\[ \ve{K}_{\lambda} = \tr{\left(\left(\frac{K_1 - \lambda_1}
                                                        {\sqrt{\lambda_1}}\right), 
                                                        \ldots, 
                                           \left(\frac{K_b - \lambda_b}{\sqrt{\lambda_b}}\right) \right)} \]
 could be approximated by a $b$-variate normal distribution (for large $N$) where each coordinate is marginally $N(0,1)$. 
 That is,  approximately,
\[ \ve{K}_{\lambda} \follows N_b(\ve{0}, \m{R}_{\ve{K}}) \]
where $\m{R}_{\ve{K}}$ is the $b \by b$ correlation matrix of $\ve{K}$ obtained from $\sm{\Sigma}_{\ve{K}}$.
Further, suppose that the rank of $\m{R}_{\ve{K}}$  is $d < b$, and let 
\[ 
\m{R}_{\ve{K}} = \m{V}\m{D}\tr{\m{V}} 
   = \left( \m{V}_d,  \m{V}_{b-d}\right)
      \left(\begin{array}{lcl} 
              \m{D}_\gamma &~~~& \m{0}_{d \by (b-d)} \\
              &&\\
               \m{0}_{(b-d) \by d}  &&  \m{0}_{(b-d) \by (b-d)} 
               \end{array} 
             \right)
     \left( \begin{array}{l}
              \tr{\m{V}}_d \\
              \\
              \tr{\m{V}}_{b-d}
              \end{array}
              \right)
\]
be the eigen decomposition of $\m{R}_{\ve{K}}$ 
having
non-zero eigen-values $\gamma_1 \geq \gamma_2 \geq \cdots \geq \gamma_d > 0$ in the diagonal matrix $\m{D}_\gamma = \diag(\gamma_1, \gamma_2, \ldots, \gamma_d)$ and the $b \by b$ matrix of orthonormal eigen-vectors $\m{V} = (\m{V}_d, \m{V}_{b-d})$ separated into those $d$ vectors in $\m{V}_d$ corresponding to the non-zero eigen-values and those of $\m{V}_{b-d}$ to the zero eigen-values.

Denoting
\[ \ve{K}_{\gamma} = {\m{D}_{\gamma}}^{-\frac{1}{2}} \tr{\m{V}} \ve{K}_\lambda = \tr{(K_{\gamma, 1}, \ldots, K_{\gamma, d})}\]
the approximation gives
\[ \ve{K}_{\gamma} \follows N_d(\ve{0}, \iden{d}) \]
and $\tr{V_{b-d}} \ve{K}_\lambda$ as a degenerate $(b-d) \by 1$ random vector having point mass at $\ve{0}$.
It follows that
\[ \tr{\ve{K}_{\gamma}} \ve{K}_{\gamma} = \sum_{i=1}^d K_{\gamma, i}^2  \approx \sum_{i=1}^d (\text{independent }~\chi^2_1) \follows \chi^2_d.\]
And,
\[ 
\begin{array}{rcl}
\bigChi 
& = & \tr{\ve{K}_\lambda} \ve{K}_\lambda \\
&&\\
& =  & \tr{\ve{K}_\lambda} \m{V}\tr{\m{V}} \ve{K}_\lambda \\
&&\\
& =  & \tr{\ve{K}_\lambda} \m{V}_d\tr{\m{V}}_d \ve{K}_\lambda\\
&&\\
& =  & \tr{\ve{K}_\gamma} \m{D}_{\gamma}  \ve{K}_\gamma\\
&&\\
&=& \sum_{i=1}^d \gamma_i K_{\gamma, i}^2,
\end{array}
\]
which suggests 
a positive linear combination of $d$ independent $\chi^2_1$ random variates.

\subsection{The corresponding ``exact'' distribution}
\label{sec:exact}
This is not an uncommon distribution for a goodness of fit statistic \cite[e.g., see][for a variety of cases]{UnifiedGeneralChiSq1975} and ``exact'' methods for calculating its probability have long been known \cite[notably,][]{Imhof1961, Davies_1980, Farebrother1984, Farebrother1990}.
These, as well as a $\chi^2$ approximation method by \cite{LiuEtAl2009} are provided by the \R package \pkg{compQuadForm}  \citep{compQuadForm} available from CRAN \cite{Rlang}.

Selected cumulative probabilities were compared to the empirical probabilities based on 10,000 replications for various sample sizes of $N$ and randomly selected block tilings of the unit square. 
The results are shown in Table \ref{table:compute_p}.
\begin{table}[htp]
\caption{Computed probabilities using the positive combination of $\chi^2_1$s.  Each run defines a single set of blocks for all of its samples.  Each sample are the $N$ paired ranks from a pair of independent uniforms and are used, with the blocks to give $\bigChi$.  This is repeated for 10,000 independent samples to give the empirical distribution of $\bigChi$ for that set of blocks and sample size.  The rank $d$ was determined numerically to be the number of $\gamma_i/\gamma_1 > 10^{-10}$.  Empirical quantiles corresponding to the cumulative probabilities $p$ were given to the methods to produce  the corresponding probabilities in each column  of the table.}

\spacingset{1.50} 

\begin{center}
{\tiny 
\begin{tabular}{|c|c|c|c|c|c|l||rrrrr|}
\hline
&&&&&&& \multicolumn{5}{c|}{Cumulative probability $p$}\\
Run & $N$ & depth & $b$& $d$ & $\tr{\gamma}$&Method &  0.900 & 0.950 & 0.975 & 0.990 & 0.999 \\
&&&&&&&&&&&\\ 
\hline
1 & 100 & 2 & 4 & 2 & (3.435, 0.565) & davies & 0.533 & 0.637 & 0.701 & 0.780 & 0.894 \\
&  &  &  &  && farebrother & 0.533 & 0.637 & 0.701 & 0.780 & 0.894 \\
 &  &  &  &  && imhof & 0.533 & 0.637 & 0.701 & 0.780 & 0.894 \\
&  &  &  &  && liu & 0.546 & 0.643 & 0.704 & 0.780 & 0.893\\
\hline
2 & 500 & 2 & 4 & 2 & (2.201, 1.799) & davies & 0.439 & 0.526 & 0.601 & 0.664 & 0.802  \\
 &  &  &  &  && farebrother & 0.439 & 0.526 & 0.601 & 0.664 & 0.802  \\
 &  &  &  &  && imhof & 0.439 & 0.526 & 0.601 & 0.664 & 0.802  \\
 &  &  &  &  && liu & 0.440 & 0.527 & 0.602 & 0.665 & 0.802  \\
 \hline
 3 & 2,000 & 2 & 4 &2 & (3.143, 0.857) & davies & 0.497 & 0.617 & 0.701 & 0.784 & 0.884  \\
 &  &  &  &  && farebrother & 0.497 & 0.617 & 0.701 & 0.784 & 0.884  \\
  &  &  &  &  && imhof & 0.497 & 0.617 & 0.701 & 0.784 & 0.884  \\
  &  &  &  &  && liu & 0.517 & 0.628 & 0.706 & 0.785 & 0.882  \\
 \hline
4 & 2,000 & 3 & 8 & 6 & (2.582, 2.055, 1.178, & davies & 0.519 & 0.622 & 0.712 & 0.800 & 0.928  \\
&&&&&1.097, 0.921, 0.166)  & farebrother & 0.519 & 0.622 & 0.712 & 0.800 & 0.928  \\
&  &  &  &  && imhof & 0.519 & 0.622 & 0.712 & 0.800 & 0.928  \\
 &  &  &  &  && liu & 0.527 & 0.626 & 0.713 & 0.798 & 0.926  \\
\hline
5  & 10,000 & 5 & 31 & 29 & (2.484, 1.649, 1.427, & davies & 0.519 & 0.644 & 0.743 & 0.828 & 0.926  \\
       &            &    &    &&    1.415, 1.359, 1.255, & farebrother & 0.519 & 0.644 & 0.743 & 0.828 & 0.926  \\
       &            &    &    &&   1.237, 1.204, 1.182, & imhof & 0.519 & 0.644 & 0.743 & 0.828 & 0.926  \\
       &            &    &    &&     1.175, 1.134, 1.124, &liu & 0.520 & 0.644 & 0.742 & 0.827 & 0.925  \\
       &            &    &    & &   1.112, 1.081, 1.071,  &&&&&&\\
       &            &    &    &&    1.063, 1.047, 1.037, &&&&&&\\
       &            &    &    &&    1.017, 1.012, 0.970, &&&&&&\\
       &            &    &    &&    0.945, 0.880, 0.859, &&&&&&\\
       &            &    &    &&    0.793, 0.750, 0.375, &&&&&&\\
       &            &    &    &&   0.283, 0.060) &&&&&&\\
       \hline
6 & 10,000 & 6 & 60 & 58 & (1.841, 1.660, 1.623,& davies & 0.487 & 0.620 & 0.731 & 0.832 & 0.943  \\
        &            &    &    & &  1.518, 1.460, 1.453, & farebrother & 0.487 & 0.620 & 0.731 & 0.832 & 0.943  \\
       &            &    &    & &   1.401, 1.371, 1.333,& imhof & 0.487 & 0.620 & 0.731 & 0.832 & 0.943  \\
         &            &    &    & &  1.309, 1.296, 1.240, & liu & 0.487 & 0.620 & 0.731 & 0.832 & 0.943  \\
        &            &    &    & &   1.227, 1.221, 1.214, &&&&&&\\
        &            &    &    & &   1.202, 1.194, 1.174, &&&&&&\\
         &            &    &    & &  \vdots &&&&&&\\
        &            &    &    & &   0.312, 0.182, 0.132, &&&&&&\\
       &            &    &    & &    0.036) &&&&&&\\
       \hline
\end{tabular}
}

\spacingset{1.75} 

\end{center}
\label{table:compute_p}
\end{table}%

As can be seen, the exact methods of \cite{Imhof1961, Davies_1980, Farebrother1984, Farebrother1990} give identical results while the $\chi^2$ approximation of \cite{LiuEtAl2009} is close. 
Though not displayed here, saddlepoint approximations \cite[e.g.,][]{Kuonen_Saddlepoint_1999} also agree with these results.
Unfortunately, none are anywhere close enough to the true cumulative probabilities given at the top of the table.
(N.B., Though of little import here, the empirical results suggest the rank $d$ might simply be $b-2$.)

The simple approach is clearly inadequate.  It would appear that the absence of independent and identically distributed $Z_{ij}$s has caused the failure of the multivariate normal approximation based only on the mean and covariance of the cell counts.

%
%
 
\newpage
\section{Approximating the null distribution of $\bigChi$}
\label{app:mle_approx}
Given sufficient computational resources, the null distribution of $\bigChi$ is  always available empirically by simulation. 
Empirical exploration of the null distribution of $\bigChi$ on the paired ranks suggests a $\chi^2$ approximation can work.  Moreover,  the degrees of freedom turn out to be (relatively) simple functions of the number of bins.

There are three cases to consider to test $\Hyp_0: X \indep Y$:
\begin{enumerate}
\item
$Y$ and $X$ are both categorical,
\item 
$Y$ and $X$ are both continuous,
\item
$Y$ is continuous and $X$ is categorical.
\end{enumerate}
The first instance is the classic contingency table having $R$ rows and $C$ columns.  In this case, $\bigChi$ has a limiting $\chi^2$ distribution on $(R-1)(C-1)$ degrees of freedom.

The other two will be investigated by examining the null distribution of $\bigChi$ via simulation.

\subsection{$Y$ and $X$ are both continuous}
This is the situation arising via random recursive binning to some depth on the lattice of paired ranks. To investigate the null distribution, the value of $\bigChi$ is simulated by forming rank pairs from independent uniform samples.

In particular, for sample sizes $N \in \{100, 500, 1000, 2000, 2500, 3000, 3500, 4000, 4500, 5000, 6000\}$, and depths from 2 to 8, the rank pairs of independent uniform samples were drawn and binned according to a single recursive configuration of ``squarified'' blocks  (N.B. the configuration was randomly produced for each depth), and the $\bigChi$ was calculated.  This was done 10,000 times for each configuration giving an empirical distribution based on 10,000 independently sampled $\bigChi$s (for each $N$ and depth pair, and that single block configuration).  This was repeated three times to give three different block configurations for each $N$ and depth pair.  A total of 108 such empirical distributions were produced.  The maximum likelihood estimate of the degrees of freedom for a $\chi^2$  fit to each empirical distribution was produced.

\begin{figure}[!ht]
\begin{center}
\includegraphics[scale = 0.45]{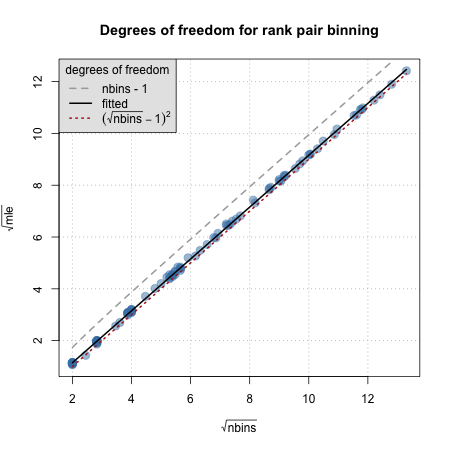}
\caption{Continuous versus continuous:  Maximum likelihood estimates of the degrees of freedom of the null distribution of $\bigChi$ under squarified random recursive binning of rank pairs. }
\label{fig:mles_cts_cts}
\end{center}
\end{figure}

Figure \ref{fig:mles_cts_cts}
shows the 108 maximum likelihood estimates plotted against the number of bins (nbins).
Square roots are shown to spread the points a little more uniformly across the range of values -- this will also lead to a simple mnemonic for approximating the degrees of freedom.
The most number of bins was 177 and the fewest was four.
The grey dashed line shows that the degrees of freedom for uniform pairs, as opposed to rank pairs, is clearly inappropriate.  

The least squares fitted line has multiple $R^2 = 99.97\%$ and predicts
\[ \sqrt{df} = -0.8576945 + 1.0015633 \sqrt{b}\]
for total number of bins (nbins) $b$.  Squaring gives a formula for the estimated degrees of freedom as a function of the number of bins. 
The coefficients could be rounded to three significant digits each since there is no evidence against the hypothesis that the intercept is -0.858 and the slope is 1.00 ($p \approx 0.13$).
This gives a simple formula for the degrees of freedom as
\begin{equation}
\label{eq:df_formula}
df \approx (\sqrt{b} - 0.858)^2.
\end{equation}
Even simpler,  and reminiscent of the contingency table case, would be the  (ballpark) estimate
\begin{equation}
\label{eq:simple_df_formula}
df \approx (\sqrt{b} - 1)^2.
\end{equation}
This is shown as the red dotted line of Figure \ref{fig:mles_cts_cts}.
As can be seen there this formula systematically underestimates the degrees of freedom (N.B.,  a test of this hypothesis would give  $p \approx 0$).  
Nevertheless, as a useful mnemonic, these degrees of freedom might be good enough in practice.

\begin{figure}[!ht]
\begin{center}
\begin{tabular}{ccc}
\includegraphics[scale = 0.3]{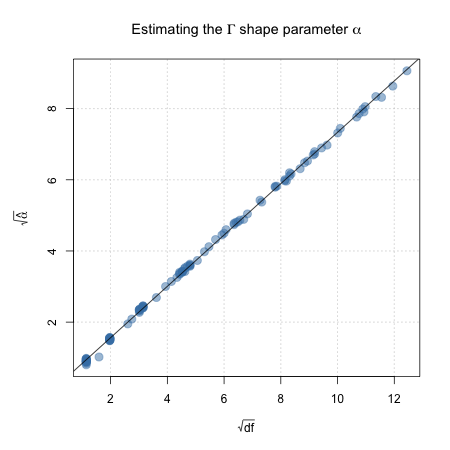}
&$~~~$&
\includegraphics[scale = 0.3]{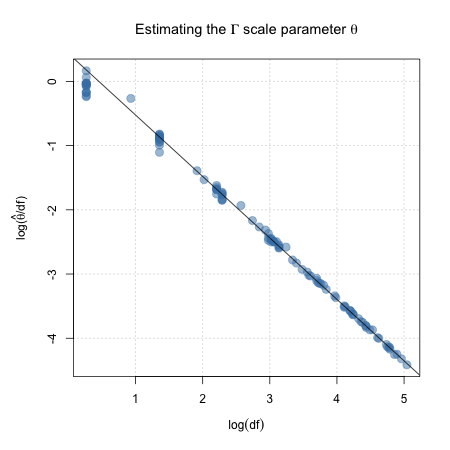}
\\
{\footnotesize
(a) The shape parameter $\alpha$}
&&
{\footnotesize
(b) The scale parameter $\theta$}
\end{tabular}
\caption{Continuous versus continuous:  Maximum likelihood estimates of the parameters of $\Gamma(\alpha, \theta)$ densities. }
\label{fig:mles_cts_cts_gamma}
\end{center}
\end{figure}

Instead of $\chi^2$, a more general approximation would be a $\Gamma(\alpha, \theta)$ density having shape parameter $\alpha$ and scale parameter $\theta$.
The maximum likelihood estimates of these parameters were also determined for the 108 null distributions described above \cite[via the \pkg{EnvStats} package of][]{EnvStats-book}.  The results are shown plotted in
Figures  \ref{fig:mles_cts_cts_gamma}(a) and (b) respectively.
There, a function of the mles is plotted against a function of the estimated degrees of freedom from equation (\ref{eq:df_formula}).  
In Figure \ref{fig:mles_cts_cts_gamma}(a), the least-squares fitted line is added, giving a formula for the estimated shape 
\begin{equation}
\label{eq:shape}
 \widehat{\alpha} = \left(0.1199774 + 0.7214124  \sqrt{df}\right)^2. 
 \end{equation}
Similarly, Figure \ref{fig:mles_cts_cts_gamma}(b) shows a robust fitted line
\cite[using default \code{lmRob()} from \R's \pkg{robust} package:][]{robust_pkg}
giving a formula for the estimated scale
\begin{equation}
\label{eq:scale}
\widehat{\theta} = \exp\left(0.4329157 + (1 - 0.9571741) \log(df)\right).
 \end{equation}
In both equations the coefficients of the fitted lines are given to several decimal places and $df$ denotes the estimated $\chi^2$ degrees of freedom from equation (\ref{eq:df_formula}).

\begin{figure}[!ht]
\begin{center}
\begin{tabular}{ccc}

\includegraphics[scale = 0.3]{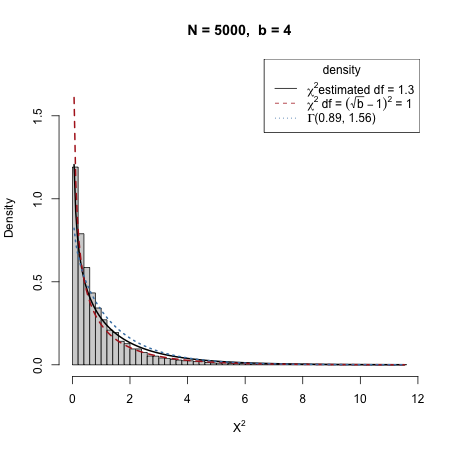
}
&
\includegraphics[scale = 0.3]{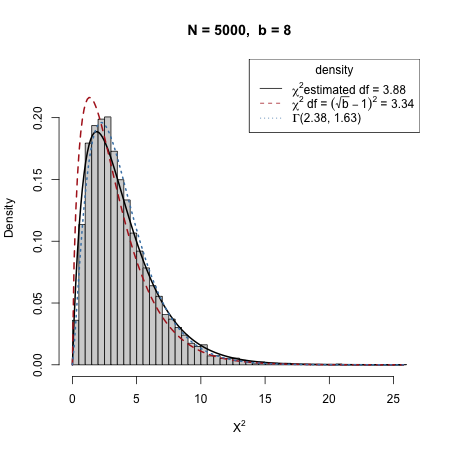
}
&
\includegraphics[scale = 0.3]{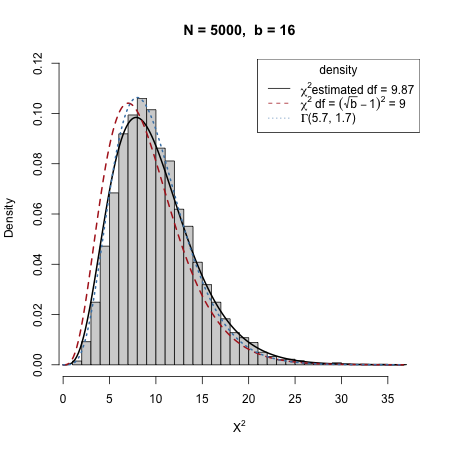
}
\\
\includegraphics[scale = 0.3]{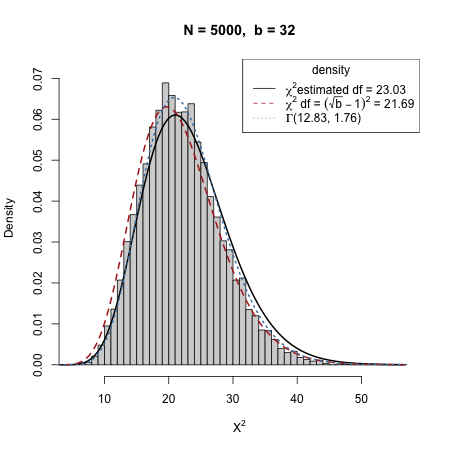
}
&
\includegraphics[scale = 0.3]{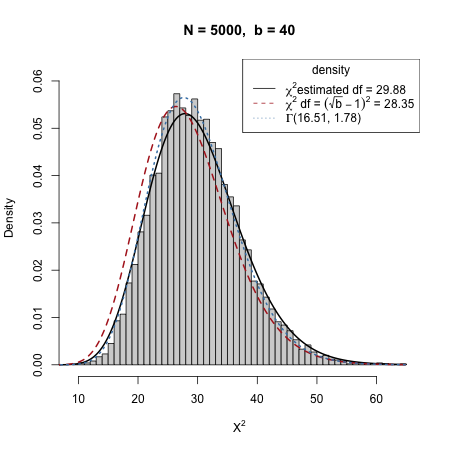
}
&
\includegraphics[scale = 0.3]{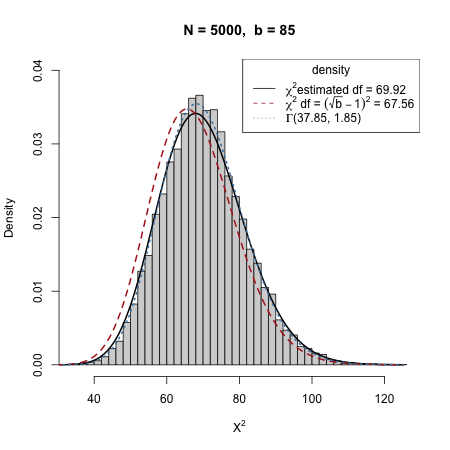
}
\\
\includegraphics[scale = 0.3]{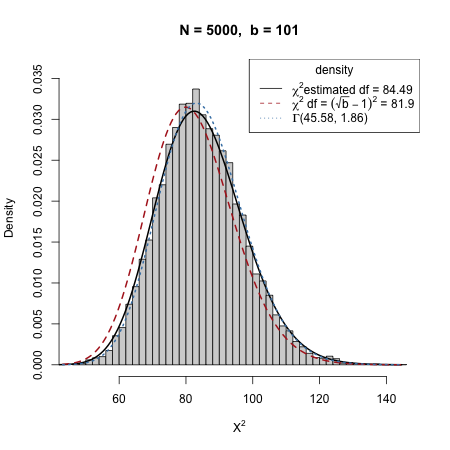
}
&
\includegraphics[scale = 0.3]{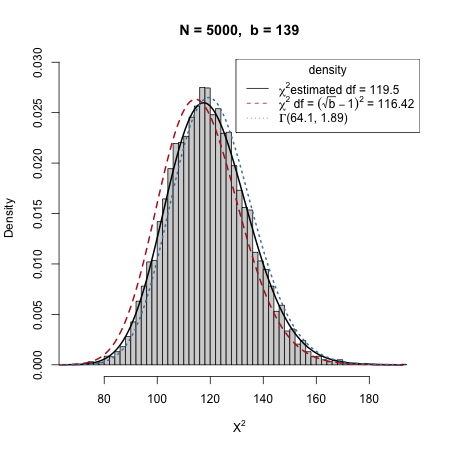
}
&
\includegraphics[scale = 0.3]{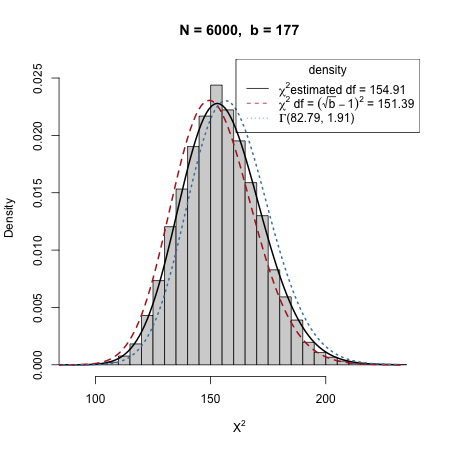
}
\\
\end{tabular}
\caption{Continuous versus continuous:  Estimating the null distribution by $\chi^2_{df}$ and $\Gamma(\alpha, \theta)$ densities. }
\label{fig:null_cts_cts_hists}
\end{center}
\end{figure}

To assess the quality of these approximations, the estimated densities were laid over the histogram (based on 10,000 $\bigChi$ values) for each of the 108 replications.  A few of these are shown in Figure \ref{fig:null_cts_cts_hists}.
The number of bins $b$ increase from left to right and top to bottom.
As can be seen, all three approximations are competitive, although the simplest $\chi^2$ with degrees of freedom $(\sqrt{b} - 1)^2$ tends to be located slightly lower and so could produce $p$-values lower than that of the empirical null distribution.

\subsubsection{Approximation test properties}
Using $\bigChi$ to test the hypothesis of independence,  results will vary depending on the approximation used.  Here, the simulated distribution based on 10,000 different repetitions for each set of blocks allows comparison in terms of the $p$-values each approximation produces.

\begin{figure}[hbt]
\begin{center}
\begin{tabular}{ccc}
\includegraphics[scale = 0.25]{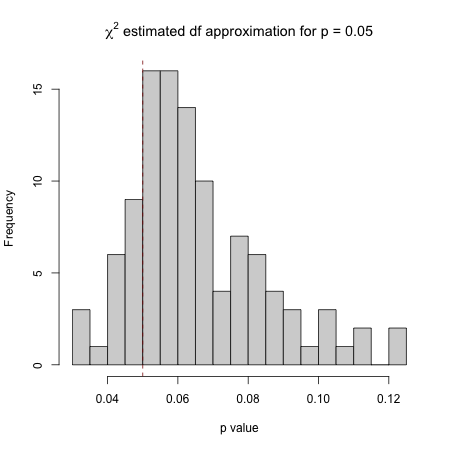} & 
\includegraphics[scale = 0.25]{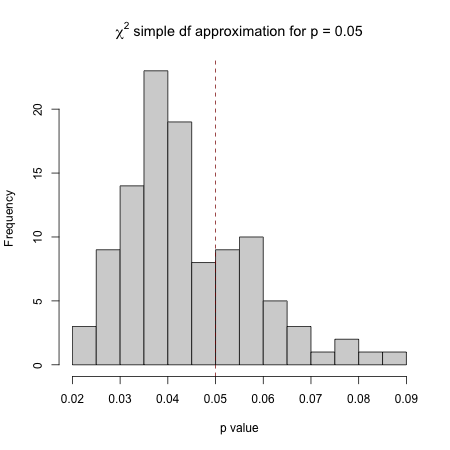} &
\includegraphics[scale = 0.25]{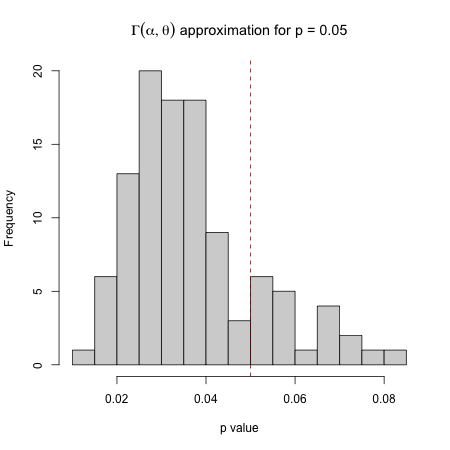} \\
\includegraphics[scale = 0.25]{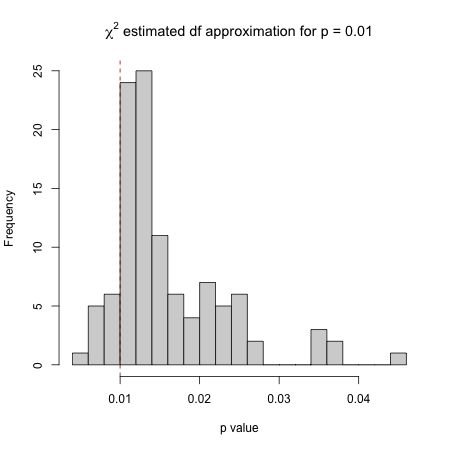} & 
\includegraphics[scale = 0.25]{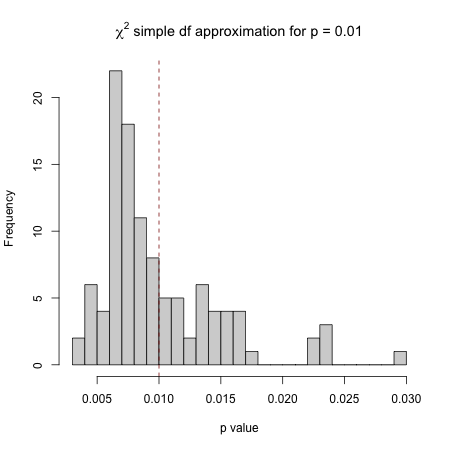} &
\includegraphics[scale = 0.25]{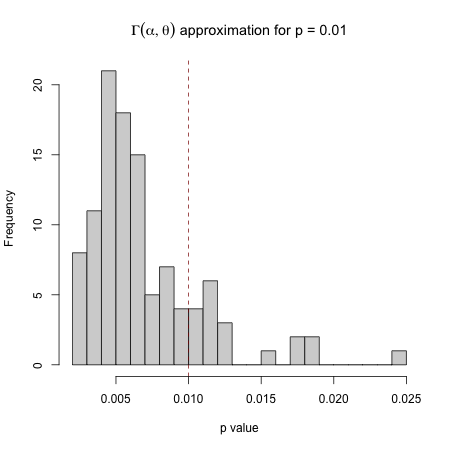} \\
{\footnotesize
(a) $\chi^2$ for df from equation (\ref{eq:df_formula})}
&
{\footnotesize
(b) $\chi^2$ for df from equation (\ref{eq:simple_df_formula})} &
{\footnotesize
(c) $\Gamma(\alpha, \beta)$ approximation }
\end{tabular}
\caption{Distributions of $p = Pr(\bigChi \geq Q_q)$ returned by different approximations.  Top row has $p = 1-q = 0.05$, bottom row $0.01$ -- each is marked by a vertical dashed red line.}
\label{fig:null_cts_cts_p_values}
\end{center}
\end{figure}

For each of the 108 different repetitions, the empirical quantiles $Q_q$ were determined for $q = 0.95$ and $q = 0.99$.  
These were then used as the value of the ``test statistic'' and the $p$-value determined as
\[ p = Pr(\bigChi \geq Q_q) \]
for each of the three approximations -- the nearer $p$ is to $1-q$, the better the approximation.
Figure \ref{fig:null_cts_cts_p_values} 
shows the resulting histograms of $p$-values for each approximation when $q = 0.95$ and when $q = 0.99$.
In both cases,  the $\Gamma(\alpha, \theta)$  approximation is the most likely to judge the $p$-value to be lower than it should be,
followed by the simple $\chi^2$ approximation.  The $\chi^2$ approximation based on the estimated degrees of freedom, on the other hand, is more likely to judge the $p$-value as higher than it truly is; this suggests it to be the most conservative of the three with the $\Gamma(\alpha, \theta)$ as the most aggressive.

Table \ref{table:cts_obs_p} 
\begin{table}[htp]
\caption{Continuous versus continuous summary statistics of $p$-values computed by each approximation.}
\begin{center}
{\footnotesize
\begin{tabular}{|l|c|c|c|c|c|c|c|c|}
\hline
&& \multicolumn{7}{c|}{Numerical summary of observed $p$-values}\\
&& \multicolumn{7}{c|}{}\\
Approximation & $p$ & minimum & $Q(0.25)$ & median & average & $Q(0.75)$ & maximum &IQR\\
\hline
$\chi^2$ estimated d.f.     & 0.05 &  0.034  & 0.053  & 0.060 &  0.065 &  0.077  & 0.122 &  0.024 \\
$\chi^2$ simple d.f.          &         & 0.022  & 0.035   &0.041  & 0.044  & 0.053 &  0.086 & 0.018\\
$\Gamma(\alpha, \theta)$&         &0.014   &  0.027   & 0.034  & 0.037  & 0.042  & 0.083 & 0.015 \\
\hline
$\chi^2$ estimated d.f.     & 0.01 &  0.006  & 0.011  & 0.013   &0.016  & 0.019 &  0.045 &  0.008\\
$\chi^2$ simple d.f.          &         & 0.003  &  0.007   & 0.008   & 0.010   & 0.012 &   0.029 &  0.005 \\ 
$\Gamma(\alpha, \theta)$&         &0.002   & 0.004 &   0.006 &   0.007  &  0.008 &   0.024&  0.004\\
       \hline
\end{tabular}
}
\end{center}
\label{table:cts_obs_p}
\end{table}%
provides numerical summaries of the histograms in Figure \ref{fig:null_cts_cts_p_values}.
As the histograms are seen to be skewed in all cases, the median is a preferred measure of centrality. 
In both cases ($p = 0.05$ and $p = 0.01$) the two  $\chi^2$ approximations have medians a little closer to $p$ than the $\Gamma(\alpha, \theta)$ approximations.  As a measure of consistency, the inter-quartile range (IQR) provides a guide that is relatively robust to outliers.  By this measure, the $\Gamma(\alpha, \beta)$ approximations have the least spread, with the simple $\chi^2$ a close second.  
Unfortunately, in neither case, the middle half of the data for  the $\Gamma*\alpha, \theta)$ approximations does not cover the value $p$.  Neither does the first $\chi^2$ approximation for $p = 0.01$. 
It would appear that the data supports  the $\chi^2$ approximation with the simple formula (\ref{eq:simple_df_formula}) for its degrees of freedom.

\begin{figure}[hbt]
\begin{center}
\begin{tabular}{ccc}
\includegraphics[scale = 0.25]{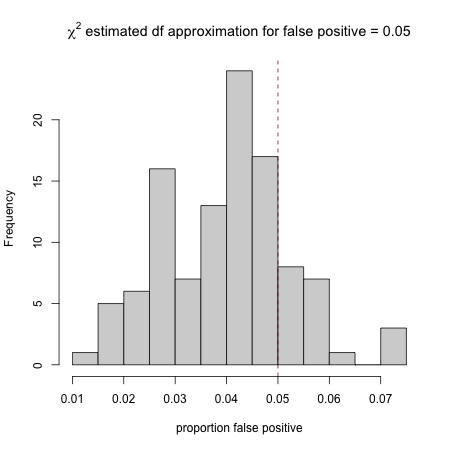} & 
\includegraphics[scale = 0.25]{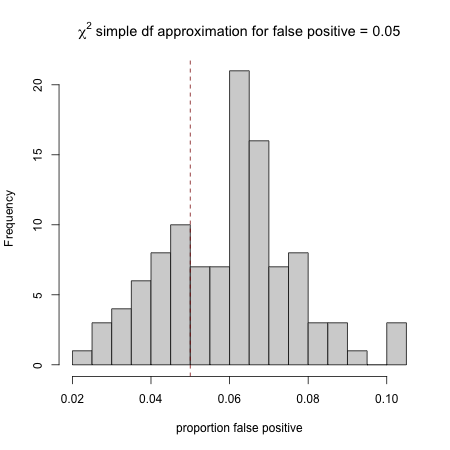} &
\includegraphics[scale = 0.25]{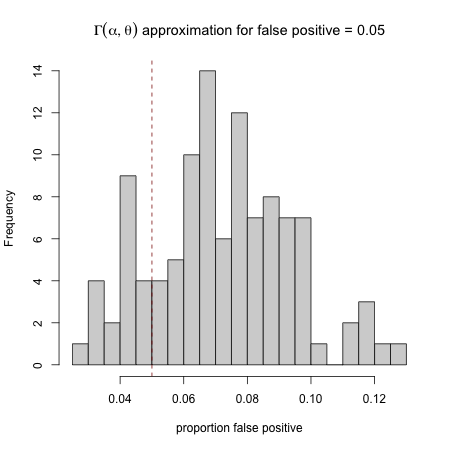} \\
\includegraphics[scale = 0.25]{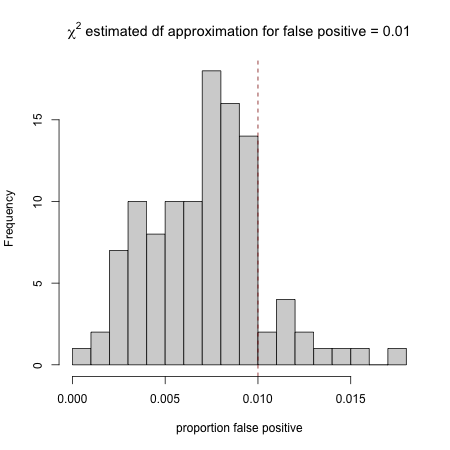} & 
\includegraphics[scale = 0.25]{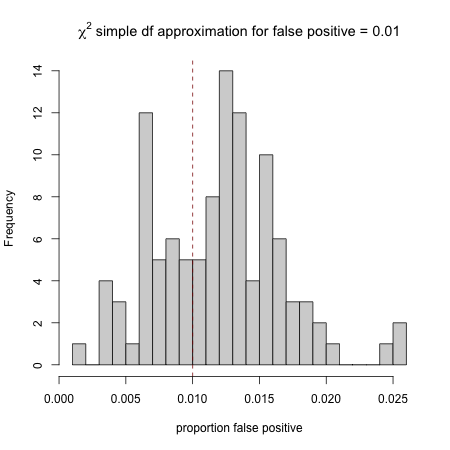} &
\includegraphics[scale = 0.25]{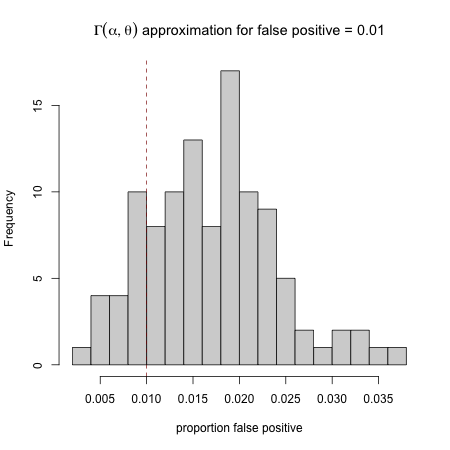} \\
{\footnotesize
(a) $\chi^2$ for df from equation (\ref{eq:df_formula})}
&
{\footnotesize
(b) $\chi^2$ for df from equation (\ref{eq:simple_df_formula})} &
{\footnotesize
(c) $\Gamma(\alpha, \beta)$ approximation}
\end{tabular}
\caption{Distributions of the proportion of false positives returned by different approximations for a true proportion $p$.  Top row has $p  = 0.05$, bottom row $0.01$ -- each is marked by a vertical dashed red line.}
\label{fig:null_cts_cts_false_positives}
\end{center}
\end{figure}

Another way to proceed would be to presume that the test would be conducted at some fixed level identified with the number of false positives desired under the null hypothesis (i.e., the size of the ``type 1 error'').  
This was fixed at two levels, 0.05 and 0.01, for each approximation and then the actual proportion of false positives determined using the simulated $\bigChi$ distributions for each of the 108 situations.  The resulting histograms of the observed proportion of false positives appear in Figure \ref{fig:null_cts_cts_false_positives}.
The histograms show that the $\chi^2$ approximation with degrees of freedom estimated by equation (\ref{eq:df_formula}) tends to produce fewer false positives than it should, and the $\Gamma(\alpha, \theta)$ approximation more false positives than it should -- for both fixed levels.
The $\chi^2$ with the simple formula (\ref{eq:simple_df_formula}) places somewhere between the two other approximations.

Table \ref{table:cts_obs_false}
\begin{table}[htp]
\caption{Continuous versus continuous summary statistics of false positives for each approximation.}
\begin{center}
{\footnotesize
\begin{tabular}{|l|c|c|c|c|c|c|c|c|}
\hline
&& \multicolumn{7}{c|}{Numerical summary of observed proportion of false positives}\\
&& \multicolumn{7}{c|}{}\\
Approximation & $level$ & minimum & $Q(0.25)$ & median & average & $Q(0.75)$ & maximum &IQR\\
\hline
$\chi^2$ estimated d.f.     & 0.05 &  0.011  &   0.030 &    0.041 &    0.040  &   0.047  &   0.073& 0.018 \\
$\chi^2$ simple d.f.          &         &  0.023  &   0.047  &   0.062  &   0.060 &    0.070  &   0.103 & 0.023 \\
$\Gamma(\alpha, \theta)$&         &0.028  &   0.058 &    0.072  &   0.072  &   0.087 &    0.128  &  0.029\\
\hline
$\chi^2$ estimated d.f.     & 0.01 & 0.000 &    0.005 &    0.008 &    0.007 &    0.009 &    0.017&   0.004\\
$\chi^2$ simple d.f.          &         & 0.002 &    0.008  &   0.012 &    0.012 &    0.015  &   0.025 &  0.007 \\ 
$\Gamma(\alpha, \theta)$&         &0.003   &  0.012  &   0.017   &  0.017  &   0.021  &   0.037 & 0.009 \\
       \hline
\end{tabular}
}
\end{center}
\label{table:cts_obs_false}
\end{table}%
provides summary statistics for these histograms.  An examination of the median values and of the inter-quartile ranges again suggests that the $\chi^2$ distribution using the simple formula (\ref{eq:simple_df_formula}) for degrees of freedom may be best of the three.

\subsection{Continuous $Y$ versus categorical $X$}
\label{sec:cts_vs_cat}
Here, paired rank data were simulated from independent marginals for sample sizes $N \in \{100, 500,$ $1000, 2000, 2500, 3000, 3500, 4000, 5000\}$. 
Binning was such that the horizontal axis was divided into some number of categories $C \in \{2, 3, 4, 6, 8, 10\}$ the vertical axis divided into bins within each category recursively to depths from 2 to 5.  For each depth, a single set of $C$ multinomial probabilities was first randomly generated for each number of categories.  Then for every sample size $N$ three binning replicates were produced via  horizontal categories given by randomly drawing a multinomial vector using the multinomial probabilities previously produced, and recursively dividing the vertical within each category to the prescribed depth (subject to bounds on the minimum expected value in each bin). 
Then, for each replicate set of bins, 10,000 samples of $N$ rank pairs drawn from independent uniforms, were allocated to the bins and the $\bigChi$ statistic determined -- giving 10,000 $\bigChi$ values for each binning replicate.  A total of 
225 such samples of empirical distributions of $\bigChi$ were produced (three replicates of 75 different combinations of number of categories, depth, and sample size).  

\begin{figure}[hbt]
\begin{center}
\begin{tabular}{ccc}
\includegraphics[scale = 0.4]{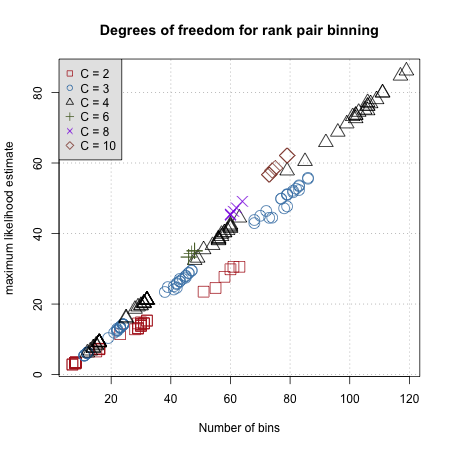}
& $~~~~$ &
\includegraphics[scale = 0.4]{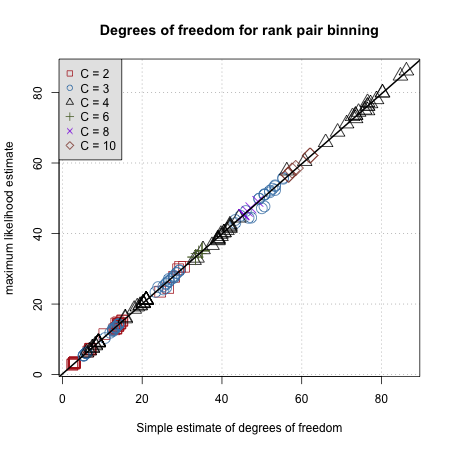}
\\
{\footnotesize
(a) MLEs versus number of bins, $b$}
& &
{\footnotesize
(b) MLES versus $(b/C -1)(C-1)$ }
\end{tabular}
\caption{Continuous versus categorical:  Estimating the null distribution by $\chi^2_{df}$. }
\label{fig:null_cts_cat_df_mles}
\end{center}
\end{figure}

For each sample of 10,000, the degrees of freedom of a $\chi^2$ distribution were fitted by maximum likelihood.  The mles are plotted against the number of bins for that replicate's binning in Figure \ref{fig:null_cts_cat_df_mles}(a).
There, each sample is distinguished by point colour and symbol according to the number of categories (ncat, or $C$) in that replicate.
As can be readily seen, replicates having the same number of categories appear to fall on a line.  Indeed, a least-squares fit of a different straight line to each set of points grouped by number of categories produced a multiple $R^2 = 99.98\%$.
The mle of the degrees of freedom appears to be a different function of the number of bins for each number of categories $C$.  A simple choice, inspired by the formula for contingency tables is
\begin{equation}
\label{eq:df_cts_cat}
df \approx  \left( \frac{b}{C} - 1\right)  \left( C -1\right) 
\end{equation}
where $C$ is the number of categories, $b$, the total number of blocks over all categories, and $b/C$ the average number of blocks per category.  
Figure  \ref{fig:null_cts_cat_df_mles}(b) plots the mle of the degrees of freedom versus this simple estimate.   A least squares fitted line is overlaid on the points showing an exceptionally strong dependence of the mle on this estimated degrees of freedom (multiple $R^2 = 99.94\%$).  Were the $y = x$ line placed on top, it would be visually indistinguishable from the fitted line. However, so tight is the least-squares fit that the strength of evidence against the hypothesis $\Hyp: y = x$ is overwhelming ($p \approx 3.9 \times 10^{-5}$).  This suggests a better estimate of the maximum likelihood estimated degrees of freedom might be
\begin{equation}
\label{eq:df_cts_cat_fit}
df \approx 
 0.201221 + 0.992706 \left( \frac{b}{C} - 1\right)  \left( C -1\right).
\end{equation}

\begin{figure}[hbt]
\begin{center}
\begin{tabular}{ccc}
\includegraphics[scale = 0.4]{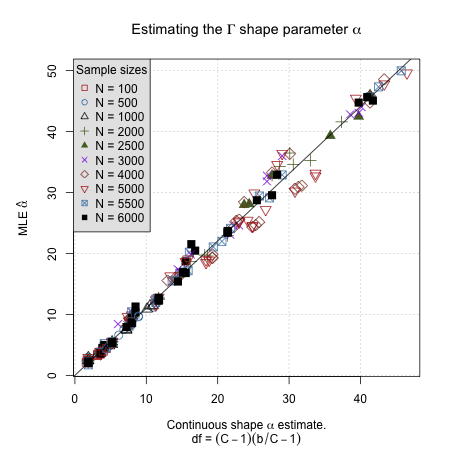}
& $~~~~$ &
\includegraphics[scale = 0.4]{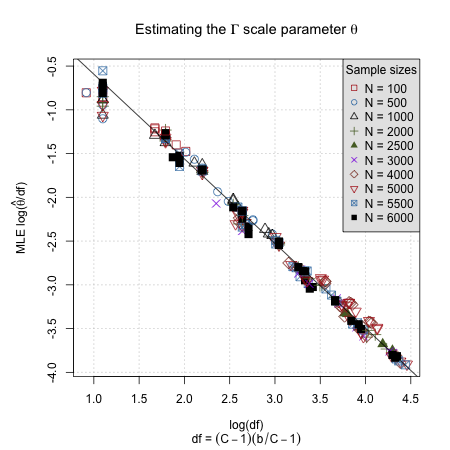}
\\
{\footnotesize
(a) MLEs of the shape $\alpha$ }
& &
{\footnotesize
(b) MLES of the scale $\theta$}
\end{tabular}
\caption{Continuous versus categorical:  Estimating the null distribution by $\Gamma(\alpha, \theta)$. }
\label{fig:mles_cts_cat_gamma}
\end{center}
\end{figure}

Alternatively, as in the continuous versus continuous case, a $\Gamma(\alpha, \theta)$ distribution could be fitted to each empirical distribution of $\bigChi$.
In Figure \ref{fig:mles_cts_cat_gamma}
the maximum likelihood estimates of the shape ($\alpha$) and scale ($\theta$) parameters of a gamma distribution.  

Anticipating that the shape estimator from Equation (\ref{eq:shape}) might be a good start, Figure  \ref{fig:mles_cts_cat_gamma}(a) plots the mles against it, this time with degrees of freedom ($df$) determined by Equation (\ref{eq:df_cts_cat}).
A straight line fitted through the origin then gives predictions
\begin{equation}
\label{eq:shape_cts_cat}
\widehat{\alpha} = 1.102814 \times  \left(0.1199774 + 0.7214124  \sqrt{df}\right)^2
~~~~\text{ where } ~~ df = \left(\frac{b}{C} -1\right) (C-1)
\end{equation}
with $b$ the total number of bins and $C$ the number of categories.

Following Figure  \ref{fig:mles_cts_cts_gamma}(b), the natural logarithms of the scale parameter mles are plotted against the logarithm of the estimated degrees of freedom in  Figure  \ref{fig:mles_cts_cat_gamma}(b).  Again, for the estimated degrees of freedom, the simple formula of Equation  (\ref{eq:df_cts_cat}) is used.  A robustly fitted line to this plot predicts the scale parameter as
\begin{equation}
\label{eq:scale_cts_cat}
\widehat{\theta} = \exp\left(0.3742961 + (1 - 0.9674642) \log(df)\right)
~~~~\text{ where } ~~ df = \left(\frac{b}{C} -1\right) (C-1).
\end{equation}
Again the intercept and slope parameters for the line of Figure \ref{eq:shape_cts_cat}(b) have been given to several decimal places.

\begin{figure}[!ht]
\begin{center}
\begin{tabular}{ccc}
\includegraphics[scale = 0.3]{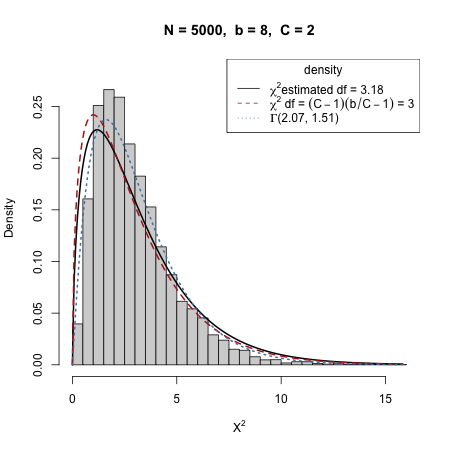
}
&
\includegraphics[scale = 0.3]{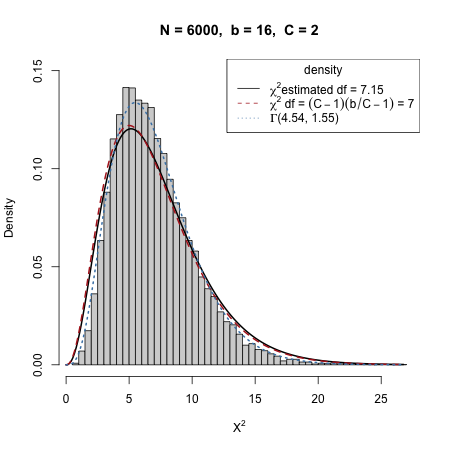
}
&
\includegraphics[scale = 0.3]{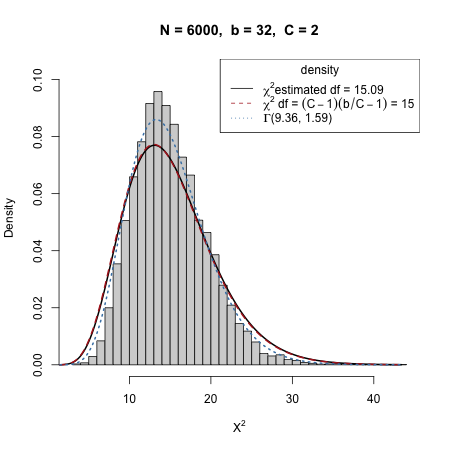
}
\\
\includegraphics[scale = 0.3]{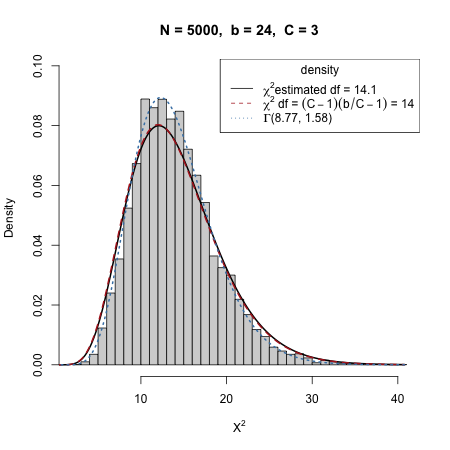
}
&
\includegraphics[scale = 0.3]{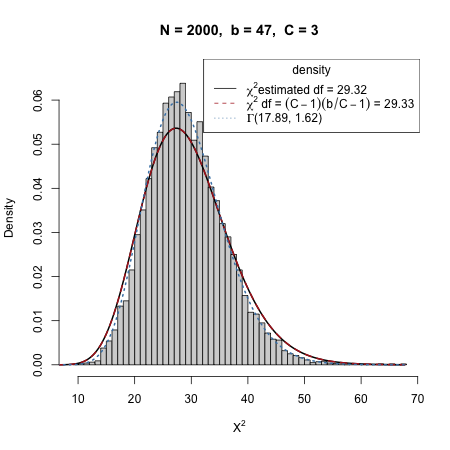
}
&\includegraphics[scale = 0.3]{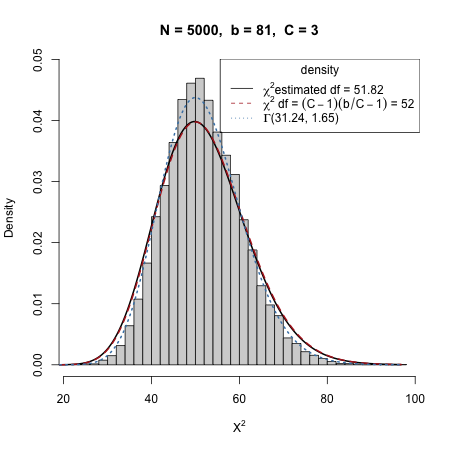
}
\\
\includegraphics[scale = 0.3]{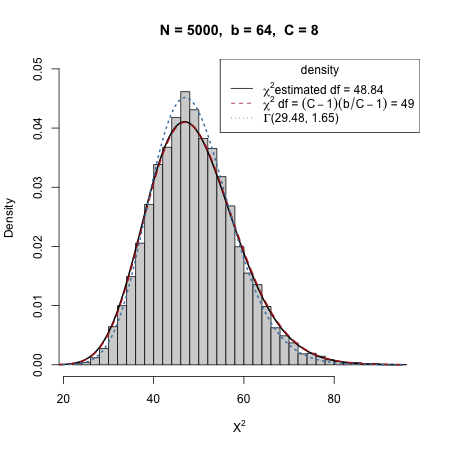
}
&
\includegraphics[scale = 0.3]{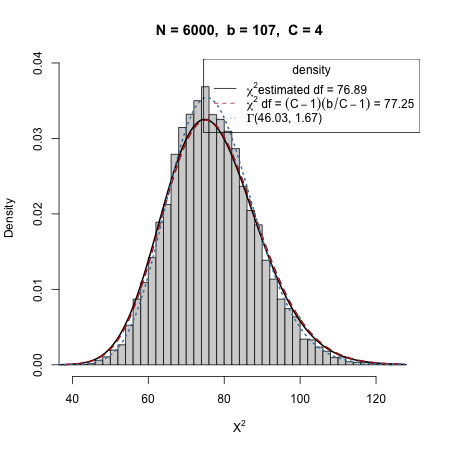
}
&
\includegraphics[scale = 0.3]{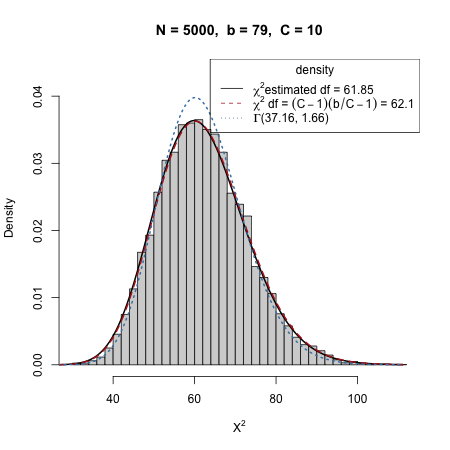
}
\\
\end{tabular}
\caption{Continuous versus categorical:  Estimating the null distribution by $\chi^2_{df}$ 
and $\Gamma(\alpha, \theta)$ densities. }
\label{fig:null_cts_cat_hists}
\end{center}
\end{figure}

To assess the quality of these approximations, the estimated densities were laid over the histogram (based on 10,000 $\bigChi$ values) for each of the 225 replications.
A varied (but haphazard) selection of nine of these are shown in Figure \ref{fig:null_cts_cat_hists}.
These are typical of the quality of the predicted distributional fits over the 225  empirical distributions generated under the hypothesis of  independence.

For the most part, the $\Gamma(\alpha, \beta)$ approximations appear to most closely match the histograms of the null distribution.  Nevertheless, as Figure \ref{fig:null_cts_cat_hists} show, the $\chi^2$s on fitted (\ref{eq:df_cts_cat_fit}) and simplified degrees of freedom (\ref{eq:df_cts_cat}) are also fairly close.

\begin{figure}[hbt]
\begin{center}
\begin{tabular}{ccc}
\includegraphics[scale = 0.3]{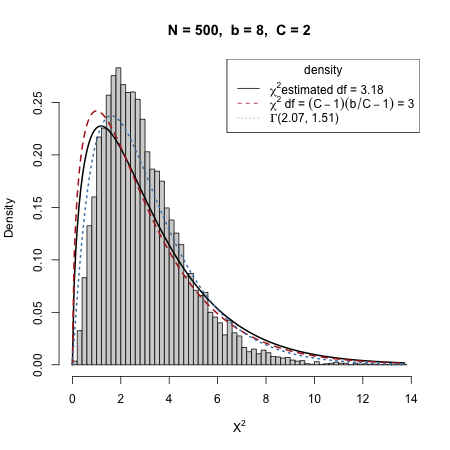} &
\includegraphics[scale = 0.3]{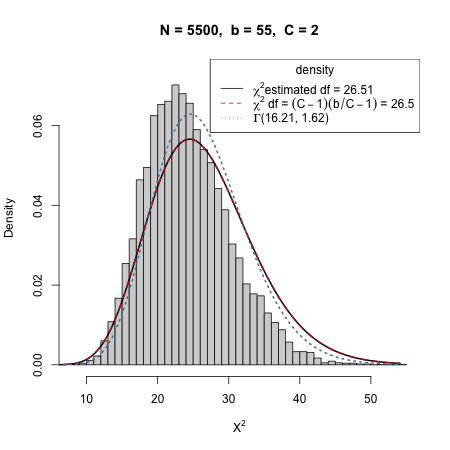} &
\includegraphics[scale = 0.3]{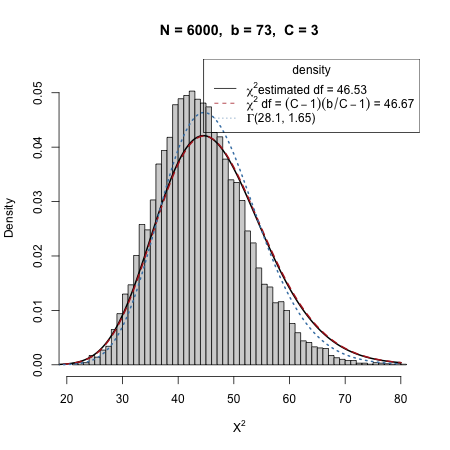}
\end{tabular}
\caption{Three of the worst approximations. About 3 more out of the 225 were similar.}
\label{fig:null_cts_cat_worst}
\end{center}
\end{figure}

In the 225 null distributions produced about six gave visibly poor approximations.  Three of these are shown in Figure \ref{fig:null_cts_cat_worst}
 as exemplars.  The right tail is particularly poorly fit by each of the approximations in these few cases.
 For a large observed value of $\bigChi$ under the null hypothesis, the $p$-value determined by any of the approximation would be much larger than it should be. 
In these cases, the poor approximations would be conservative, in the sense that they would trigger fewer results as significant than they should.
 
\subsubsection{Approximation test properties}
As in the continuous versus continuous case, the behaviour of the approximations are now examined in two
different ways.  First, by taking the quantile $Q_q$ (for $q = 0.95$ and $0.99$) from the empirical null distribution and observing what value $p = 1- q$ each approximation reports for its upper tail probability.
Second, for different fixed $p$ (viz., $0.05$ and $0.01$) now representing the size of the test (the type one error, or false positive rate) is fixed for each approximation and the actual proportion of the null distribution which would be declared positive is determined.

\begin{figure}[hbt]
\begin{center}
\begin{tabular}{ccc}
\includegraphics[scale = 0.25]{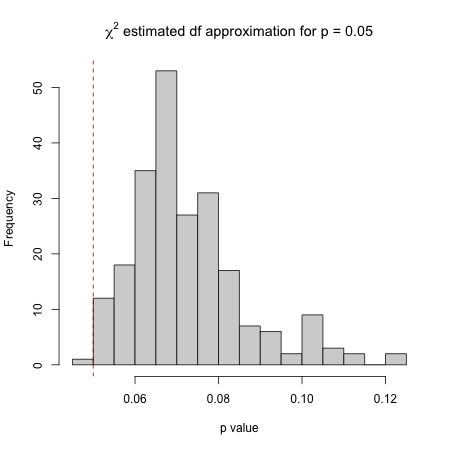} & 
\includegraphics[scale = 0.25]{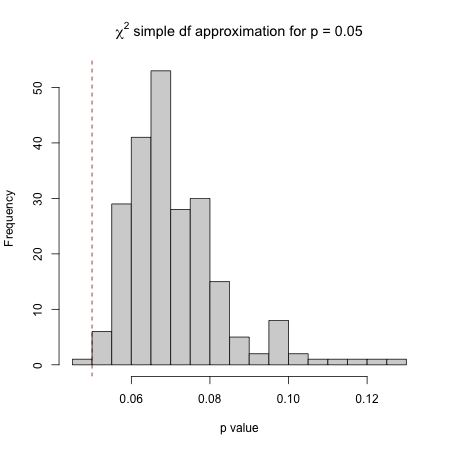} &
\includegraphics[scale = 0.25]{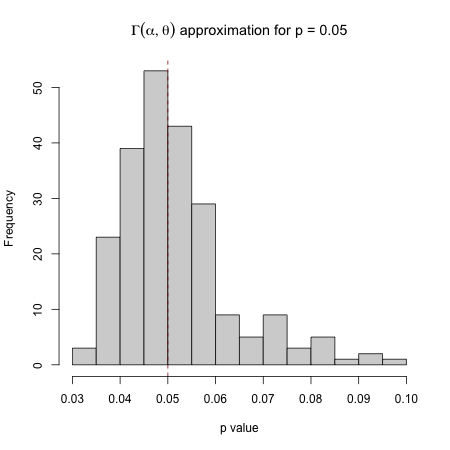} \\
\includegraphics[scale = 0.25]{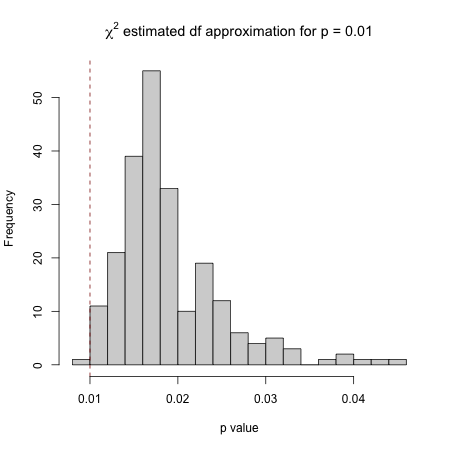} & 
\includegraphics[scale = 0.25]{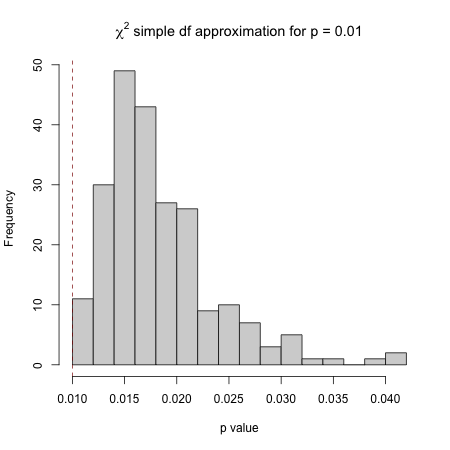} &
\includegraphics[scale = 0.25]{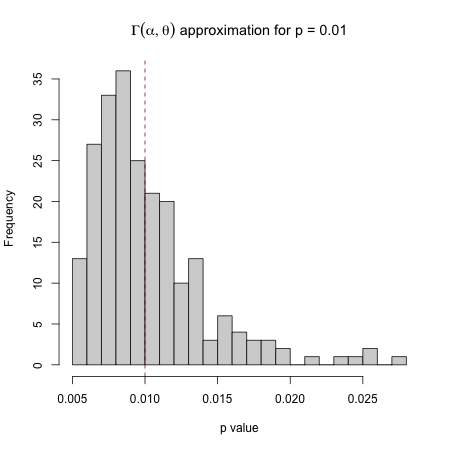} \\
{\footnotesize
(a) $\chi^2$ for df from equation (\ref{eq:df_cts_cat_fit})}
&
{\footnotesize
(b) $\chi^2$ for df from equation (\ref{eq:df_cts_cat})} &
{\footnotesize
(c) $\Gamma(\alpha, \beta)$ approximation }
\end{tabular}
\caption{Distributions of $p = Pr(\bigChi \geq Q_q)$ returned by different approximations.  Top row has $p = 1-q = 0.05$, bottom row $0.01$ -- each is marked by a vertical dashed red line.}
\label{fig:null_cts_cat_p_values}
\end{center}
\end{figure}

Figure \ref{fig:null_cts_cat_p_values} 
shows the $p$-values for the first assessment.
Both $\chi^2$ approximations produce $p$ values much larger than they should.  Again, this could be interpreted as being conservative in terms of triggering significance. 
In contrast, the $\Gamma(\alpha, \theta)$ approximation produces $p$  values on either side of the correct value.

The summary statistics of Table \ref{table:cat_obs_p} 
\begin{table}[h!]
\caption{Continuous versus categorical summary statistics of $p$-values computed by each approximation.}
\begin{center}
{\footnotesize
\begin{tabular}{|l|c|c|c|c|c|c|c|c|}
\hline
&& \multicolumn{7}{c|}{Numerical summary of observed $p$-values}\\
Approximation & $p$ & minimum & $Q(0.25)$ & median & average & $Q(0.75)$ & maximum &IQR\\
\hline
$\chi^2$ estimated d.f.     & 0.05 &  0.046 &    0.064 &    0.069  &   0.073  &   0.078 &    0.122 &   0.013\\
$\chi^2$ simple d.f.          &         &  0.047  &   0.063 &    0.068 &    0.071 &    0.076  &   0.125  &  0.013\\
$\Gamma(\alpha, \theta)$&         &0.032  &   0.044  &   0.049 &    0.052  &   0.056  &   0.098 &   0.012\\
\hline
$\chi^2$ estimated d.f.     & 0.01 &  0.010  &   0.015  &   0.017 &   0.019  &   0.021  &   0.044 &  0.006\\
$\chi^2$ simple d.f.          &         &  0.010 &    0.015  &   0.017  &   0.018  &   0.021 &    0.042 &   0.006\\ 
$\Gamma(\alpha, \theta)$&         & 0.005  &   0.007  &   0.009  &   0.010  &   0.012  &   0.028 &  0.004\\
       \hline
\end{tabular}
}
\end{center}
\label{table:cat_obs_p}
\end{table}%
reinforce this, showing the median of the $\Gamma(\alpha, \theta)$ to be close to the target $p$.
This approximation also has the smallest inter-quartile range.  In contrast, the middle half of the values produced by either 
$\chi^2$ approximation exclude the correct $p$;
indeed, when $p = 0.10$,  both $\chi^2$ approximations have 0.10 at their minimum.  There is not much to choose between the two $\chi^2$ approximations.

\begin{figure}[hbt]
\begin{center}
\begin{tabular}{ccc}
\includegraphics[scale = 0.25]{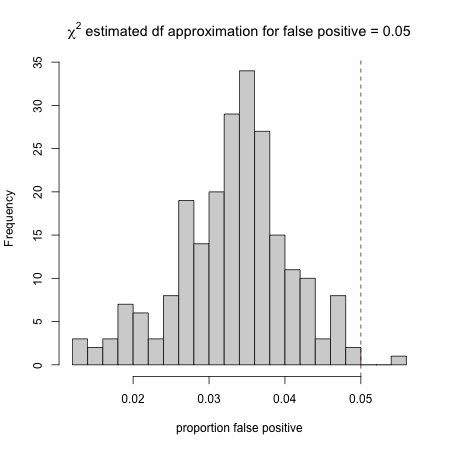} & 
\includegraphics[scale = 0.25]{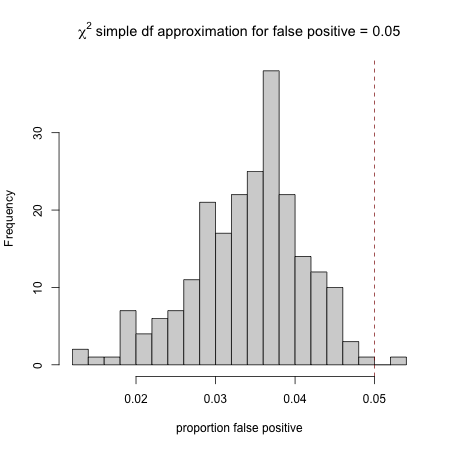} &
\includegraphics[scale = 0.25]{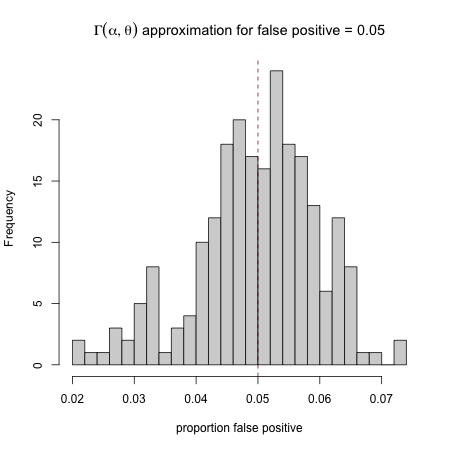} \\
\includegraphics[scale = 0.25]{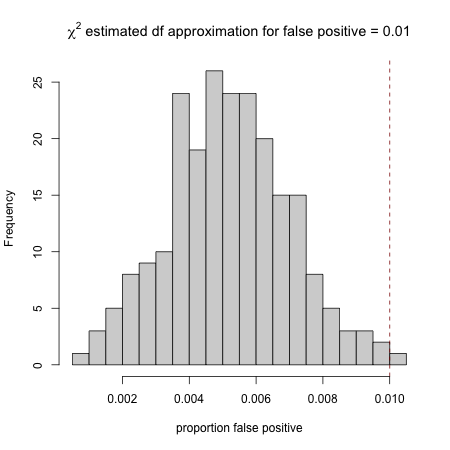} & 
\includegraphics[scale = 0.25]{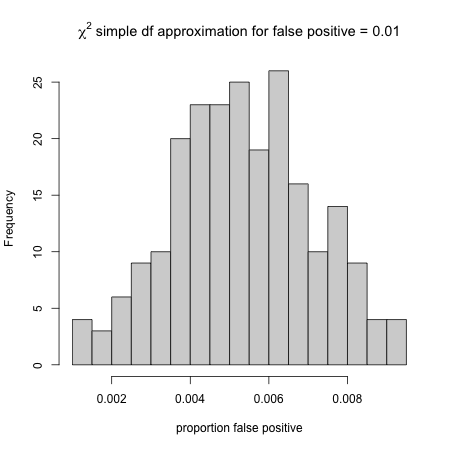} &
\includegraphics[scale = 0.25]{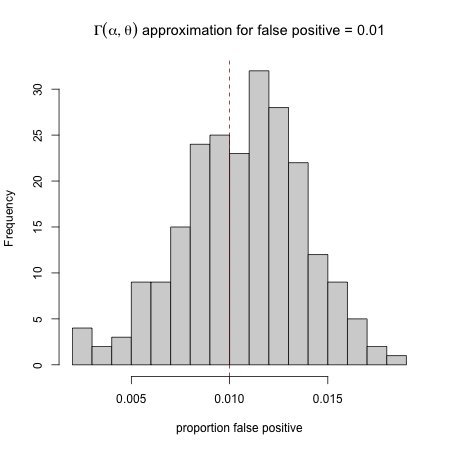} \\
{\footnotesize
(a) $\chi^2$ for df from equation (\ref{eq:df_cts_cat_fit})}
&
{\footnotesize
(b) $\chi^2$ for df from equation (\ref{eq:df_cts_cat})} &
{\footnotesize
(c) $\Gamma(\alpha, \beta)$ approximation}
\end{tabular}
\caption{Distributions of the proportion of false positives returned by different approximations for a true proportion $p$.  Top row has $p  = 0.05$, bottom row $0.01$ -- each is marked by a vertical dashed red line.}
\label{fig:null_cts_cat_false_positives}
\end{center}
\end{figure}

Figure \ref{fig:null_cts_cat_false_positives}
illustrates the values obtained by the approximation for the two false positive proportions.
Now the two $\chi^2$ approximations nearly always under reported the false positive rates.
In the case of the $1\%$ rate and the simple degrees of freedom formula (\ref{eq:df_cts_cat}), no values 
reach that of the true false positive proportion.

These observations are reinforced by the summary statistics of Table \ref{table:cat_obs_false}.
\begin{table}[h!]
\caption{Continuous versus categorical summary statistics of false positives for each approximation.}
\begin{center}
{\footnotesize
\begin{tabular}{|l|c|c|c|c|c|c|c|c|}
\hline
&& \multicolumn{7}{c|}{Numerical summary of observed proportion of false positives}\\
Approximation & $level$ & minimum & $Q(0.25)$ & median & average & $Q(0.75)$ & maximum &IQR\\
\hline
$\chi^2$ estimated d.f.     & 0.05 & 0.013 &   0.029 &   0.034 &   0.033 &   0.038 &   0.054 &  0.008\\
$\chi^2$ simple d.f.          &          &0.013   & 0.030  &  0.035  &  0.034 &   0.038  &  0.052 & 0.009\\
$\Gamma(\alpha, \theta)$&          &0.022 &   0.045 &   0.051  &  0.050 &   0.056  &  0.073 &  0.012\\
\hline
$\chi^2$ estimated d.f.     & 0.01 &  0.001  &  0.004  &  0.005 &   0.005&    0.006 &   0.010 &   0.002 \\
$\chi^2$ simple d.f.          &         & 0.001  &  0.004  &  0.005  &  0.005 &   0.007 &   0.010 &  0.002\\ 
$\Gamma(\alpha, \theta)$&         & 0.003  &  0.009  &  0.011  &  0.011 &   0.013  &  0.018 &   0.004\\
       \hline
\end{tabular}
}
\end{center}
\label{table:cat_obs_false}
\end{table}%
The median (and average) values for the $\Gamma(\alpha, \theta)$ approximation are very close to the true false positive rate from the null distribution and the inter-quartile ranges are very small (essentially only produces changes in the next decimal place).
The $\chi^2$ are located off the mark (lower than they should be) and are effectively indistinguishable from one another.  They do enjoy an even smaller inter-quartile range (in the case of  $1\%$ false positive, half the size)  than the $\Gamma(\alpha, \theta)$ approximation and so would produce more consistent sets of values across the different numbers of blocks and categories. 
%

\newpage
\section{Colouring of the departure display}
\label{app:colouring}
In the departure display, each bin is coloured according to the sign and magnitude of the standardized Pearson residual for that bin -- red indicates a positive residual, blue a negative one.
Only those that are of a large enough magnitude will be coloured, the remainder will be white.

The  $k$th bin's contribution to $\bigChi$ is the square of the Pearson residual, 
\[\frac{o_k - e_k}{\sqrt{e_k}} \]
which uses the Poisson approximation and its standard deviation to give an approximately $N(0,1)$ random variate (e.g., see Appendix \ref{app:simple_approx}). 
A better approximation is had using the standardized Pearson residual 
\[\frac{o_k - e_k}{\sqrt{e_k}} 
            \times \left(
                 \left(\frac{N}{N-1}\right)
                 \left( 1 - \frac{r}{N} \right)
                 \left(1 - \frac{c}{N} \right)
                 \right)^{-\frac{1}{2}} 
                 \]
which standardizes $(o_k - e_k)$ by the variance of $o_k$ as determined for the count in 
Appendix \ref{app:paired_ranks}.

Colouring is only applied when the standardized Pearson residual is greater than 2, and full saturation is applied to all Pearson residuals greater than the $1 - \frac{0.001}{K}$ quantile of the $N(0,1)$ distribution. The lower end of this range is approximately the 0.95 quantile of the $N(0,1)$ distribution while the upper end is the 0.999 quantile with a Bonferroni correction conservatively accounting for the number of bins. Light shading therefore indicates bins which have residuals that may indicate dependence, should enough occur, while bins with darker shading indicate bins that warrant inspection on their own.
The gradient is created using \code{colorRampPalette} in \R (\cite{Rlang}) with the number of breaks supplied by the user.
\newpage
\section{Recursive maximal score binning}
\label{app:RecursiveMaxScoreBinning}

For recursive binning, rather than split bins randomly,  bins could be split to maximize the measure of departure, $\bigChi$.
The direction of the split (horizontal or vertical) and its cut point (always one of the ranks, see Proposition \ref{prop:convscore}, Appendix \ref{app:proof}) 
are chosen to produce the largest $\bigChi$ at that step.

\begin{figure}[!h]
\begin{center}
\begin{tabular}{ccc}
\includegraphics[scale = 0.6]{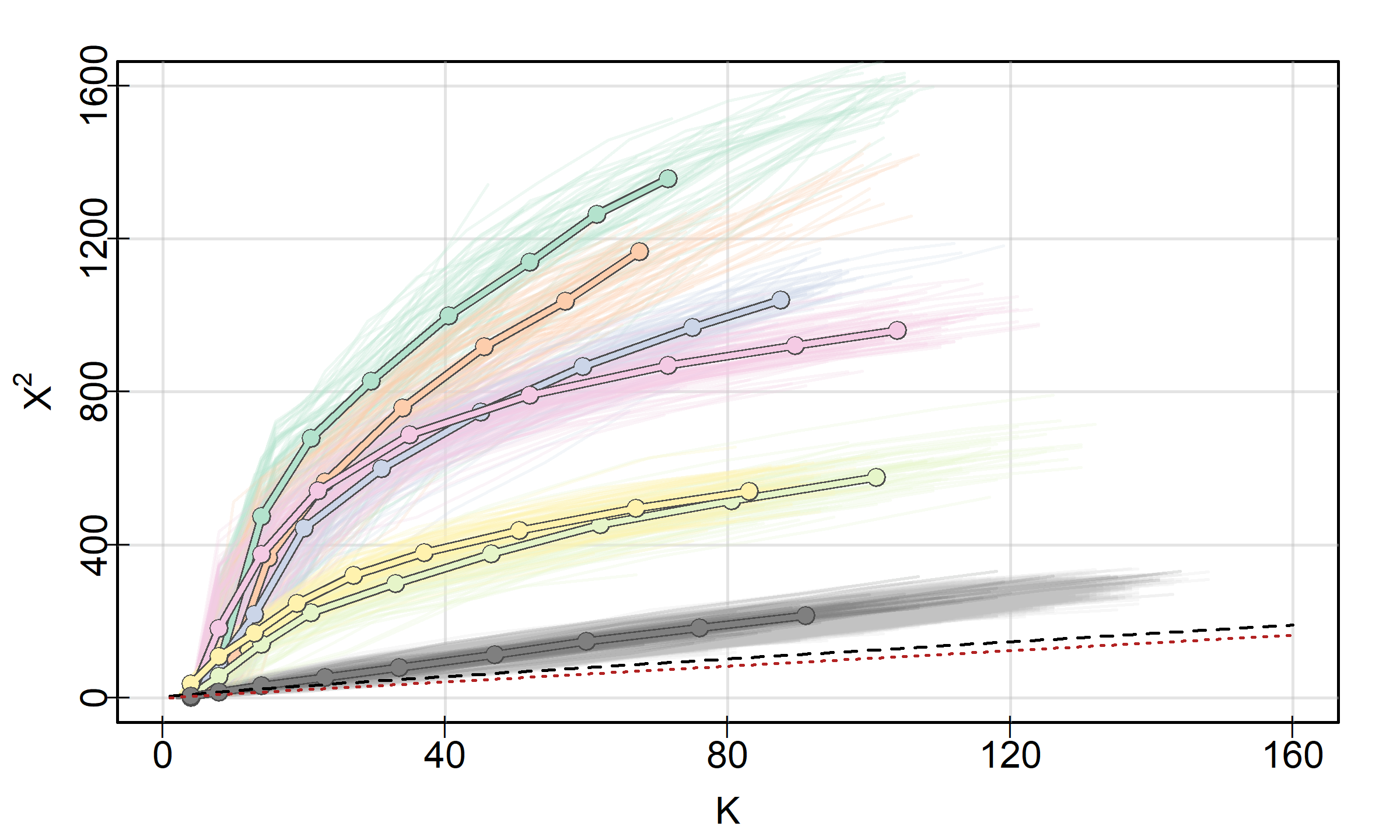}
& $~~~$ & 
\includegraphics[scale = 0.6]{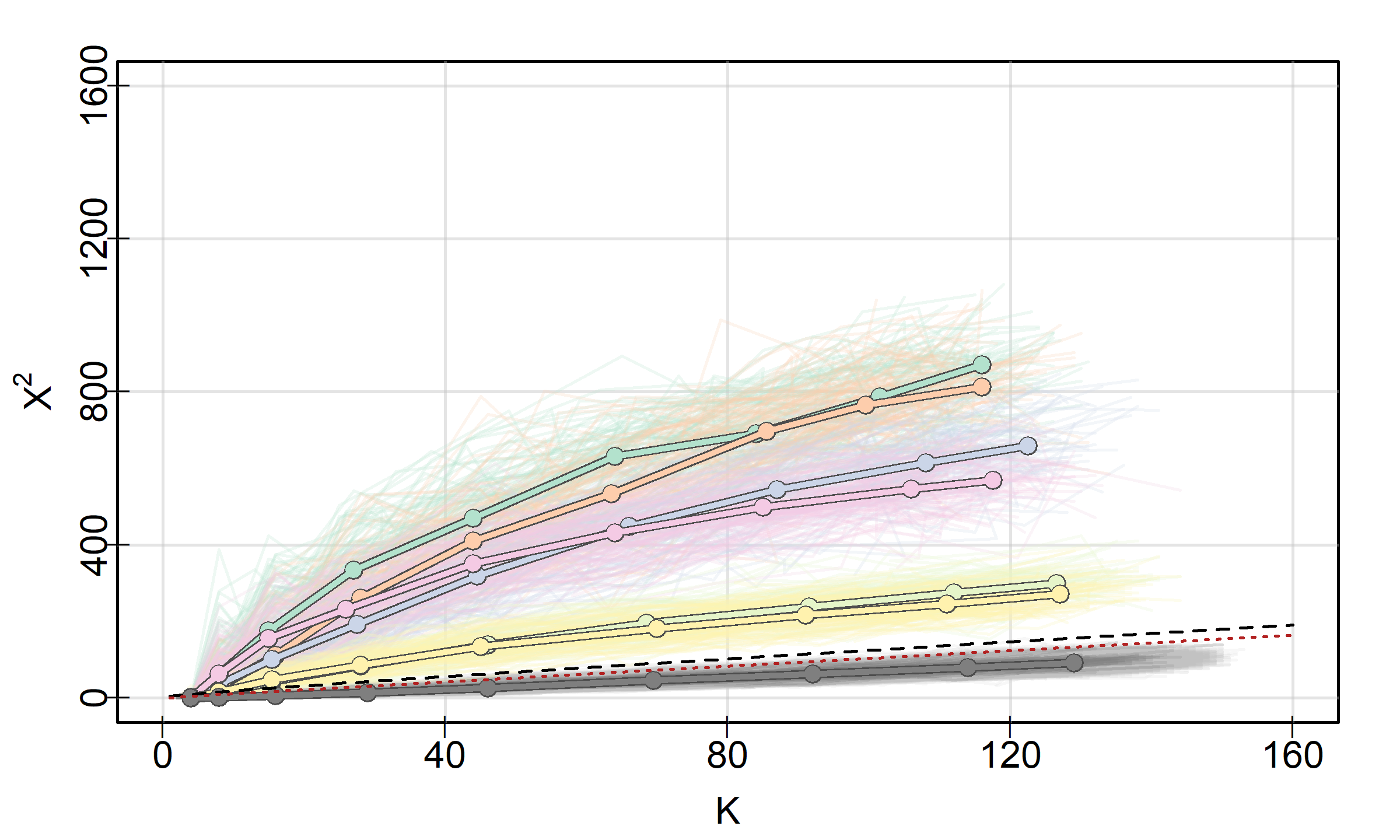} \\
{\footnotesize (a) Maximal  $\bigChi$ splitting} 
&&
{\footnotesize (b) Random splitting} 
\end{tabular}

\caption{{\footnotesize Recursive maximal score binning of exemplars.  Shown colour coded are 100 replications as paths for each of the seven exemplar data patterns.  The median path of each exemplar is plotted with a thicker line width.  Dashed and dotted curves delineate the 0.95 quantile of a $\chi^2_{K-1}$ and of a 
 $\chi^2_{(\sqrt{K} -1)^2}$.}}
\label{fig:simDataBinMaxPaths}
\end{center}
\end{figure}

Compared to its random, and data-independent,  counterpart,  
recursive maximal $\bigChi$ binning will necessarily amplify any departures from $\Hyp_0$.  This is readily seen in Figure \ref{fig:simDataBinMaxPaths}(a)
where the $\bigChi$ values are generated for 100 simulations of each exemplar using the maximal binning.
For comparison, Figure \ref{fig:simDataBinMaxPaths}(b) reproduces the random splitting curves of Figure \ref{fig:simDataBinAllRandPaths}  on the same vertical scale.
Maximal binning  produces much higher values of $\bigChi$ and better separates the data configurations.
Moreover, because there is greedy maximization at each step, the individual paths within a configuration are smoother.

\begin{figure}[!h]
\begin{center}
\begin{tabular}{ccc}
\includegraphics[scale = 1]{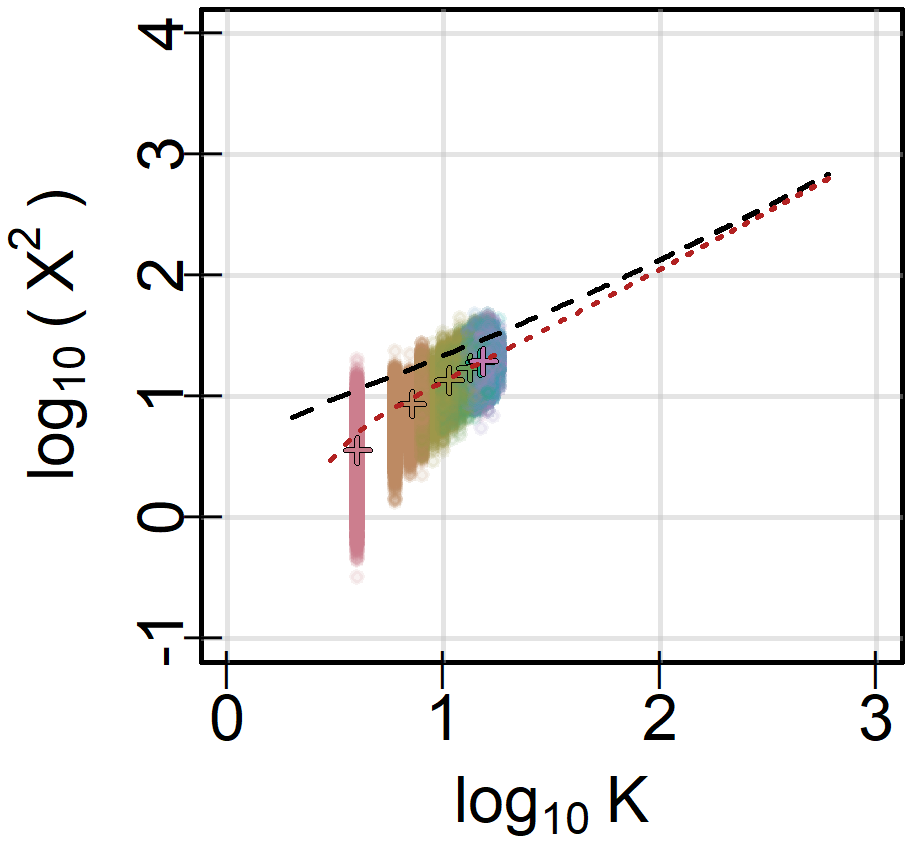} & \includegraphics[scale = 1]{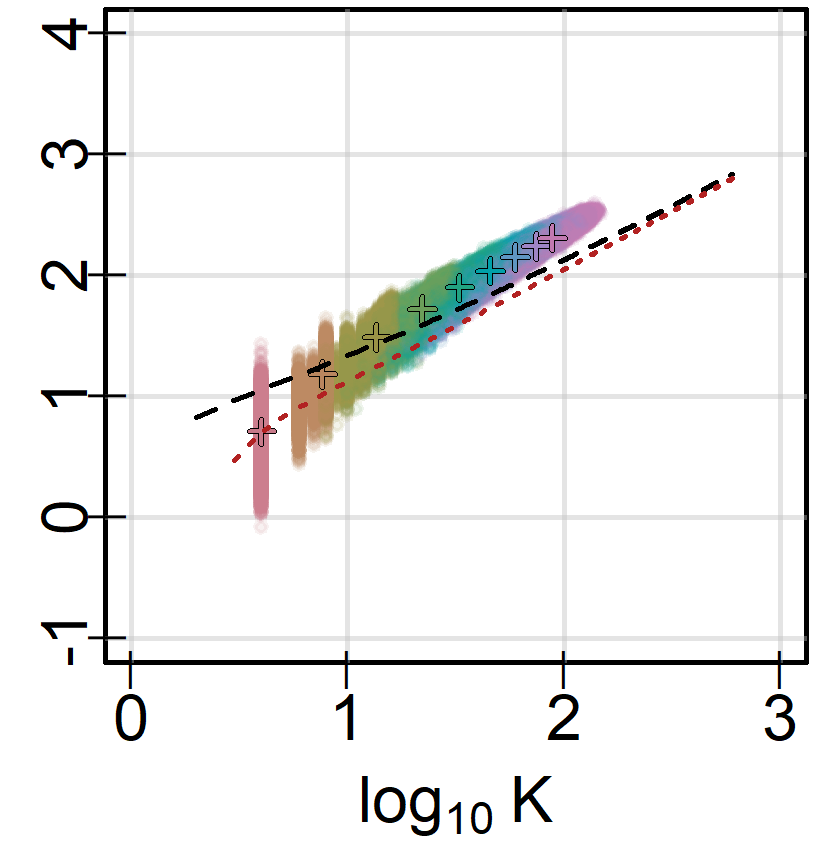} & \includegraphics[scale = 1]{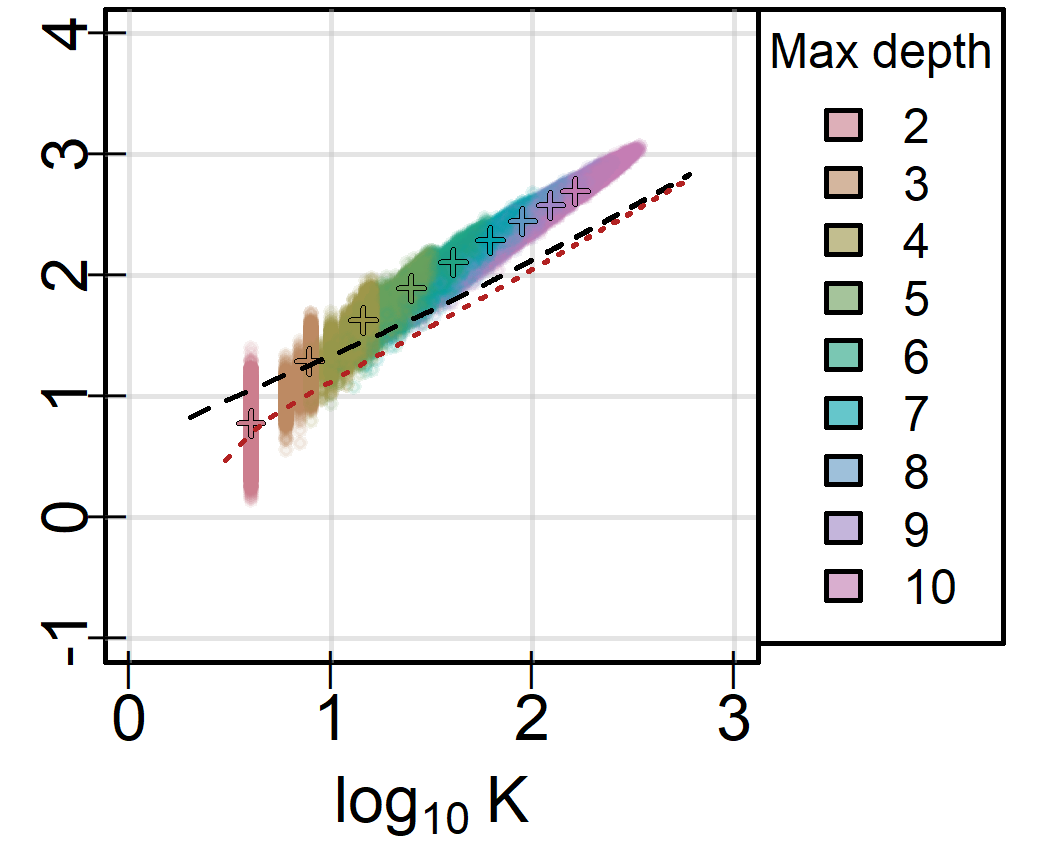} \\
{\footnotesize (a) $N$ = 100} & {\footnotesize (b) $N$ = 1,000} & {\footnotesize (c) $N$ = 10,000} \\
\end{tabular}
	\caption{{\footnotesize Recursive maximal score binning under $\Hyp_0$. 
	For each depth/colour, 10,000 independent replications are plotted.
	Crosses identify average $K$ and $\bigChi$ for each depth/colour. Dashed  and dotted curves follows the 0.99 quantile for $\chi^2_{K-1}$ and $\chi^2_{(\sqrt{K} -1 )^2}$, respectively.}}
\label{fig:depthBySampleSizeMaxChi}
\end{center}
\end{figure}

The null configuration paths are also raised, unfortunately now appearing above both $\chi^2$
approximations -- using these cutoffs would produce nearly all false positives (e.g., for large sample sizes and numbers of bins).
Figure \ref{fig:depthBySampleSizeMaxChi}
shows just how poorly these approximations fare under $\Hyp_0$ when $\bigChi$ is calculated using maximal $\bigChi$ binning.
Compared to the equivalent display Figure \ref{fig:depthBySampleSizeRand} and comparing the two shows how maximal binning results in a tighter point cloud that is both located higher vertically and with a higher slope. 

The crosses identify the pair averages of $(K, \bigChi)$ for each maximal depth which, within each plot, smoothly change as a function of  $K$. 
Unlike the random binning results of Figure \ref{fig:depthBySampleSizeRand}, the pattern of crosses in Figure \ref{fig:depthBySampleSizeMaxChi} change dramatically with sample size.
From Figure \ref{fig:depthBySampleSizeMaxChi} (a) to (c), $N$ changes by an order of magnitude each time and the slope of the curve of crosses increases. 
Any approximation would have to be function of both $K$ and $N$.

\subsection{Maximal $\bigChi$ departure displays}
\label{sec:maxgraphic}
By greedily maximizing $\bigChi$ when choosing which bin to split, the resulting bins should also 
have their values amplified in the departure display.  Again, this should emphasize non-null patterns at the cost of introducing more false positives.

\begin{figure}[!h]
	\begin{center}
		\begin{tabular}{c}
			\includegraphics[scale = 0.8]{redRanks} \\
			\includegraphics[scale = 0.8]{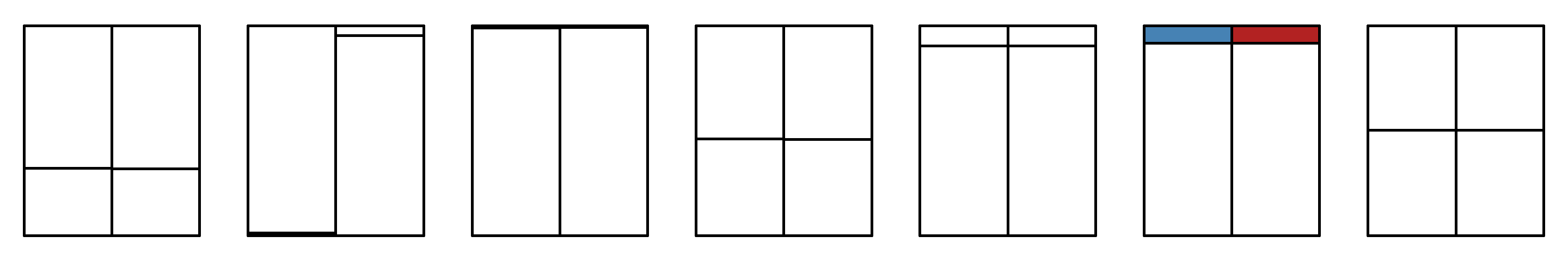} \\
			\includegraphics[scale = 0.8]{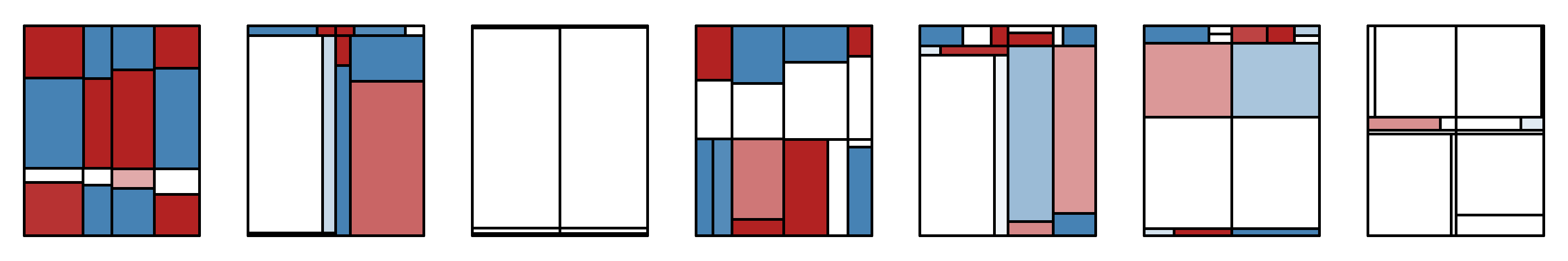} \\
			\includegraphics[scale = 0.8]{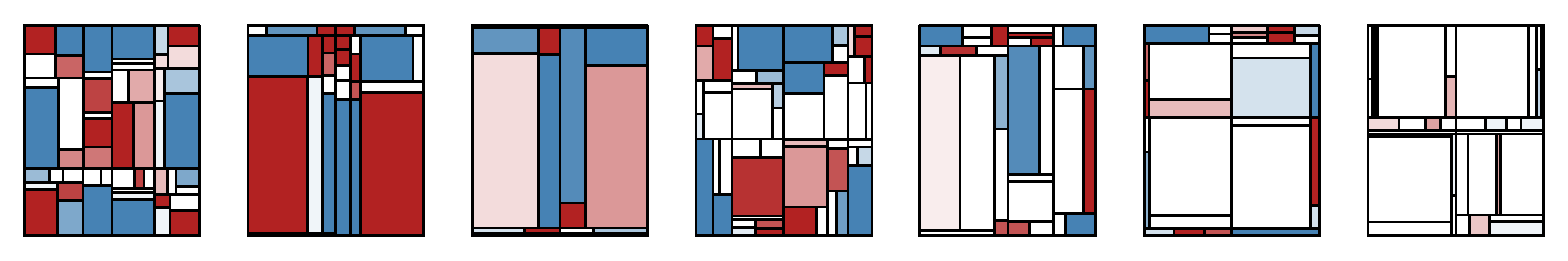} \\
			\includegraphics[scale = 0.8]{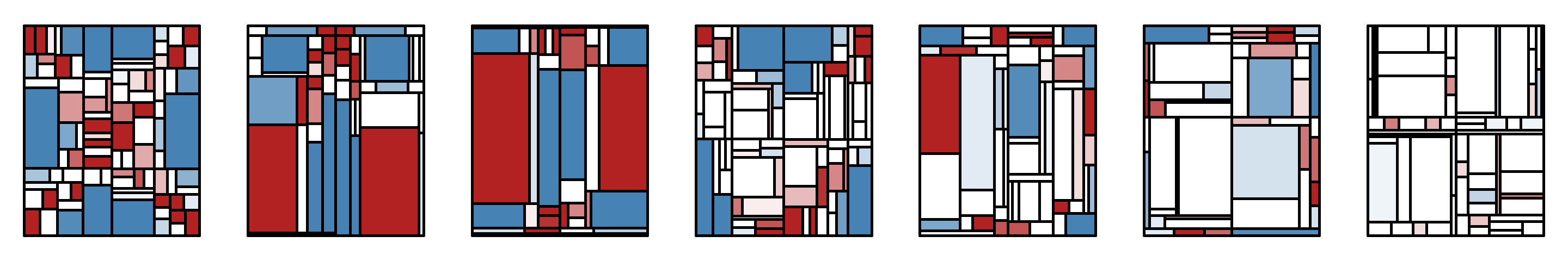} \\
			\includegraphics[scale = 0.8]{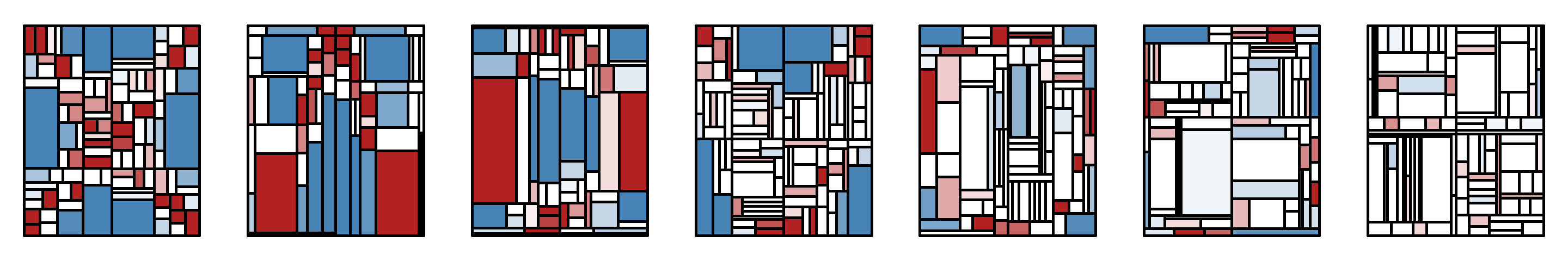} \\
			\includegraphics[scale = 0.8]{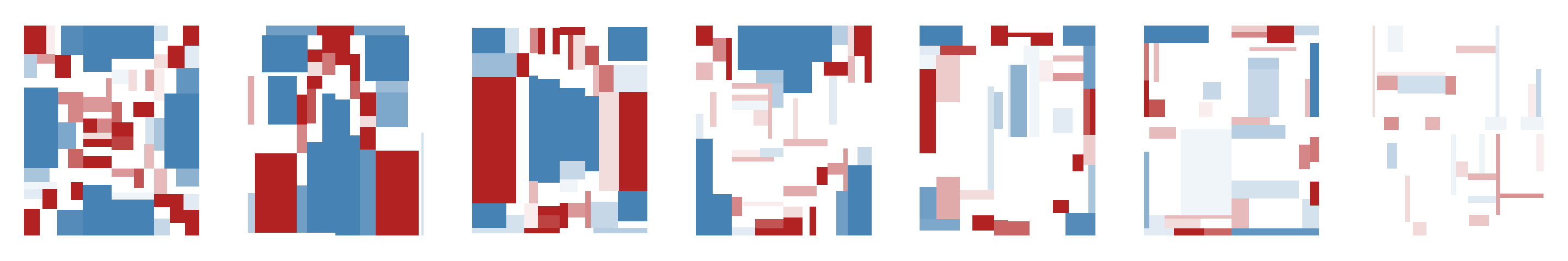} \\
		\end{tabular}
	\end{center}
	\caption{{\footnotesize Departure displays from recursive maximal $\bigChi$ binning.  Rank patterns are shown in the first row. 
	Binning depths of 2, 4, 6, 8, and 10 appear within each column for that configuration; the last row is again depth 10 but without bin boundaries.}}	
\label{fig:simDataMaxEvol}
\end{figure}

The visual effect on the display of the exemplar configuration patterns of Figure \ref{fig:simulatedPatterns} is shown in Figure \ref{fig:simDataMaxEvol}
 in the same way as those of Figure \ref{fig:simDataRandEvol}.
 Since, greedy maximization is being used, the ``squarify'' option is not used, thus allowing the maximization to split the shorter dimension if that leads to a greater increase in $\bigChi$.
 
As might be expected, amplification has resulted in non-null configurations being apparent earlier, for fewer recursive divisions when compared to the random binning.
Even for $d = 4$ (i.e., row 3 of Figure \ref{fig:simDataMaxEvol}), a few of the non-null configurations are discernible in the maximal $\bigChi$ departure display. 
In the last two rows, where $d = 10$ ($K \approx 2^{10}$),  the general structure of each non-null configurations is easily discerned even with bin borders drawn.
Even without the ``squarify'' option, the non-null configurations are revealed.

The cost is revealed by comparing the last column, the null configuration, in Figures  \ref{fig:simDataRandEvol} and \ref{fig:simDataMaxEvol}.
There, the maximal binning triggers many more false positives than does the random binning.

Maximal binning would be preferred to random binning as it will be more sensitive to departures from independence and its departure display will be good at revealing the underlying pattern of dependence.  However, unless its null distribution is determined, either by permutation or by a good approximation, it cannot be generally recommended.

As the random binning appears to also reveal the patterns, though possibly at the cost of greater depth binning, and does not suffer as much from false positives, it is generally recommended over maximal binning.  Maximal binning, should it be used at all, should be used with caution and only for relative comparisons between pairs of variates (i.e., not for absolute assessment).
%

\newpage
\section{Algorithms}
\label{app:algorithms}

Though presented as recursion, the recursive binning outlined above is implemented iteratively in \code{AssocBin}.
Two separate pieces,  Algorithms \ref{algo:binner} and \ref{algo:maxSplit} provide the details -- Algorithm \ref{algo:binner} describes an outer function managing the splitting and stopping while Algorithm \ref{algo:maxSplit} details the inner splitting procedure.

\begin{algorithm}[H]
  \caption{Iterative binning of the ranks}
  \label{algo:binner}
  \begin{algorithmic}[5]
{\footnotesize    \Structures
    \State {\footnotesize $b_i$ (a \emph{bin}): rectangular region of $[0,n]^2$ defined by boundaries, contained observations $\ve{s}_i, \ve{t}_i$ of length $o_i$, and depth (number of splits from initial bin to $b_i$); $\mathcal{B}$ (a \emph{bin list}): partition of $[0,n]^2$ into bins $b_1, b_2, \dots, b_K$; $\ve{s}, \ve{t}$ (marginal ranks): permutations of $\tr{(1, 2, \dots, n)}$; $\ve{e}$ (stop vector): vector $\tr{(e_1, e_2, \dots, e_K)} \in \{0,1\}^K$, $1$ at position $i$ if $b_i$ meets the stop criteria}
    \EndStructures

    \vspace{8pt}
    
    \Procedure{binner}{$\ve{s}$, $\ve{t}$, $checkStopCrit$, \textsc{maxSplit}}
    \State \textbf{construct} $b_0$: bounds $[0,n \leftarrow length(s)]$ on both margins; depth 0; containing $\ve{s}$, $\ve{t}$
    \State \textbf{split} $b_0$ randomly in half vertically or horizontally into bins $b_1$, $b_2$
    \State $\mathcal{B}$ $\leftarrow$ $\{b_1, b_2\}$; $\ve{e} \leftarrow$ $checkStopCrit( \mathcal{B})$; $K \leftarrow 2$ \Comment{Initialize}
    \While {$sum(\ve{e}) < K$} \Comment{Go until all bins meet stop criteria}
    \For {$i \in \{i: e_i = 0\}$}
    \State $\mathcal{B} \leftarrow \{b_1, b_2, \dots, b_{i-1}, b_{i+1}, \dots, b_K\} \union$ \textsc{maxSplit}$(b_i)$
    \EndFor
    \State $K \leftarrow 2K - sum(\ve{e})$; $\ve{e} \leftarrow checkStopCrit(\mathcal{B})$ \Comment{Update $K$ before $\ve{e}$}
    \EndWhile
    \State \textbf{return} $\mathcal{B}$
    \EndProcedure
}
  \end{algorithmic}
\end{algorithm}

Note that the case of tied scores must be addressed in Algorithm \ref{algo:maxSplit}. If all splits have the same score along a margin, there is no clear maximizing split. In this case, a reasonable choice is to halve the bin at the ceiling of the average of its upper and lower bounds in the hope of breaking the tie for subsequent score calculations. This works, for example, on the first split when both margins contain the complete vector of ranks and every split gives observed counts that match expectation exactly.

\begin{algorithm}[H]
  \caption{Score maximizing splitter}
  \label{algo:maxSplit}
  \begin{algorithmic}[5]
{\footnotesize    \Structures
    \State {\footnotesize $b_i$ (a \emph{bin}): rectangular region of $[0,n]^2$ defined by boundaries, contained observations $\ve{s}_i, \ve{t}_i$ of length $o_i$, and depth (number of splits from initial bin to $b_i$); $\ve{d}_s, \ve{d}_t$ (score vectors): vectors $\tr{[(d_s)_1, (d_s)_2, \dots, (d_s)_{o_i+1}]} \in \Reals^{o_i+1}$ giving the score values for potential splits}
    \EndStructures

    \vspace{8pt}
    
    \Procedure{maxSplit}{$b_i$, $scoreFun$}
    \State $\ve{s}_i^{\prime} \leftarrow \ve{s}_i \union \{\min(\ve{s}_i) - 1\}$; $\ve{t}_i^{\prime} \leftarrow \ve{t}_i \union \{\min(\ve{t}_i) - 1\}$ \Comment{Allow empty splits below data}
    \State $\ve{d}_s \leftarrow scoreFun(\ve{s}_i^{\prime}, b_i)$; $\ve{d}_t \leftarrow scoreFun(\ve{t}_i^{\prime}, b_i)$ \Comment{Score all potential splits of $b_i$}
    \If {$({d}_s)_j = \max(\ve{d}_s)$ and  $({d}_t)_j = \max(\ve{d}_t)$ $\forall j \in \{1, \dots, o_i+1\}$}
       \If {$\max(\ve{d}_s) > \max(\ve{d}_t)$} \Comment{If all scores on both margins are tied...}
         \State \textbf{split} $b_i$ in half on $s$ to get bins $b_1, b_2$ \Comment{...halve on the larger margin}
       \ElsIf {$\max(\ve{d}_s) < \max(\ve{d}_t)$}
         \State \textbf{split} $b_i$ in half on $t$ to get bins $b_1, b_2$
       \Else
       \State \textbf{split} $b_i$ in half randomly on $s$ or $t$
       \EndIf
    \ElsIf {$\max(\ve{d}_s) \geq \max(\ve{d}_t)$} \Comment{Ties go to $s$ if not all scores are equal}
      \State \textbf{split} $b_i$ on $s$ at $({s}_i^{\prime})_j$ where $j = \min\{j : (d_s)_j = \max(\ve{d}_s)\}$ to get bins $b_1, b_2$
    \Else
      \State \textbf{split} $b_i$ on $t$ at $({t}_i^{\prime})_j$ where $j = \min\{j : (d_t)_j = \max(\ve{d}_t)\}$ to get bins $b_1, b_2$
    \EndIf
    \State \textbf{return} $\{b_1, b_2\}$
    \EndProcedure}

  \end{algorithmic}
\end{algorithm}

Note that Algorithm \ref{algo:maxSplit} also supports random splits. If the score function generates independent and identically distributed $U[0,1]$ realizations, the coordinate of the maximum score value will be uniformly distributed among potential split coordinates.

\newpage
\section{Score maximizing splits}
\label{app:proof}

To maximize $d(o_k, e_k)$, denote the $o_k$ pairs of ranks in bin $k$ as $\{(s_{k1}, t_{k1}), (s_{k2}, t_{k2}), \dots, (s_{ko_k}, t_{ko_k})\}$ and its $s$ bounds $(l_s, u_s]$ and $t$ bounds $(l_t, u_t]$. Bin $k$ can be split either by a vertical edge at $c_s \in (l_s, u_s)$ or a horizontal edge at $c_t \in (l_t, u_t)$ to give two new bins with two new scores. Denote the observed and expected values for the bin above $c_s$ as $o_{k+}(c_s)$ and $e_{k+}(c_s)$ respectively (analogously, those above $c_t$ as $o_{k+}(c_t)$ and $o_{k+}(c_t)$), and use the subscript $k-$ in the same way for the bin below the new edge. A split at $c$ therefore changes the total score measured by $d(\cdot, \cdot)$ for the region $(l_s, u_s] \times (l_t, u_t]$ by
\begin{equation} \label{eq:propMeas:scoreChange}
  \delta_k(c, d) = d \Big ( o_{k+}(c), e_{k+}(c) \Big ) + d \Big ( o_{k-}(c), e_{k-}(c) \Big ) - d(o_k, e_k),
\end{equation}
and so the maximizing split coordinate along a given dimension is
$$c^*_i = \argmax_{c} \delta_k(c, d) = \argmax_{c} \Big [ d \big ( o_{k+}(c), e_{k+}(c) \big ) + d \big ( o_{k-}(c), e_{k-}(c) \big ) \Big ].$$
Though $e_{k+}(c)$ and $e_{k-}(c)$ vary continuously in the split coordinate $c$, both of $o_{k+}(c)$ and $o_{k-}(c)$ change only when $c$ is the coordinate of an observation in bin $k$, in other words when $c_s \in \{s_{k1}, s_{k2}, \dots, s_{ko_k}\}$ or $c_t \in \{t_{k1}, t_{k2}, \dots, t_{ko_k}\}$. This has an important consequence.

\begin{proposition}[The score-maximizing split] \label{prop:convscore}
	If $d(x,y)$ is continuous and convex in $y$, that is $\frac{d^2}{dy^2} d(x,y) \geq 0,$ then the split coordinate $c$ maximizing $\delta_k(c, d) = d \big ( o_{k+}(c), e_{k+}(c) \big ) + d \big ( o_{k-}(c), e_{k-}(c) \big ) - d(o_k, e_k)$ is the coordinate of one of the points within bin $k$.
\end{proposition}
\begin{proof}
	Without loss of generality, consider a split at $c_s \in (s_{kj}, s_{k(j+1)})$ between the $j$ and $j+1$ horizontal coordinates in bin $k$. This implies the number of observations above is constant at $o_{k+} = o_k - j$ and below is constant at $o_{k-} = j$ by definition. Furthermore $e_{k+}(c_s) = (u_s - c_s)(u_t - l_t)/n$ and $e_{k-}(c_s) = (c_s - l_s)(u_t - l_t)/n$ so that
	\begin{equation*}
		\delta_k(c_s, d) = d \left (o_k - j, \frac{(u_s - c_s)(u_t - l_t)}{n} \right ) + d \left (j, \frac{(c_s - l_s)(u_t - l_t)}{n} \right ) - d( o_k, e_k ).
	\end{equation*}
	If $\delta_k(c_s, d)$ is convex and continuous in $c_s$, then its maximum must occur at one of $s_{kj}$ or $s_{k(j+1)}$ (see Figure \ref{fig:proofImage}).
	
	\begin{figure}[!h]
		\begin{center}
			\includegraphics[scale = 0.3]{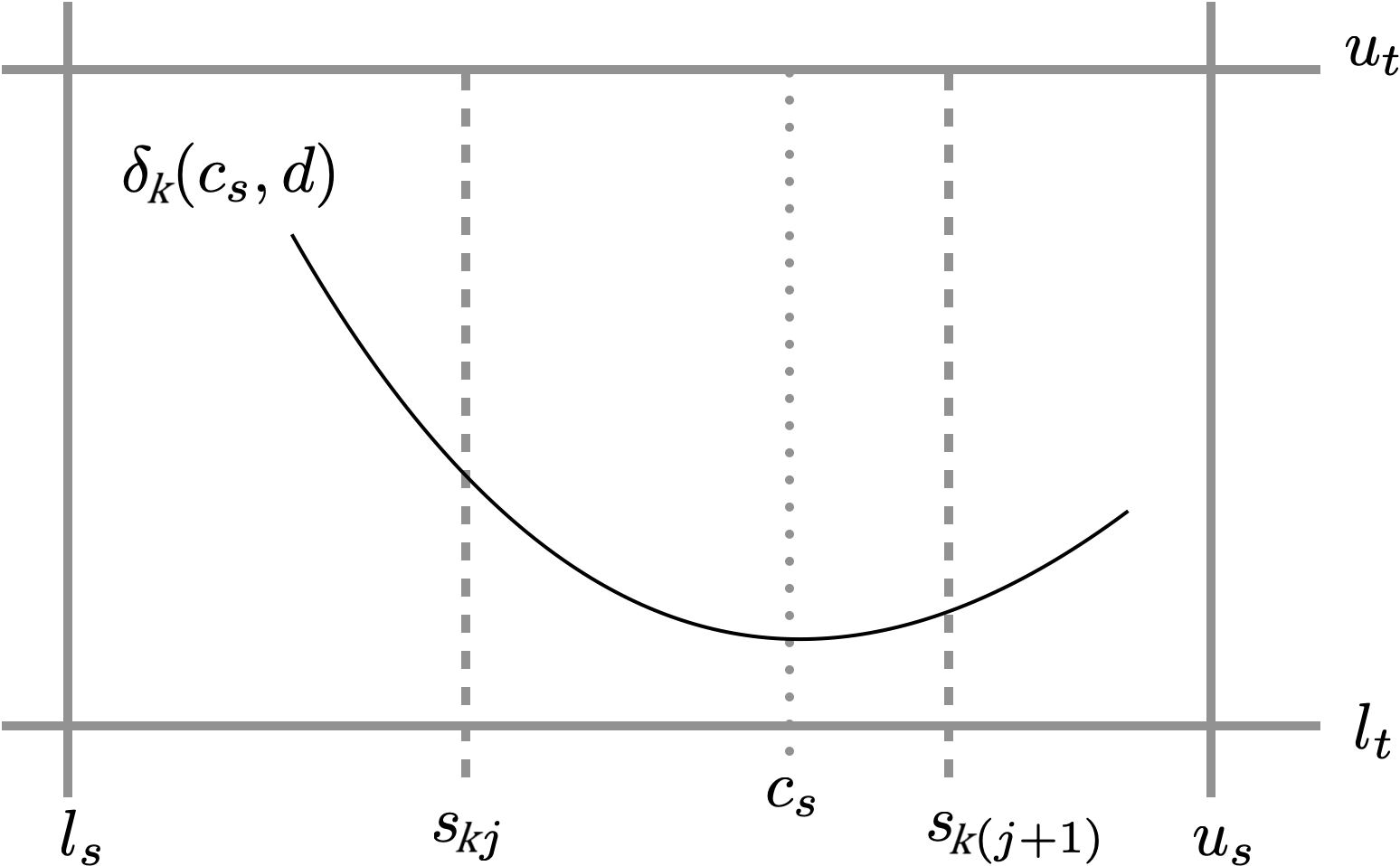}
			\caption{{\footnotesize A sketch of $\delta_k(c_s,d)$ in bin $k$.}}
			\label{fig:proofImage}
		\end{center}
	\end{figure}

	Therefore, we only need to prove the convexity of $\delta_k(c_s,d)$ to prove that its maximum occurs at an observation coordinate. Consider the sign of its second derivative, given by
	\begin{equation*}
		\frac{d^2}{dc_s^2} \delta_i(c_s, d)  = \frac{(u_t - l_t)^2}{n^2} \left [ \frac{d^2}{de_{k-}^2} d \left (j, e_{k-} \right ) + \frac{d^2}{de_{k+}^2} d \left (o_k - j, e_{k+} \right ) \right ].
	\end{equation*}
	This is greater than or equal to zero if $\frac{d^2}{dy^2} d(x,y) \geq 0$ for all $x \in \{0, 1, \dots, o_k\}$.
	
	In this case, $\delta_k(c_s, d)$ is concave up between the horizontal coordinates of the points within a bin so that any optimum within these bounds must be minimum. As these are continuous functions, this means the maximum must occur at one of the boundaries of the interval $(s_{kj}, s_{k(j+1)})$. The same argument holds for every interval horizontally and vertically.
\end{proof}

In particular, note that $\frac{d^2}{dy^2} (x - y)^2/y = 2 x^2 / y^3 \geq 0$ for all $x \geq 0$ and $y \geq 0$.
This means that splits only need to be considered at the points in $\{(s_{k1}, t_{k1}), (s_{k2}, t_{k2}), \dots,$ $(s_{ko_k}, t_{ko_k})\}$ to maximize $\bigChi$. The same is true for the mutual information, which has $d(x,y) = x \log ( x / y )$.
\newpage

\section{The \pkg{AssocBin} package}
\label{app:AssocBin}

Recursive rank binning under user-specified stop criteria and splitting logic is implemented in the \R{} package \pkg{AssocBin}available on CRAN (\citealp{AssocBin, Rlang}). 
Calling the \code{inDep} function on an \R{} \code{data.frame}, random recursive binning is performed on every pair of variates and an object of class \code{inDep} is produced. Optionally, the user can specify how bins are to be split and when splitting should be stopped in this call.
Alongside the bins resulting from the specified recursive binning on all pairs, the value of $\bigChi$ and the corresponding $\chi^2$ $p$-value are included for each pair in every \code{inDep} instance.
In the wine example of Section \ref{sec:wineEx}, \code{inDep} was called on the data frame, with random ``squarified'' binning, to create the binnings for the 78 possible variate pairs (both categorical and continuous).
As the data contain both continuous and categorical data, the degrees of freedom for each pair follow from the number of restrictions for that pair.

Methods for the generic \code{summary} and \code{plot} functions support the quick interrogation of the prevalence of relationships and the strongest pairwise relationships in the data. The \code{summary} method reports the total number of pairs, the count of each type combination, and the number significant at 5\% and 1\%.
Calling \code{summary} on the resulting \code{inDep} data structure for the wine data, indicated that 73 of the pairs were significant at the 5\% and 1\% levels, suggesting strong dependence between all variates in the data.

The \code{plot} method produces an array of plots provided a set of indices giving the ranks of log $p$-values to display.
So, for example, specifying indices 1, 2, and 3 shows the three pairs with the smallest $p$-values, and presumably the strongest relationships.
For each index, a row of three subplots is produced.
First come two scatterplots of the raw data and the marginal ranks with random tie-breaking. To visually indicate categorical variates and distinguish their levels in the first scatterplot, lines and white space are placed proportionally to the distribution of the different levels and the points within are jittered. In the second, the white space is removed but the lines remain.
The third plot is a departure display 
shaded according to the (asymptotic) significance of the standardized Pearson residuals, coloured as described in Appendix \ref{app:colouring}.

\end{document}